\documentclass[]{interact}

\usepackage{pdflscape}
\usepackage{epstopdf}% To incorporate .eps illustrations using PDFLaTeX, etc.
\usepackage{subfigure}% Support for small, `sub' figures and tables

\usepackage[numbers,sort&compress]{natbib}% Citation support using natbib.sty
\bibpunct[, ]{[}{]}{,}{n}{,}{,}% Citation support using natbib.sty
\usepackage{color}
\usepackage{placeins}
\theoremstyle{plain}% Theorem-like structures

\theoremstyle{definition}

\theoremstyle{remark}

\begin{document}

%\jvol{00} \jnum{00} \jyear{2015} \jmonth{January}

%\articletype{GUIDE}

\title{Phase field modelling of precipitate morphologies in systems with tetragonal interfacial free energy anisotropy}

\author{
\name{Arijit Roy and M. P. Gururajan$^{\ast}$\thanks{$^\ast$Corresponding author. Email: gururajan.mp@gmail.com}}
\affil{Department of Metallurgical Engineering and Materials Science, Indian Institute of Technology Bombay, 
Powai, Mumbai 400076 INDIA.}
\received{July 2017}
}

\maketitle

\begin{abstract}

A wide variety of morphologies arise due to the tetragonal anisotropy in interfacial free energy.
In this paper, we report on a family of Extended Cahn-Hilliard (ECH) models for 
incorporating tetragonal anisotropy in interfacial free energy. We list the non-zero and independent 
parameters that are introduced in our model and list the constraints on them. For appropriate choice 
of these parameters, our model can produce a many of the morphologies seen in tetragonal systems 
such as di-pyramids, rods, plates and their truncated variants. We analyse these morphologies and 
show that they indeed are equilibrium morphologies consistent with the Wulff construction.

\end{abstract}


\section{Introduction} \label{Intro}

A wide variety of precipitate and crystalline morphologies are reported 
in tetragonal systems, during phase transformations 
and crystal growth, respectively. The physiochemical properties of 
such particles depends on their morphologies; for example, in the case
of TiO$_2$, the properties of the crystallites can be tuned by engineering 
its facets~\cite{LiuEtAl2011}; and, in the case of tin oxide whiskers, the
gas sensing properties differ depending on the morphology~\cite{Egashira1998}.
Hence, understanding the formation of such faceted morphologies is of great 
interest both from a scientific and application point of view.

In some cases, such as partially 
stabilized ZrO$_2$ and tetragonal ZrO$_2$ precipitates in
cubic ZrO$_2$ solid solution~\cite{Lanteri1986}, and
Cu precipitates in Fe-Cu system~\cite{Othen1994,Monzen2004}, 
the elastic energy plays
a key role (apart from the tetragonal crystal structure
of the systems). However, we do not consider the 
elastic stress effects in this paper (though, our model
can be extended to include the elastic stress effects). 

On the other hand, even in the absence of 
elastic stresses, the interphase interfacial free 
energy anisotropy that arises from the tetragonal
crystal structure of the participating phases could 
be important in determining the crystallite / precipitate / second 
phase morphologies; see for example,  metallic whiskers of 
$\beta$-Sn~\cite{Baskes1997,Boettinger2005, Mittemeijer2008}; PdO formed
by internal oxidation of Pd in SiO$_2$~\cite{ChenSchmidt1979}; single
crystal PbTiO$_3$ nanorods synthesised by solid state reaction~\cite{DengEtAl2005};
SnO$_2$ nano plates obtained using hydrothermal oxidation of SnS$_2$~\cite{WangXiao2009};
SnO thin plates obtained using oxidation in an aqueous solution~\cite{UchiyamaEtAl2006}
and plasma processing~\cite{SaitoEtAl2012};
faceted, short Sn whiskers grown on Sn finish surfaces~\cite{TuLi2005};
faceted morphology of In crystallites deposited on cleaved graphite surface~\cite{HeyraudMetois1986}
and on potassium chloride~\cite{Yanagihara1982};
tetragonal nano-rods and nano-tubes with cubic cross-section in $\alpha$-MnO$_2$
synthesised via the hydrothermal route~\cite{LuoEtAl2008}; and, faceted, tetragonal CeO$_2$ nanocrystals 
obtained via the hydrothermal synthesis~\cite{KanekoEtAl2007}.
Specifically, the studies  on tetragonal TiO$_2$ deserve special mention; a wide variety of morphologies such 
as tetragonal di-pyramids and their truncated versions are predicted based on interfacial
free energy calculations (at times based on first principle calculations) and faceted nanorods, 
plates, di-pyramids and their truncated versions
are obtained experimentally~\cite{BarnardEtAl2005,Liu2009,LiuEtAl2011,YangEtAl2012}.

Further, it is known that during solid-solid phase transformations, the relative crystalline 
 symmetry of the phases determines the symmetry of the interface~\cite{Kikuchi1962,Kikuchi1979}; 
for example, if the phases are $L1_2$-ordered (FCC derivative structure), the antiphase boundaries reflect 
tetragonal symmetry~\cite{Kikuchi1979}. Thus, even in non-tetragonal crystal systems, the
interfacial free energy can have tetragonal symmetry. 

Our aim in this paper is to study the morphology of precipitates in systems with tetragonal
interfacial free energy anisotropy -- using phase field models. 
Phase field models are ideal for the study of morphology of precipitates and crystallites;
recently, we have used the Extended Cahn-Hilliard (ECH) model to study the precipitate morphologies
in systems with cubic and hexagonal interfacial free energy 
anisotropy~\cite{NaniGururajan,ArijitEtAl,ArijitThesis}. 
Some aspects of the tetragonal symmetry (distinction between $c$ and $a$,$b$ axes) can be introduced 
using the classical Cahn-Hilliard equation
with second rank gradient free energy coefficient; and ECH models are not  
necessary~\cite{BraunEtAl,WangBanerjeeSuKhachaturyan,Vaithyanathan2002}.
However, using such second rank tensors, it is not possible to obtain some of the tetragonal
morphologies observed in the experiments such as plates and rods with square cross-section,
bi-pyramids and their truncated versions. Hence, in this paper, we use the ECH
model for systems with tetragonal interfacial free energy anisotropy and show that our phase field model
can indeed produce crystalline morphologies observed / predicted in these systems.

\section{Formulation}\label{Formulation}

In this paper, we briefly describe the ECH model~\cite{AbiHaider,LowengrubEtAl,NaniGururajan,ArijitEtAl}
to study the morphological evolution of precipitates in systems with tetragonal interfacial free energy anisotropy; 
the detailed formulation can be found elsewhere~\cite{ArijitThesis}.
Our description is based on a scalar, compositional order parameter ($c$); however, the extension
to non-conserved order parameters and to combinations of conserved and non-conserved order parameters
is straightforward. 

Let us consider a binary alloy with composition $c$; we define the the gradient $c_i$ (a vector), curvature $c_{ij}$
  (a second rank tensor) and aberration $c_{ijk}$ (a third rank tensor) of the (scalar) composition field as follows:
$c_i = \frac{\partial c}{\partial x_i}$; $c_{ij} = \frac{\partial^2 c}{\partial x_i \partial x_j}$;
$c_{ijk} = \frac{\partial^3 c}{\partial x_i \partial x_j \partial x_k}$. 

The free energy $F$, can be written as follows~\cite{ArijitThesis}:
\begin{eqnarray}
F = N_V \int_V f dV, \label{E:F_ch}
\end{eqnarray}  
where $N_V$ is the number of atoms / molecules in the given volume $V$ and $f$ is the free energy density given
by
\begin{eqnarray}
f &=& f_0(c) + P^{I}_{ij} c_{i} c_{j} + M^{I}_{ijkl} c_{i} c_{j} c_{k} c_{l}  + Q^{II}_{ijkl} c_{ij} c_{kl} \notag \\
&+& N^{I}_{ijklmn} c_i c_j c_k c_l c_m c_n + R^{I}_{ijklmn} c_{ijk} c_{lmn},  \label{E:f_extended}
\end{eqnarray}
where,$P^{I}_{ij} = \frac{1}{2!}\frac{\partial^2 f}{\partial c_{i} \partial c_{j}}\bigg|_{c_{0}}$ is a second
rank (coefficient) tensor; 
$M^{I}_{ijkl} = \frac{1}{4!}\frac{\partial^4 f}{\partial c_{i} \partial c_{j} \partial c_{k} \partial c_{l}}\bigg|_{c_0}$ and $Q^{II}_{ijkl} = \frac{1}{2!}\frac{\partial^2 f}{\partial c_{ij} \partial c_{kl}}\bigg|_{c_0}$ are 
fourth rank (coefficient) tensors; and,  $N^{I}_{ijklmn} = \frac{1}{6!} \frac {\partial^6 f}{\partial c_{i} \partial c_{j} \partial c_{k} \partial c_{l} \partial c_{m} \partial c_{n} }\bigg|_{c_0}$, $N^{II}_{ijklmn} = \frac{1}{5!} \frac {\partial^5 f}{\partial c_{ij} \partial c_{k}
\partial c_{l} \partial c_{m} \partial c_{n} }\bigg|_{c_0}$, and,  $R^{I}_{ijklmn} = \frac{1}{2!} \frac {\partial^2 f}{\partial c_{ijk} \partial c_{lmn} }\bigg|_{c_0}$ are the sixth rank (coefficient) tensors.
 
In writing the above expression, we have assumed that there are no elastic stresses and that the underlying
crystalline continuum is centrosymmetric; we have used Gauss theorem and positive-definiteness arguments; the
details of the derivation is similar to that of ECH for systems with cubic~\cite{AbiHaider,LowengrubEtAl} and 
hexagonal~\cite{NaniGururajan,ArijitEtAl} symmetries and can be found in~\cite{ArijitThesis}.

As described by Nye in his classic text~\cite{Nye}, using intrinsic symmetries (such as,
for sufficiently smooth $f$,  
$\frac{\partial^2 f}{\partial c_{i} \partial c_{j}} = \frac{\partial^2 f}{\partial c_{j} \partial c_{i}}$) and
commutative properties (such as $c_i c_j = c_j c_i$), it is possible to reduce the number of independent
components of the various tensors. Specifically, we can reduce the independent components of 
$P^{I}_{ij}$ from 9 to 6, that of $M^{I}_{ijkl}$ from 81 to 15, that is $Q^{II}_{ijkl}$ from 81 to 21,
that of $N^{I}_{ijklmn}$ from 729 to 28, and, that of $R^{I}_{ijklmn}$ from 729 to 55 (see below for
the enumeration for sixth rank tensors).

It is easier to identify and enumerate the independent components and their numbers, respectively, for the fourth
and sixth rank tensors if we represent them in the reduced matrix notation. Such notation, known as Voigt notation
(represented as matrices) is available
in Nye for fourth rank tensors; here we show a similar notation for sixth rank tensors.

Since both $N^{I}_{ijklmn}$ and and $R^{I}_{ijklmn}$ are invariant under the exchange of 
$i$, $j$ and $k$, or $l$, $m$ and $n$ indices, we collect all possible independent ways in which 
$i$, $j$ and $k$ indices can appear and reduce them to a single index as follows:
\begin{align}
&111 \to 1&  &222 \to 4&  &333 \to 7& &123 \to 0. \notag \\  
&221 \to 2&  &112 \to 5&  &113 \to 8 \label{E:u_ijk}\\  
&331 \to 3&  &332 \to 6&  &223 \to 9 \notag
\end{align}
Since each pair of $ijk$-indices can be arranged in 10 independent ways, 
a sixth rank tensor with two such pair of indices, can be expressed in a $10\times 10$ matrix form.

Thus,  we can write $N^{I}_{ijklmn}$ 
in $N^{I}_{\alpha \beta}$ (or similarly $R^{I}_{ijklmn}$ in $R^{I}_{\alpha \beta}$) form as follows:
\FloatBarrier
\begin{table}[h]
\centering
\begin{tabular}{c c c c} 
&&&\\
 && $ \begin{bmatrix}
N^{I}_{11}\enspace &N^{I}_{12}\enspace    &N^{I}_{13}\enspace    &N^{I}_{14}\enspace   &N^{I}_{15}\enspace   &N^{I}_{16}\enspace   &N^{I}_{17}\enspace    &N^{I}_{18}\enspace    &N^{I}_{19}\enspace    &N^{I}_{10}\enspace      \\[1em]
\bullet\enspace &N^{I}_{22}\enspace    &N^{I}_{23}\enspace    &N^{I}_{24}\enspace   &N^{I}_{25}\enspace   &N^{I}_{26}\enspace   &N^{I}_{27}\enspace   &N^{I}_{28}\enspace   &N^{I}_{29}\enspace   &N^{I}_{20}\enspace    \\[1em]
\bullet\enspace &\bullet\enspace    &N^{I}_{33}\enspace    &N^{I}_{34}\enspace   &N^{I}_{35}\enspace   &N^{I}_{36}\enspace   &N^{I}_{37}\enspace   &N^{I}_{38}\enspace   &N^{I}_{39}\enspace   &N^{I}_{30}\enspace    \\[1em]
\bullet\enspace &\bullet\enspace    &\bullet\enspace    &N^{I}_{44}\enspace   &N^{I}_{45}\enspace   &N^{I}_{46}\enspace   &N^{I}_{47}\enspace   &N^{I}_{48}\enspace   &N^{I}_{49}\enspace   &N^{I}_{40}\enspace    \\[1em]
\bullet\enspace &\bullet\enspace    &\bullet\enspace    &\bullet\enspace   &N^{I}_{55}\enspace   &N^{I}_{56}\enspace   &N^{I}_{57}\enspace   &N^{I}_{58}\enspace   &N^{I}_{59}\enspace   &N^{I}_{50}\enspace   \\[1em]
\bullet\enspace &\bullet\enspace    &\bullet\enspace    &\bullet\enspace   &\bullet\enspace   &N^{I}_{66}\enspace   &N^{I}_{67}\enspace      &N^{I}_{68}\enspace      &N^{I}_{69}\enspace      &N^{I}_{60}\enspace     \\[1em]
\bullet\enspace &\bullet\enspace    &\bullet\enspace    &\bullet\enspace   &\bullet\enspace   &\bullet\enspace   &N^{I}_{77}\enspace   &N^{I}_{78}\enspace   &N^{I}_{79}\enspace   &N^{I}_{70}\enspace    \\[1em]
\bullet\enspace &\bullet\enspace    &\bullet\enspace    &\bullet\enspace   &\bullet\enspace   &\bullet\enspace   &\bullet\enspace   &N^{I}_{88}\enspace   &N^{I}_{89}\enspace   &N^{I}_{80}\enspace    \\[1em]
\bullet\enspace &\bullet\enspace    &\bullet\enspace    &\bullet\enspace   &\bullet\enspace   &\bullet\enspace   &\bullet\enspace   &\bullet\enspace   &N^{I}_{99}\enspace   &N^{I}_{90}\enspace   \\[1em]
\bullet\enspace &\bullet\enspace    &\bullet\enspace    &\bullet\enspace   &\bullet\enspace   &\bullet\enspace   &\bullet\enspace      &\bullet\enspace      &\bullet\enspace      &N^{I}_{00}\enspace        \end{bmatrix}$  &  \\
&&& \\
\end{tabular}\\
\end{table}
\FloatBarrier
The lower diagonal components satisfying intrinsic symmetry are represented by $\bullet$. Thus, the total number of independent 
components (of $N^{I}_{ijklmn}$ and $R_{ijklmn}$) obtained by adding the diagonal and upper diagonal (as lower diagonal terms are 
numerically equal to upper diagonal terms by intrinsic symmetry) term of the $10\times10$ matrix is 55. The tensor 
$N^{I}_{ijklmn}$ has an additional symmetry: it is invariant under the exchange of all the $i$, $j$, $k$, $l$, $m$ and $n$; 
since there are 27 such components which are invariant under the exchange of indices, the total number of independent components 
for $N^{I}_{ijklmn}$ is $55-27=28$. 

Crystalline symmetry arguments are very powerful in reducing the total 
number of non-zero components and independent components; see Nani and Gururajan~\cite{NaniGururajan} 
for isotropic, cubic and hexagonal systems. In this paper, we use the direct inspection method of Nye~\cite{Nye} 
(which is valid for all crystal classes except trigonal and hexagonal) to reduce the the total number
of non-zero and independent components of the coefficient tensors assuming tetragonal symmetry. 

\begin{figure}[htpb]
\centering
\includegraphics[height=2.5in,width=2.5in]{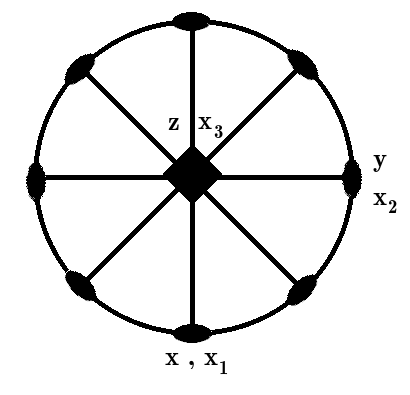}
\caption{Choice of axes and the allowed symmetry operations for ditetragonal-dipyramidal crystal class ($4/mmm$).}\label{F:4mmm}
\end{figure}

The characteristic symmetry for the tetragonal system is 4-fold~\cite{Nye}. There are seven tetragonal crystal classes;
in this paper, we present the results for the ditetragonal-dipyramidal crystal class (represented by the symbol $4/mmm$ 
in International Tables and by $\mathrm{D}_{4h}$ in Schoenflies). The Fig.~\ref{F:4mmm} shows the choice of axes and 
the symmetry operations for $4/mmm$.
There exist three predominant axes of symmetry; namely, two 2-fold axes of symmetry or diad axes and one 4-fold axis of 
symmetry or tetrad axis. Using these symmetry operations and demanding that the tensors remain invariant under these
operations, the independent and non-zero components for various tensors can be deduced. 

In the case of $\mathbf{P^I}$ tensor, the only non-zero surviving components are $P^I_{11}$, $P^I_{22}$ and $P^I_{33}$.
Of these, there are only two independent components, namely, $P^I_{11}$ and $P^I_{33}$. Let us call these two 
components as $p_1$ and $p_2$:
\begin{eqnarray}
p_1 &=& P^{I}_{11} = P^{I}_{22} \notag \\
p_2 &=& P^{I}_{33} \label{E:P_reduced_components}
\end{eqnarray}

The free energy contribution originating from the second rank gradient tensorial term can be obtained by the double contraction
(two inner multiplications) $P^{I}_{ij} c_i c_j$. Let us call the resultant scalar as $P^2$. 
\begin{eqnarray}
P^2 &=& p_1 (c_1^2 + c_2^2) + p_3 c_3^2 . \label{E:secondP_tensor_pol}
\end{eqnarray}
Thus, we have reduced our second rank tensor term of the free energy expansion into a polynomial in the gradient terms.

Similarly, the total number of independent components for a fourth rank coefficient tensor in systems with tetragonal symmetry 
is reduced to 6: $M^I_{11}$, $M^I_{33}$, $M^I_{44}$, $M^I_{66}$, $M^I_{12}$, and, $M^I_{13}$. The fourth rank curvature 
coefficient tensor $Q^{II}_{ijkl}$ also has the same non-zero tensor components as $M^{I}_{ijkl}$. Similar to the case of second 
rank gradient tensor $\mathbf{P^I}$, we rename the independent and non-zero components 
of $\mathbf{M^{I}}$ and $\mathbf{Q^{II}}$ 
as listed in Table.~\ref{T:ind_nonzero_MQ}.
\begin{table}[tbh] 
\centering
\caption{\label{T:ind_nonzero_MQ} Independent and non-zero components of fourth rank gradient tensor $\mathbf{M^{I}}$ and 
curvature tensor $\mathbf{Q^{II}}$.}
\begin{tabular}{|c|c|c|}
\hline \hline
{\bf Components } & {\bf for $\mathbf{M^I}$} & {\bf for $\mathbf{Q^{II}}$ } \\
\hline \hline
$M^{I}_{11} =M^{I}_{22}$ & $m_1$&$q_1$ \\
\hline \hline
$M^{I}_{33} $ & $m_2$&$q_2$ \\
\hline \hline
$M^{I}_{12}  (\equiv M^I_{21})$ & $m_3$&$q_3$ \\
\hline \hline
$M^{I}_{13} = M^{I}_{23} (\equiv M^I_{31}, M^I_{32})$ & $m_4$&$q_4$ \\
\hline \hline
$M^{I}_{44} = M^{I}_{55} $ & $m_5$&$q_5$ \\
\hline \hline
$M^{I}_{66} $ & $m_6$&$q_6$ \\
\hline \hline
\end{tabular}
\end{table}

Given the independent and non-zero components of these tensors, it is straightforward to obtain the 
free energy contributions stemming from the inner products $M^{I}_{ijkl}c_{i}c_{j}c_{k}c_{l}$ (say, $P^4$)
and $Q^{II}_{ijkl}c_{ij}c_{kl}$ (say, $Q$) in Eq.~\ref{E:f_extended}; this results in the polynomial form as follows (see
Table~\ref{T:MQ_phi_deriv} for the detailed enumeration).
\begin{eqnarray}
P^4 &=& m_1 (c^4_{1} + c^4_{2}) + m_2 c^4_{3} \notag \\
&+& 2 m_3 c^2_{1} c^2_{2} + 2 m_4 c^2_{3} (c^2_{1} + c^2_{2}) \notag \\
&+& m_5 c^2_{3} (c^2_{1} + c^2_{2}) + m_6 c^2_{1} c^2_{2}. \label{E:fourM_tensor_pol0}
\end{eqnarray}
Note that the factors of 2 come from the symmetry of the matrix (off diagonal terms on the lower half of the matrix form).
Clubbing the coefficients for $c^2_{1} c^2_{2}$ and $c^2_{3} (c^2_{1} + c^2_{2})$, we can rewrite Eq.~\ref{E:fourM_tensor_pol0} as follows:
\begin{eqnarray}
P^4 &=& m_1 (c^4_{1} + c^4_{2}) + m_2 c^4_{3} \notag \\
&+& (2 m_3 + m_6) c^2_{1} c^2_{2} + (2 m_4 + m_5) c^2_{3} (c^2_{1} + c^2_{2}).
\end{eqnarray}
Thus, without loss of generality, we can assume that there are only four constants are needed for $\mathbf{M^{I}}$; the terms
$m_3$ and $m_6$ and $m_4$ and $m_5$ can be clubbed together; to keep the notation simple, we use $m_3$ to
denote $2 m_3 + m_6$ and $m_4$ to denote $2 m_4 + m_5$; thus, 
we obtain
\begin{equation}
P^4 = m_1 (c^4_{1} + c^4_{2}) + m_2 c^4_{3} + m_3 c^2_{1} c^2_{2} + m_4 c^2_{3} (c^2_{1} + c^2_{2}). \label{E:fourM_tensor_pol}
\end{equation}
\begin{eqnarray}
Q &=& q_1 (c_{11}^2 + c_{22}^2) + q_2 c_{33}^2 \notag \\ 
&+& 2 q_3 c_{11} c_{22} + 2 q_4 (c_{11} + c_{22}) c_{33} \notag \\
&+& q_5 (c_{13}^2 + c_{23}^2) + q_6 c_{12}^2. \label{E:fourQ_tensor_pol}
\end{eqnarray}
As opposed to $\mathbf{M^{I}}$, we need all the six constants for $\mathbf{Q}$ which is a consequence of the lower
intrinsic symmetry of $\mathbf{Q}$, namely, all possible exchanges of the indices are not allowed.

\begin{table}[tbh] 
\centering
\caption{\label{T:MQ_phi_deriv} Listing of $c$-derivatives that multiply $\mathbf{M^{I}}$ and $\mathbf{Q^{II}}$ tensor coefficients.}
\begin{tabular}{|c|c|c|c|}
\hline \hline
{\bf Component} & {\bf Corresponding }&{\bf Component} & {\bf Corresponding }  \\
 {\bf of $\mathbf{M^I}$} & {\bf $c$-derivatives} & {\bf of $\mathbf{Q^I}$} & {\bf $c$-derivatives}\\
\hline \hline
$m_1$& $c_{1}^4$ &$q_1$& $c_{11}^2$  \\
	 & $c_{2}^4$ &	  & $c_{22}^2$  \\
\hline \hline
$m_2$& $c_{3}^4$ &$q_2$& $c_{33}^2$  \\
\hline \hline
$m_3$& $c_{1}^2c_{2}^2$ &$q_3$& $c_{11}c_{22}$  \\
\hline \hline
$m_4$& $c_{1}^2c_{3}^2$ &$q_4$& $c_{11}c_{33}$  \\
	 & $c_{2}^2c_{3}^2$ &	  & $c_{22}c_{33}$  \\
\hline \hline
$m_5$& $c_{2}^2c_{3}^2$ &$q_5$& $c_{23}c_{23}$  \\
	 & $c_{1}^2c_{3}^2$ &	  & $c_{13}c_{13}$  \\
\hline \hline
$m_6$& $c_{1}^2c_{2}^2$ &$q_6$& $c_{12}c_{12}$  \\
\hline \hline
\end{tabular}
\end{table}

The sixth rank tensor $\mathbf{N^{I}}$ is reduced to the form shown in Table~\ref{T:N_ijklmn_2}, and, the non-zero and independent 
components of the sixth rank tensorial coefficient are listed in Table.~\ref{T:ind_nonzero_NR}. The non-zero and independent components of $\mathbf{R^{I}}$ are same as $\mathbf{N^{I}}$ tensor. Thus, after employing the crystalline symmetry arguments, 
we are able to reduce the independent components of $\mathbf{N^{I}}$ and $\mathbf{R^{I}}$ tensors from 55 to 11.  

\begin{table}[htbp]\caption{\label{T:N_ijklmn_2} Non-zero and independent components of the sixth rank tensor $\mathbf{N^{I}}$ in matrix form.}
\centering
\begin{tabular}{c c} 
&\\
$  \begin{array}{c} 
  \\[1em]
    \\[1em]
 \mathbf{N^{I}} = \\[1em]
   \\[1em]
     \\[1em]
  \\[1em]
 \end{array}$& $ \begin{bmatrix}
N^{I}_{11}\enspace &N^{I}_{12}\enspace    &N^{I}_{13}\enspace    & \enspace   & \enspace   & \enspace   & \enspace    & \enspace    & \enspace    & \enspace      \\[1em]
\bullet\enspace &N^{I}_{22}\enspace    &N^{I}_{23}\enspace    & \enspace   & \enspace   & \enspace   & \enspace   & \enspace   & \enspace   & \enspace    \\[1em]
\bullet\enspace &\bullet\enspace    &N^{I}_{33}\enspace    & \enspace   & \enspace   & \enspace   & \enspace   & \enspace   & \enspace   & \enspace    \\[1em]
 \enspace & \enspace    & \enspace    &N^{I}_{11}\enspace   &N^{I}_{12}\enspace   &N^{I}_{13}\enspace   & \enspace   & \enspace   & \enspace   & \enspace    \\[1em]
 \enspace & \enspace    & \enspace    &\bullet\enspace   &N^{I}_{22}\enspace   &N^{I}_{23}\enspace   & \enspace   & \enspace   & \enspace   & \enspace   \\[1em]
 \enspace & \enspace    & \enspace    &\bullet\enspace   &\bullet\enspace   &N^{I}_{33}\enspace   & \enspace      & \enspace      & \enspace      & \enspace     \\[1em]
 \enspace & \enspace    & \enspace    & \enspace   &\enspace   & \enspace   &N^{I}_{77}\enspace   &N^{I}_{78}\enspace   &N^{I}_{78}\enspace   & \enspace    \\[1em]
 \enspace & \enspace    & \enspace    & \enspace   & \enspace   & \enspace   &\bullet\enspace   &N^{I}_{88}\enspace   &N^{I}_{89}\enspace   & \enspace    \\[1em]
 \enspace & \enspace    & \enspace    & \enspace   & \enspace   & \enspace   &\bullet\enspace   &\bullet\enspace   &N^{I}_{88}\enspace   & \enspace   \\[1em]
 \enspace & \enspace    & \enspace    & \enspace   & \enspace   & \enspace   & \enspace      & \enspace      & \enspace      &N^{I}_{00}\enspace        \end{bmatrix}$    \\
& \\
\end{tabular}\\
\end{table}

\begin{table}[tbh] 
\centering
\caption{\label{T:ind_nonzero_NR} Independent and non-zero components of sixth rank gradient tensor $\mathbf{N^{I}}$ and curvature tensor $\mathbf{R^I}$.}
\begin{tabular}{|c|c|c|c|c|c|}
\hline \hline
{\bf Components } & {\bf for $\mathbf{N^I}$} & {\bf for $\mathbf{R^I}$ } 
& {\bf Components } & {\bf for $\mathbf{N^I}$} & {\bf for $\mathbf{R^I}$ }\\
\hline \hline
$N^{I}_{11} =N^{I}_{44}$ & $n_1$&$r_1$ &
$N^{I}_{77} $ & $n_2$&$r_2$ \\
\hline \hline
$N^{I}_{22} = N^{I}_{55}$ & $n_3$&$r_3$&
$N^{I}_{89} $ & $n_4$&$r_4$ \\
\hline \hline
$N^{I}_{13} =N^{I}_{46}$ & $n_5$&$r_5$ &
$N^{I}_{23} =N^{I}_{56}$ & $n_6$&$r_6$ \\
\hline \hline
$N^{I}_{88} = N^{I}_{99}$ & $n_7$&$r_7$&
$N^{I}_{33} =N^{I}_{66}$ & $n_8$&$r_8$ \\
\hline \hline
$N^{I}_{78} =N^{I}_{79}$ & $n_9$&$r_9$ &
$N^{I}_{12} =N^{I}_{45}$ & $n_{10}$&$r_{10}$ \\
\hline \hline
$N^{I}_{00} $ & $n_{11}$&$r_{11}$ &&&\\
\hline \hline
\end{tabular}
\end{table}

We obtain the scalar contribution to the free energy from the two sixth rank tensorial coefficients 
$N^{I}_{ijklmn}$ and $R^{I}_{ijklmn}$ by the inner product of these tensors with the gradients 
($c_{i} c_{j} c_{k} c_{l} c_{m} c_{n}$) and aberrations ($c_{ijk} c_{lmn}$)
respectively. Let the free energy contribution originating from gradient term  be called $P^6$ and that 
from aberration term be called as $R$. The detailed enumeration of these terms is shown in 
Table.~\ref{T:NR_phi_deriv}.
\begin{table}[ptbh] 
\centering
\caption{\label{T:NR_phi_deriv} Listing of $c$-derivatives that multiply $\mathbf{N^{I}}$ and $\mathbf{R^{I}}$ tensor coefficients.}
\begin{tabular}{|c|c|c|c|}
\hline \hline
{\bf Component} & {\bf Multiplied }&{\bf Component} & {\bf Multiplied }  \\
 {\bf of $\mathbf{N^I}$} & {\bf $c$-derivatives} & {\bf of $\mathbf{R^I}$} & {\bf $c$-derivatives}\\
\hline \hline
$n_1$& $c_{1}^6$ &$r_1$& $c_{111}^2$  \\
	 & $c_{2}^6$ &	  & $c_{222}^2$  \\
\hline \hline
$n_2$& $c_{3}^6$ &$r_2$& $c_{333}^2$  \\
\hline \hline
$n_3$& $c_{1}^2c_{2}^4$ &$r_3$& $c_{221}^2$  \\
	 & $c_{1}^4c_{2}^2$ &	  & $c_{112}^2$  \\
\hline \hline
$n_4$& $c_{1}^2c_{2}^2c_{3}^2$ &$r_4$& $c_{113}c_{223}$  \\
\hline \hline
$n_5$& $c_{1}^4c_{3}^2$ &$r_5$& $c_{111}c_{331}$  \\
	 & $c_{2}^4c_{3}^2$ &	  & $c_{222}c_{332}$  \\
\hline \hline
$n_6$& $c_{1}^2c_{2}^2c_{3}^2$ &$r_6$& $c_{221}c_{331}$  \\
	 & $c_{1}^2c_{2}^2c_{3}^2$ &	  & $c_{112}c_{332}$  \\
\hline \hline
$n_7$& $c_{1}^4c_{3}^2$ &$r_7$& $c_{113}c_{113}$  \\
	 & $c_{2}^4c_{3}^2$ &	  & $c_{223}c_{223}$  \\
\hline \hline
$n_8$& $c_{1}^2c_{3}^4$ &$r_8$& $c_{331}^2$  \\
	 & $c_{2}^2c_{3}^4$ &	  & $c_{332}^2$  \\
\hline \hline
$n_9$& $c_{1}^2c_{3}^4$ &$r_9$& $c_{333}c_{113}$  \\
	 & $c_{2}^4c_{3}^4$ &	  & $c_{333}c_{223}$  \\
\hline \hline
$n_{10}$& $c_{1}^4c_{2}^2$ &$r_{10}$& $c_{111}c_{221}$  \\
	 & $c_{1}^2c_{2}^4$ &	  & $c_{222}c_{112}$  \\
\hline \hline
$n_{11}$& $c_{1}^2c_{2}^2c_{3}^2$ &$r_{11}$& $c_{123}^2$  \\
\hline \hline
\end{tabular}
\end{table}

Using Table.~\ref{T:NR_phi_deriv}, we can easily build the free energy in polynomial form. 
To derive the sixth order gradient free energy polynomial ($P^6$), we multiply the elements of column 
1 with those in 2 and sum all such terms:
\begin{eqnarray}
P^6 &=& n_1\ (c^6_{1} + c^6_{2}) + n_2\ c^6_{3} \notag \\
&+& 2 n_{10}\ c^2_1 c^2_2(c^2_1 + c^2_2) +  2 n_5\ c^2_3 (c^4_1 + c^4_2) + n_3\ c^2_1 c^2_2(c^2_1 + c^2_2) \notag \\
&+& 4 n_6\ c^2_1 c^2_2 c^2_3 + n_8\ c^4_3 (c^2_1 + c^2_2) + 2 n_9\ c^4_3 (c^2_1 + c^2_2) + n_7\ c^2_3 (c^4_1 + c^4_2) \notag \\ 
&+& 2 n_4\ c^2_1 c^2_2 c^2_3 + n_{11}\ c^2_1 c^2_2 c^2_3. \label{E:sixN_tensor_pol0}
\end{eqnarray}

Clubbing the coefficients of $c^2_1 c^2_2(c^2_1 + c^2_2)$, $c^2_3 (c^4_1 + c^4_2)$, $c^4_3 (c^2_1 + c^2_2)$ and $c^2_1 c^2_2 c^2_3$, we get
\begin{eqnarray}
P^6 &=& n_1\ (c^6_{1} + c^6_{2}) + n_2\ c^6_{3} \notag \\
&+& (2 n_{10} + n_3)\ c^2_1 c^2_2(c^2_1 + c^2_2) +  (2 n_5 + n_7)\ c^2_3 (c^4_1 + c^4_2) \notag \\
&+& (n_8 + 2 n_9)\ c^4_3 (c^2_1 + c^2_2) \notag \\ 
&+& (4 n_6 + 2 n_4 + n_{11})\  c^2_1 c^2_2 c^2_3. \label{E:sixN_tensor_pol}
\end{eqnarray}
Thus, without loss of generality, we can assume that six independent components are needed for $\mathbf{N^I}$.

Similarly, the aberration term $R$ is written as follows:
\begin{eqnarray}
R &=& r_1\ (c^2_{111} + c^2_{222}) + r_2\ c^2_{333} \notag \\
&+& 2 r_{10}\ (c_{111} c_{221} + c_{222} c_{112}) +  2 r_5\ (c_{111} c_{331} + c_{222} c_{332})  \notag \\
&+& r_3\ (c_{221} c_{221} + c_{112} c_{112}) + 2 r_6\ (c_{221} c_{331} + c_{112} c_{332})  \notag \\ 
&+& r_8\ (c_{331} c_{331} + c_{332} c_{332}) + 2 r_9\ (c_{333} c_{113} + c_{333} c_{223})  \notag \\
&+& r_7\ (c_{113} c_{113} + c_{223} c_{223}) + 2 r_4\ c_{113} c_{223} + r_{11}\ c_{123} c_{123}. 
\label{E:sixR_tensor_pol0}
\end{eqnarray}
Thus, in contrast to $\mathbf{N^{I}}$, we need eleven independent components to describe $\mathbf{R^{I}}$.

Thus, we have managed to express the free energy in polynomial form in terms of the gradient, 
curvature and aberration components:
\begin{eqnarray} \label{FinalFreeEnergy}
f & = &  f_0 + P(c_1,c_2,c_3) + Q(c_{11},c_{22},c_{33},c_{12},c_{23},c_{13}) \\  \nonumber
& & + R(c_{111},c_{222},c_{333},c_{112},c_{113},c_{221},c_{223},c_{331},c_{332},c_{123}),
\end{eqnarray} 
where $f_0$ is the bulk free energy density (typically assumed to be a double well potential, 
namely, $A c^2 (1-c)^2$), 
and, $P$, $Q$ and $R$ are homogeneous polynomials; further,    
$P$ consists of three parts: homogeneous polynomials of orders 2 (denoted by P$^2$), 4 (denoted by P$^4$) and 
6 (denoted by P$^6$); $Q$ and $R$ are homogeneous polynomials of order 2. 
In Table~\ref{Table1}, we list the forms of these polynomials. The coefficients of these polynomials are 
assumed to be constants, and, as indicated below,
by choosing them appropriately, we incorporate the tetragonal anisotropy in interfacial energy. 
\begin{table}[tbh] 
\centering
\caption{\label{Table1} Polynomials $P$, $Q$ and $R$.}
\begin{tabular}{|c|c|c|}
\hline \hline
{\bf Polynomial } & {\bf Order} & {\bf Form} \\
\hline \hline
$P^2$ & 2&$p_1 (c_1^2 + c_2^2) + p_2 c_3^2$ \\
\hline
$P^4$& 4 & $m_1 (c_1^4+c_2^4) + m_2 c_3^4 + 2 m_3 c_1^2 c_2^2  + 2 m_4 (c_1^2 + c_2^2) c_3^2$ \\
\hline
$P^6$& 6 & $n_1 (c_1^2+c_2^2)^3 + n_2 c_3^6$ \\ 
&&+$n_3 (c_1^2+c_2^2)(c_1^2 c_2^2) + n_4(c_1^2 +c_2^2)^2 c_3^2 $ \\
&&+$ n_5 (c_1^2+c_2^2) c_3^4 + n_6 c_1^2 c_2^2 c_3^2 $ \\
\hline \hline
$Q$ & 2 & $q_1 (c_{11}^2 +c_{22}^2) + q_2 c_{33}^2$ \\
 && $+ 2 q_{3} c_{11}c_{22} + 2 q_4 (c_{11} + c_{22}) c_{33}$  \\
 && $+q_5 (c_{13}^2 + c_{23}^2) +  q_6 c_{12}^2$ \\
\hline \hline
$R$ & 2 & $r_1 (c_{111}^2 + c_{222}^2) + r_2 c_{333}^2 + r_{3} (c_{112}^2 + c_{221}^2)$ \\
&& $+ 2 r_4 c_{113} c_{223} + 2 r_5 (c_{111}c_{331} + c_{222} c_{332}) $ \\
&& $+ 2 r_6 (c_{112} c_{332} + c_{221}  c_{331}) + r_7 (c_{113}^2+ c_{223}^2)$ \\
&&  $ + r_8 (c_{331}^2 + c_{332}^2) + 2 r_9 c_{333}(c_{113} + c_{223}) $ \\
&& $+ 2 r_{10}  (c_{111}c_{221} + c_{222}c_{112}) + r_{11}  c_{123}^2$ \\
\hline \hline
\end{tabular}
\end{table}

The polynomials $P$, $Q$ and $R$ represent the contribution of interfacial free energy; hence,
we demand their term-wise positive definiteness. Such a demand helps us derive
the constraints on the independent components. Once again, such an exercise for
second and fourth rank tensors have been carried out in Nye~\cite{Nye} and we have
extended the algebra to sixth rank tensors. The constraints on
the independent constants for the various tensor coefficients are summarised in Table~\ref{ConstraintsTable}.
\begin{table}[ptbh]
\centering
\caption{\label{ConstraintsTable} Constraints on the coefficients of the polynomials listed in Table~\ref{Table1}: tetragonal anisotropy.}
\begin{tabular}{|c|c|}
\hline \hline
 & {\bf Tetragonal} \\
\hline \hline
$P^2$  & $p_1$,$p_2$ $\geq 0$  \\
\hline
$P^4$ & $m_1 \geq 0$;$m_2 \geq 0$ \\
 & $m_3 \ge |m_1|$;$m_4 \ge |\sqrt{m_2(m_1 + m_3)}|$\\
\hline
$P^6$ &$n_1 \ge 0$;$n_2\ge 0$;$n_3\ge 0$;$n_4\ge 0$; \\
&$n_5\le |\sqrt{\frac{1}{2}n_2 n_4}|$;$n_6\le n_4$;\\
\hline \hline
$Q$ &$q_1 \ge 0$;$q_2\ge 0$;$q_5 \ge 0$;$q_6\ge 0$ \\
 & $q_3 \ge |q_1|$;$q_4 \ge |\sqrt{q_2(q_1 + q_3)}|$\\
\hline \hline
$R$ & $r_1 \ge 0$;$r_2\ge 0$;$r_3 \ge 0$;$r_7\ge 0$;$r_8\ge 0$;$r_{11}\ge 0$\\
&$r_5$ and $r_6$ should have same sign;\\
& $r_4\le r_7$;$r_9\le |\sqrt{\frac{1}{2}r_2 (r_4 + r_7)}|$\\
& $r_{10}\le |\sqrt{r_1 r_3}|$;$r_{10} r_5 r_6 \ge  \frac{1}{2} \Big[ r_3 \left( r_5\right)^2 + r_1\left( r_6\right)^2\Big]$\\
& $r_{12}=r_1$;$r_{13} = r_{3}$;$r_{14} = r_{10}$\\
\hline \hline
\end{tabular}
\end{table}

\subsection{Integrity basis and polynomial form}

Nani and Gururajan~\cite{NaniGururajan} show that the polynomials in gradients can be directly written down using
the integrity basis approach of Smith et al~\cite{Smith1962} for any of the 32 crystal classes -- using the
building blocks of these polynomials called integrity basis. However, for the higher order polynomials composed 
of curvature terms ($c_{ij} c_{kl}$), or the polynomials composed of aberration terms ($c_{ijk} c_{lmn}$), 
there is no integrity basis and our approach outlined above is to be used. As an example, we show the formulation
of the $P^4$ polynomial using the integrity basis approach; more details of the approach and the formulation
of the $P^6$ polynomial can be found in~\cite{ArijitThesis}.

Let us consider the tetragonal-ditetragonal-dipyramidal crystal class. From Smith et al~\cite{Smith1962}, 
the integrity bases are as follows: $c_1 ^2 + c_2^2$, $c_3^2$ and $c_1 ^2 c_2^2$. For the fourth order gradient contribution,
then, the free energy polynomial is given as
\begin{eqnarray}
P^4 &=&m_1^{'} (c_1^2 + c_2^2)^2 + m_2^{'} c_3^4  + m_3^{'} c_1^2 c_2^2 + m_4^{'} (c_1^2 + c_2^2) c_3^2 \notag \\
&=& m_1^{'} (c_1^4 + c_2^4) + m_2^{'} c_3^4  + (2 m_1^{'} + m_3^{'}) c_1^2 c_2^2+ m_4^{'} (c_1^2 + c_2^2) c_3^2,
 \label{E:tetra4_free-eng-final-0}
\end{eqnarray}
where $m_1^{'}$, $m_2^{'}$, $m_3^{'}$ and $m_4^{'}$ are the coefficients of the free energy polynomial. By comparing Eq.~\ref{E:fourM_tensor_pol} with Eq.~\ref{E:tetra4_free-eng-final-0}, we can see the following relationships:
\begin{align}
&m_{1}^{'}=m_1&  & 2 m_{1}^{'} + m_{3}^{'}= m_3 \notag \\  
&m_{2}^{'}=m_2&  & m_{4}^{'}= m_4  \notag
\end{align}
As $m_1^{'}$, $m_2^{'}$ and $m_4^{'}$ are equal to $m_1$, $m_2$ and $m_4$ respectively, constraints are also identical 
to non-primed components. Further, knowing the constraint on $m_1^{'}$ and $m_3$, we can obtain the constraint
on $m_3^{'}$. 

Similarly, we can build the sixth order polynomial in tetragonal symmetry.

\subsection{Anisotropy of the higher order polynomials} \label{S:Ch3-note_higher_poly}

As indicated elsewhere (Roy et al~\cite{ArijitEtAl} and Roy~\cite{ArijitThesis}), by plotting the 
polynomials listed in Table.~\ref{Table1} (after normalising -- in real space in the case of polynomials 
based on gradients and in reciprocal space in the case of polynomials based on curvature and aberration terms --
with the primes indicating normalisation) for 
various choices of the parameters, the anisotropy that would be incorporated for the given parameters
can be better understood. Note that though these polynomials are built by considering the symmetry of the free 
energy, they are also useful in describing any direction dependent property. 

\begin{figure}[htpb]
\centering
\subfigure[]{\includegraphics[height=2.16in,width=2.16in]{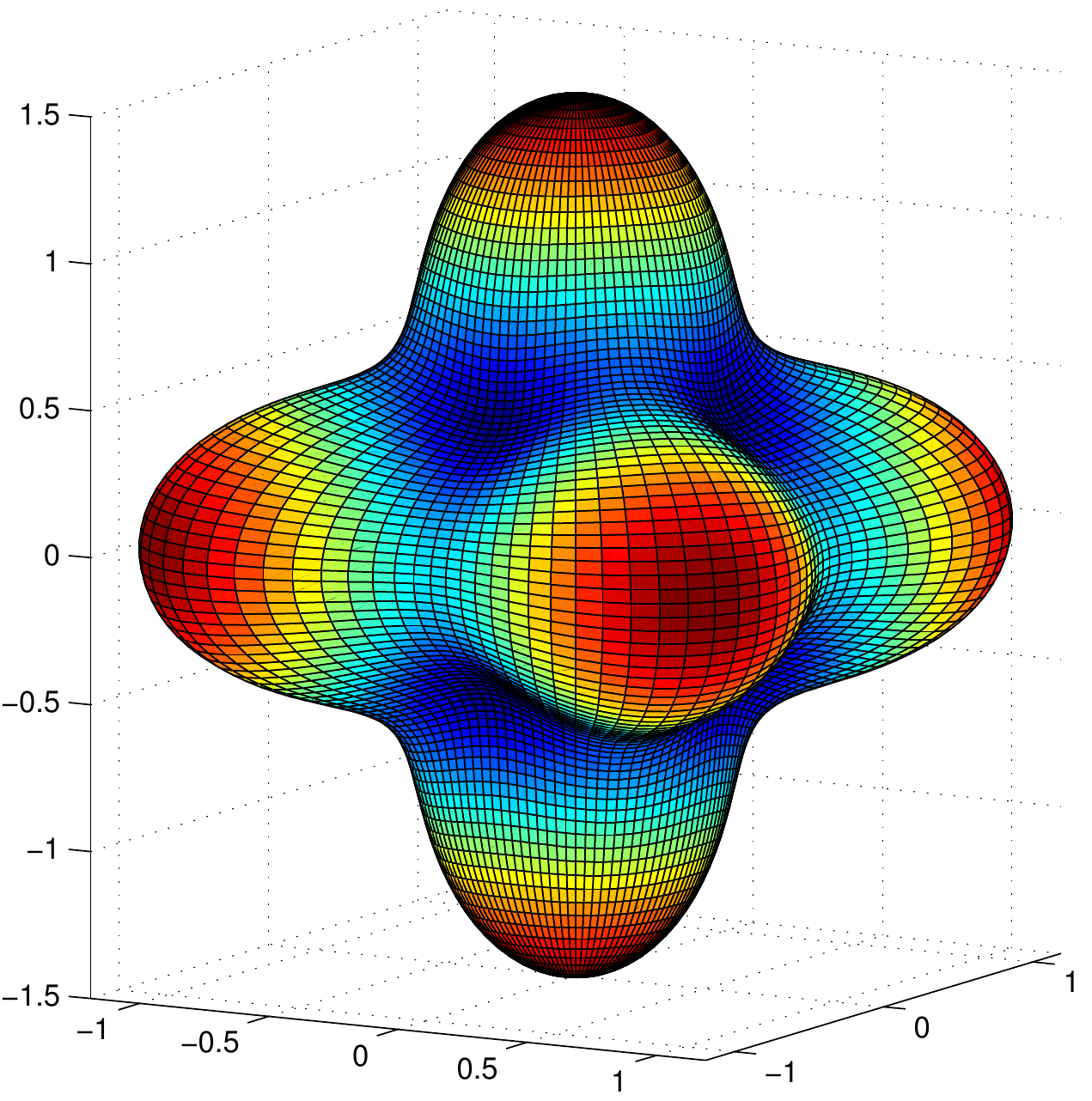}}
\subfigure[]{\includegraphics[height=2.16in,width=2.16in]{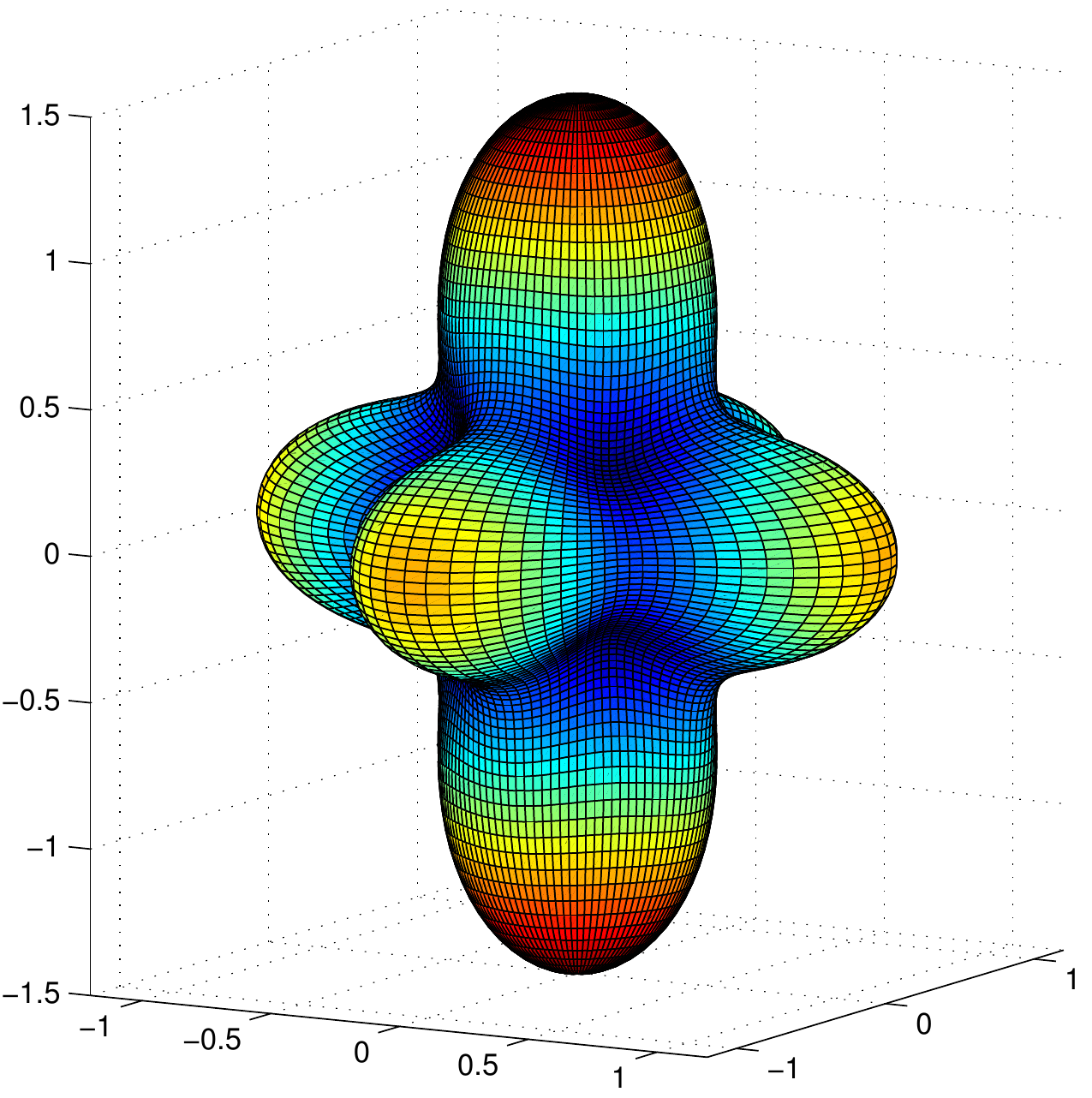}} 
\caption{3-D polar plots in a tetragonal anisotropic system obtained using $[P^4]^{\prime}$. The choice of
the polynomial coefficients are such that either (a) $\langle 100 \rangle$ planes or (b) $\langle 111 \rangle$ planes are the preferred planes.}\label{F:tetra_P4_100}
\end{figure}

For example, for the choice of $m_1 = 1.0$, $m_2 = 1.5$, $m_3 = 2$ and $m_4 = 0.3$ the polar plot
is as shown in Fig.~\ref{F:tetra_P4_100} (a) -- indicating a preference of $\langle 100 \rangle$ plane over $\langle 110 
\rangle$. Similarly, to make $\langle 110 \rangle$ planes favourable, we use $m_1 = 1.2$, $m_2 = 1.5$, $m_3 = 0.2$ 
and $m_4 = 0.1$ and resulting polar plot is shown in Fig.~\ref{F:tetra_P4_100} (b). 

\begin{figure}[htpb]
\centering
\subfigure[]{\includegraphics[height=1.8in,width=2.56in]{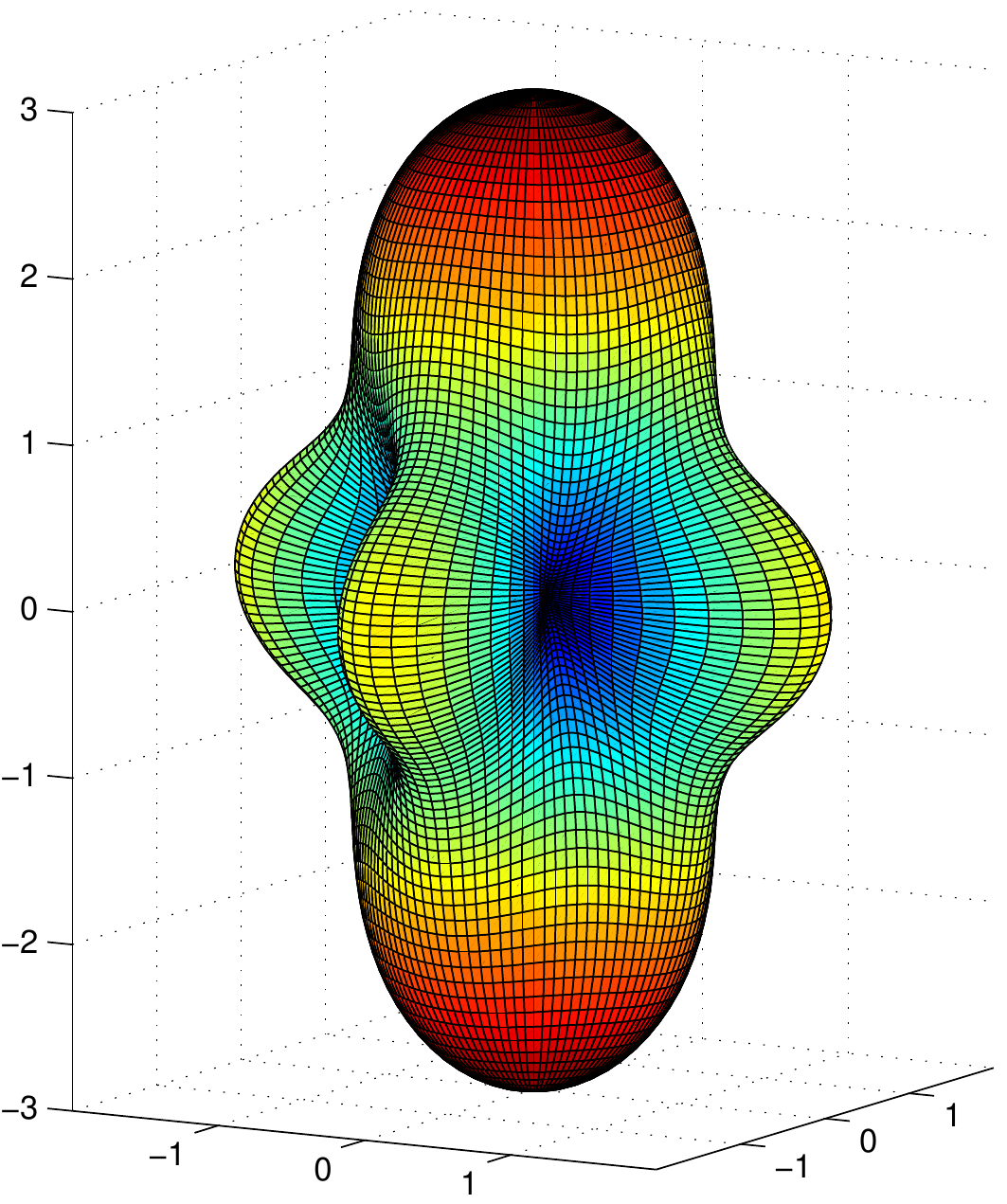}}
\subfigure[]{\includegraphics[height=1.8in,width=2.1in]{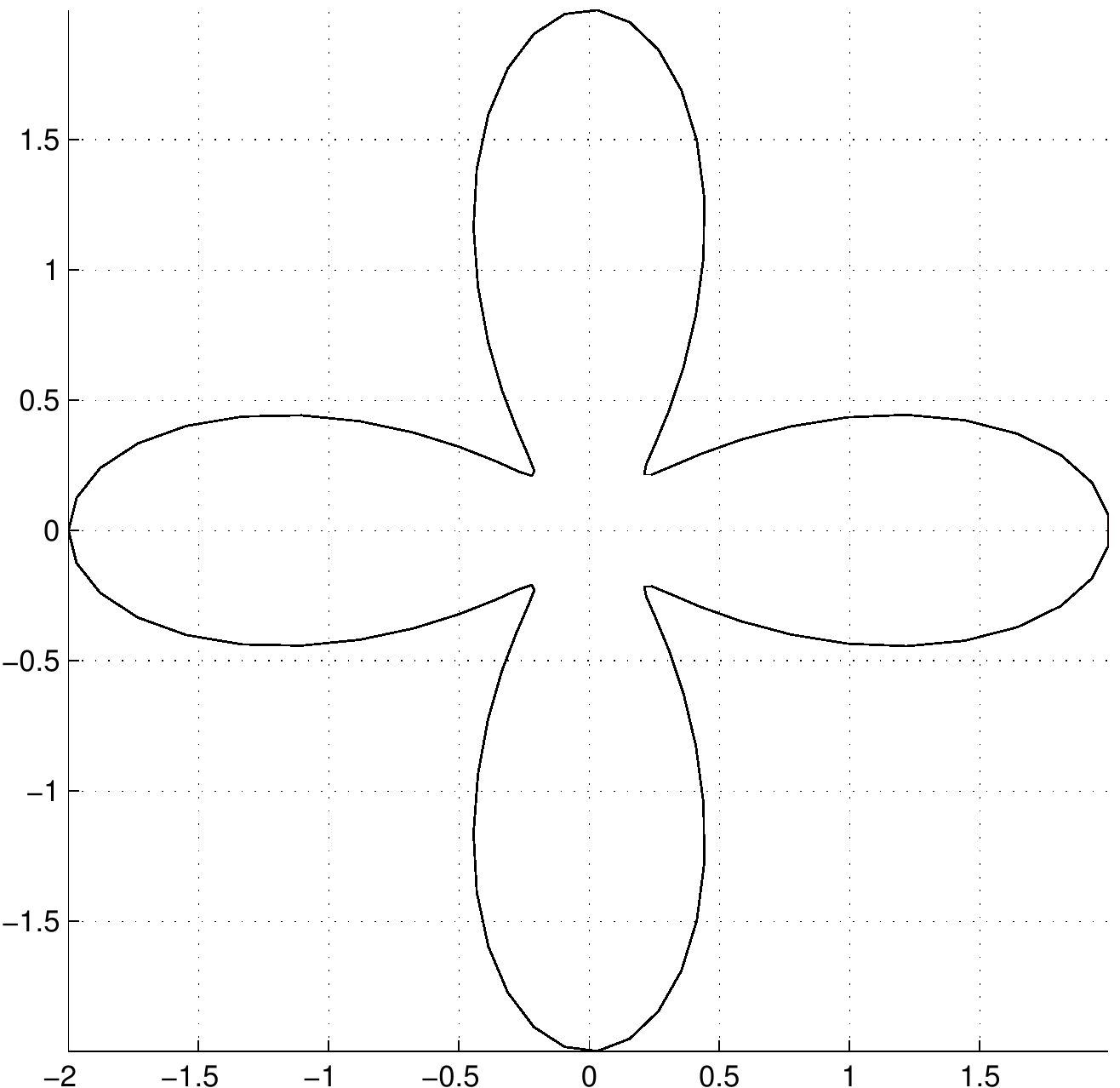}}
\subfigure[]{\includegraphics[height=1.8in,width=0.6in]{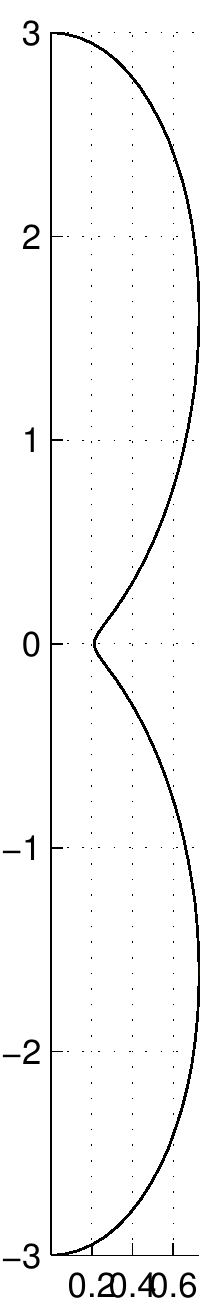}} 
\caption{(a) 3-D polar plot along with (b) $xy$ and (c) $xz$ sections in a tetragonal anisotropic system for prism like equilibrium morphology; obtained using $[P^4]^{\prime}$.}\label{F:tetra_P4_prism}
\end{figure}

\begin{figure}[htpb]
\centering
\subfigure[]{\includegraphics[height=1.8in,width=2.56in]{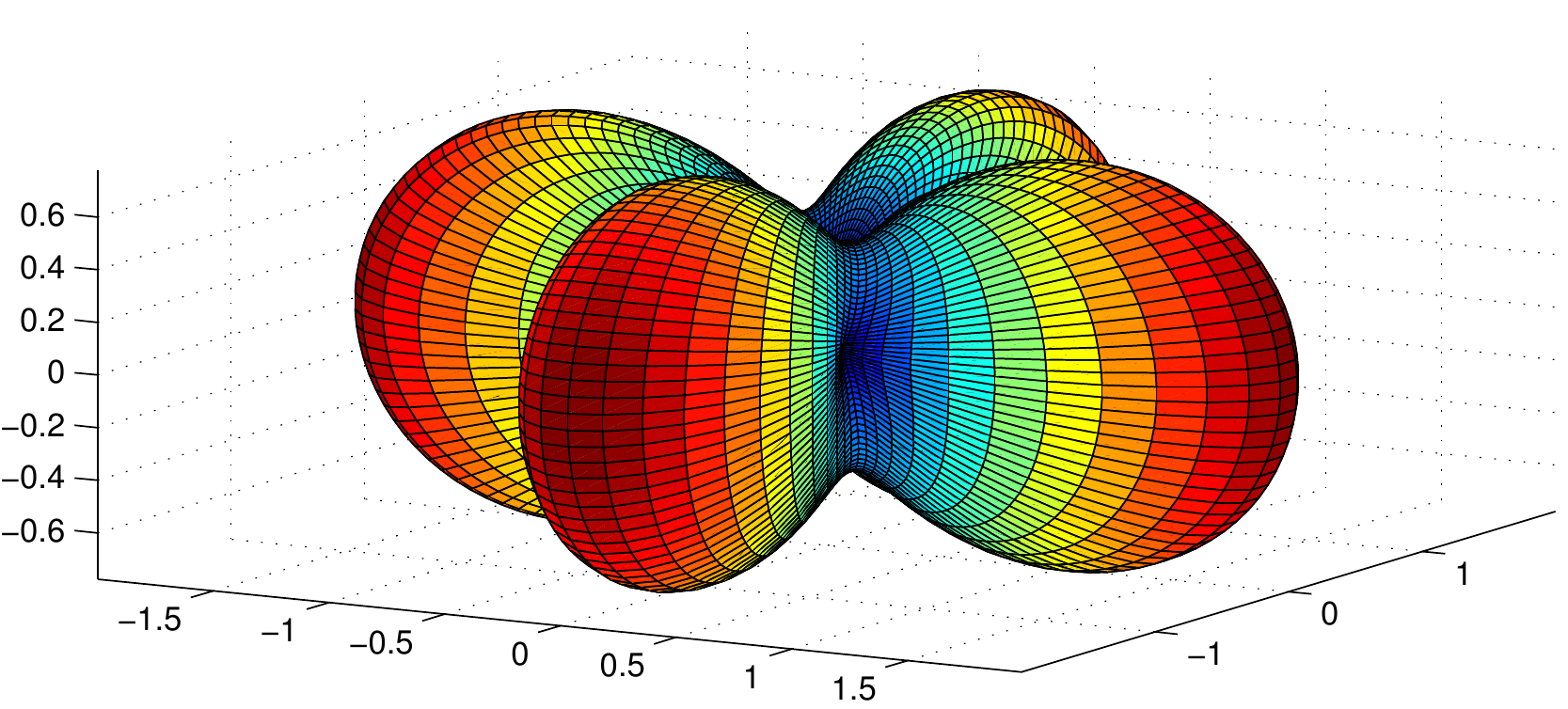}}
\subfigure[]{\includegraphics[height=1.8in,width=2.1in]{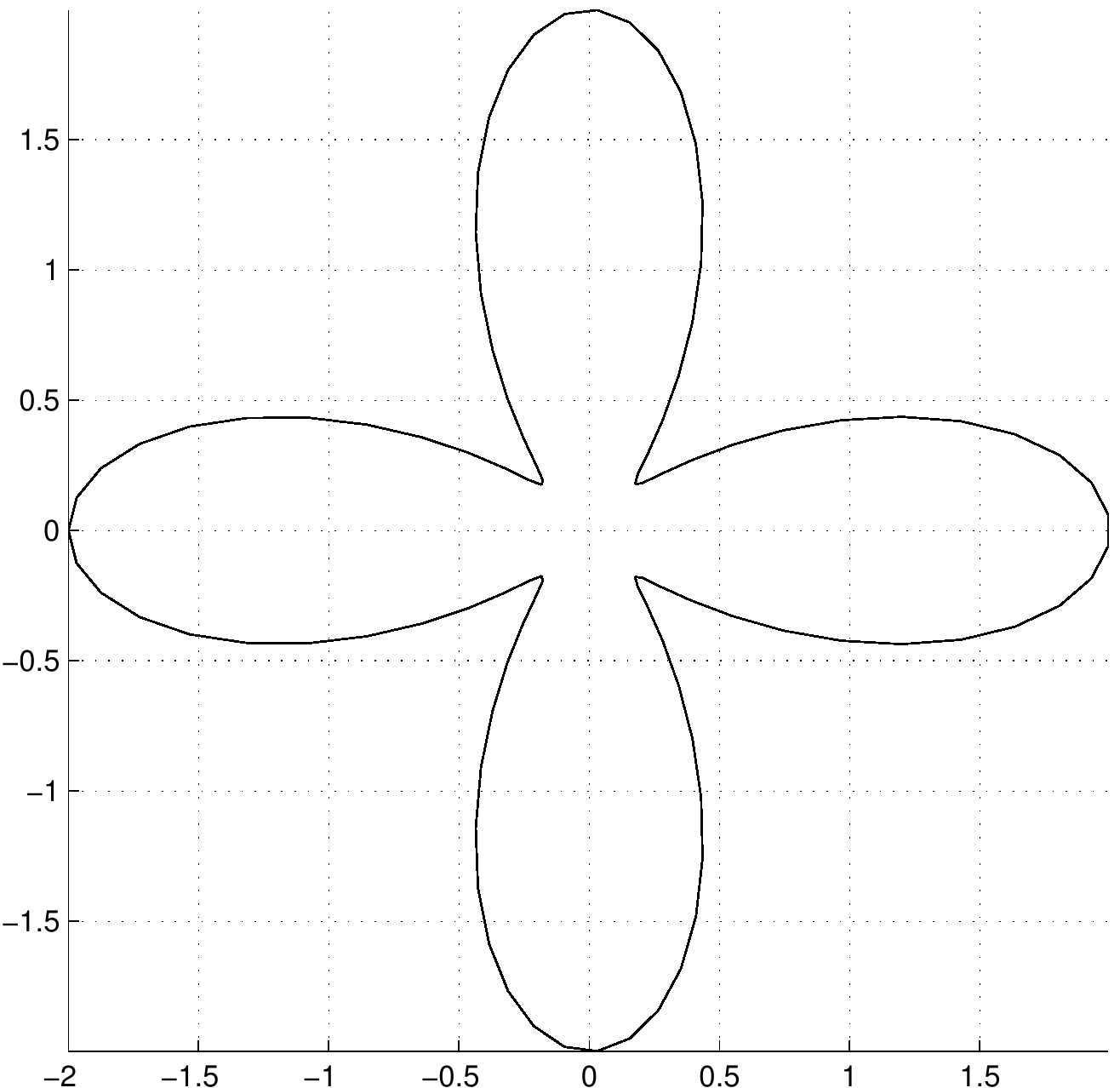}}
\subfigure[]{\includegraphics[height=1.8in,width=0.6in]{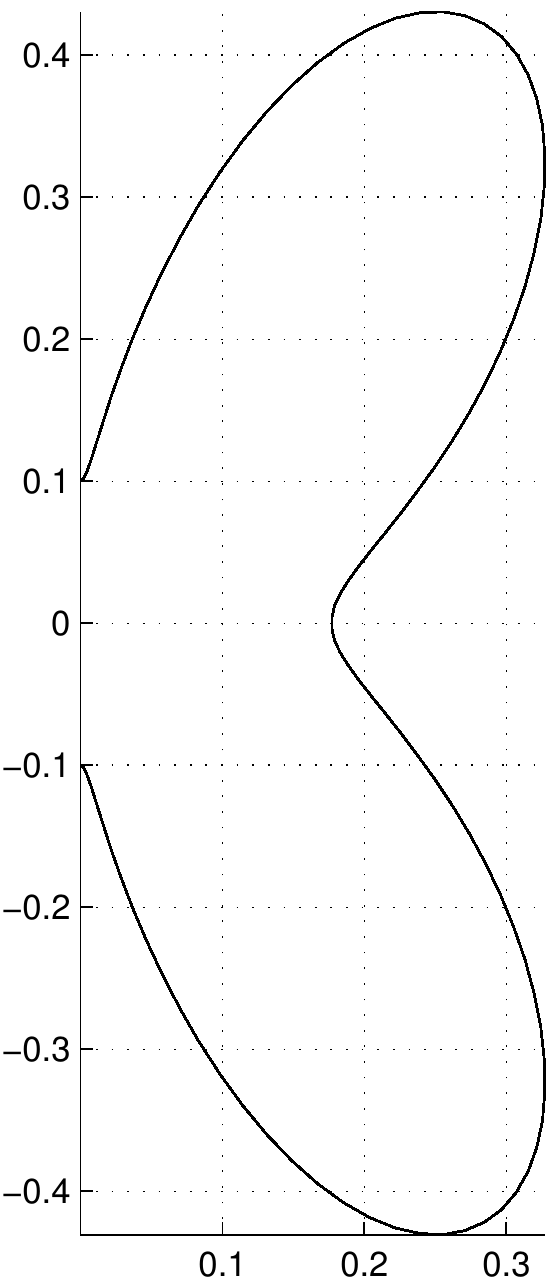}} 
\caption{(a) 3-D polar plot along with (b) $xy$ and (c) $xz$ sections in a tetragonal anisotropic system for plate like equilibrium morphology; obtained using $[P^4]^{\prime}$.}\label{F:tetra_P4_plate}
\end{figure}

If we choose $m_1 = 2.0$, $m_2 = 3$, $m_3 = -1.4$ and $m_4 = 0.9$, we can see in Fig.~\ref{F:tetra_P4_prism} (a), that 
there is no minima along the z-direction and hence $\langle 001 \rangle$ planes will not be present. Fig.~\ref{F:tetra_P4_prism} (b)
shows the $xy$-section of the 3-D polar plot and one can clearly see the four fold symmetry. In Fig.~
\ref{F:tetra_P4_prism} (c), the $xz$-section is shown, which has a dip in the $xy$ plane. From these three plots, if we imagine 
the equilibrium shape of the precipitate, we can easily see that it leads to a tetragonal prism. Similarly, if we choose 
$m_1 = 2.0$, $m_2 = 0.1$, $m_3 = -1.5$ and $m_4 = 1.0$, we can obtain a dip along the z-direction and hence 
$\langle 001 \rangle$ planes will form in the precipitate morphology. We show the 3-D polar plot, $xy$ and $xz$-sections 
in Fig.~\ref{F:tetra_P4_plate} (a), (b) and (c) respectively. We can see that the equilibrium shape of the 
precipitate that results from these polar plots and their sections is a tetragonal plate like morphology.
 
\begin{figure}[htpb]
\centering
\subfigure[]{\includegraphics[height=2.16in,width=2.16in]{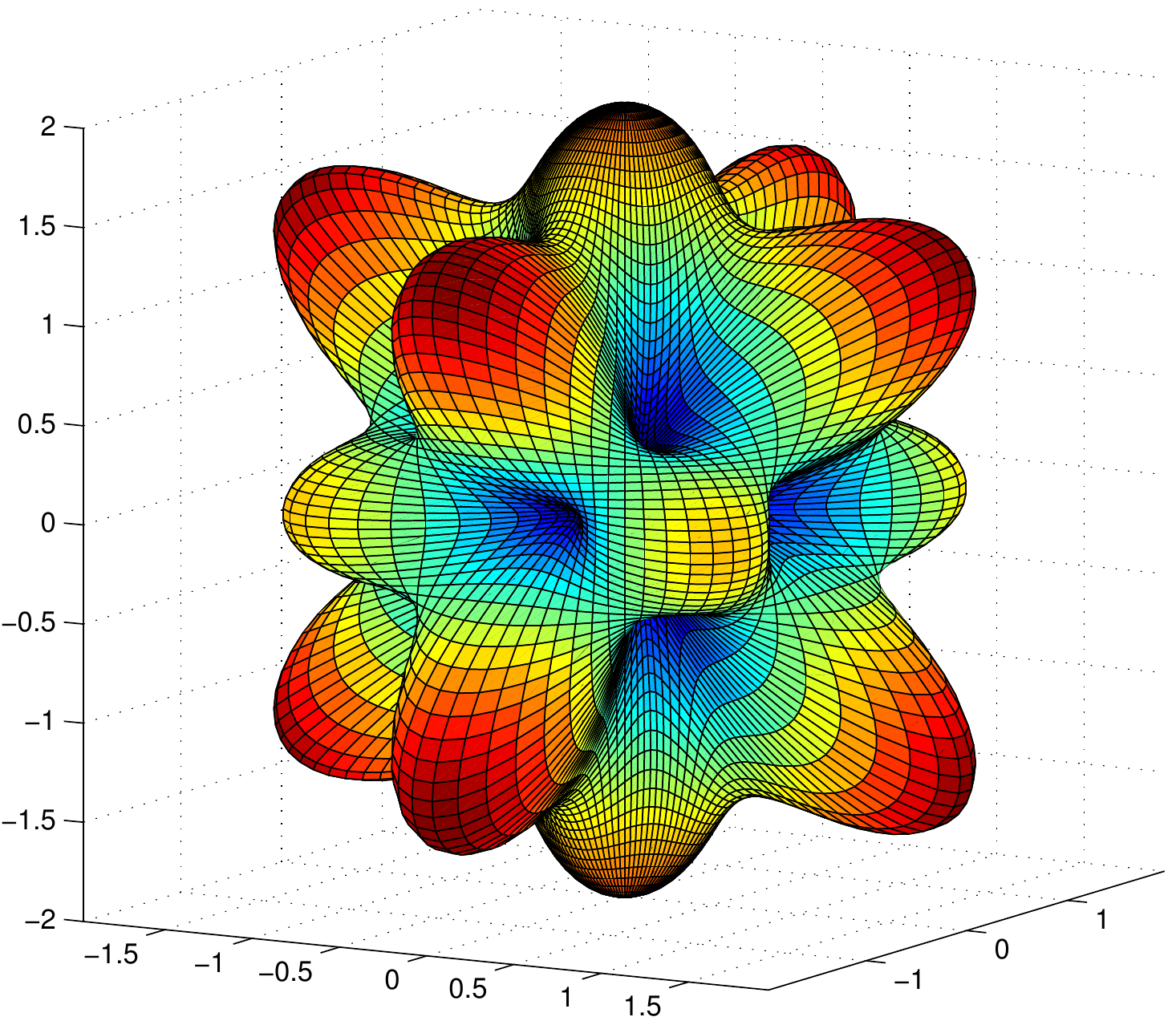}}
\subfigure[]{\includegraphics[height=2.16in,width=2.16in]{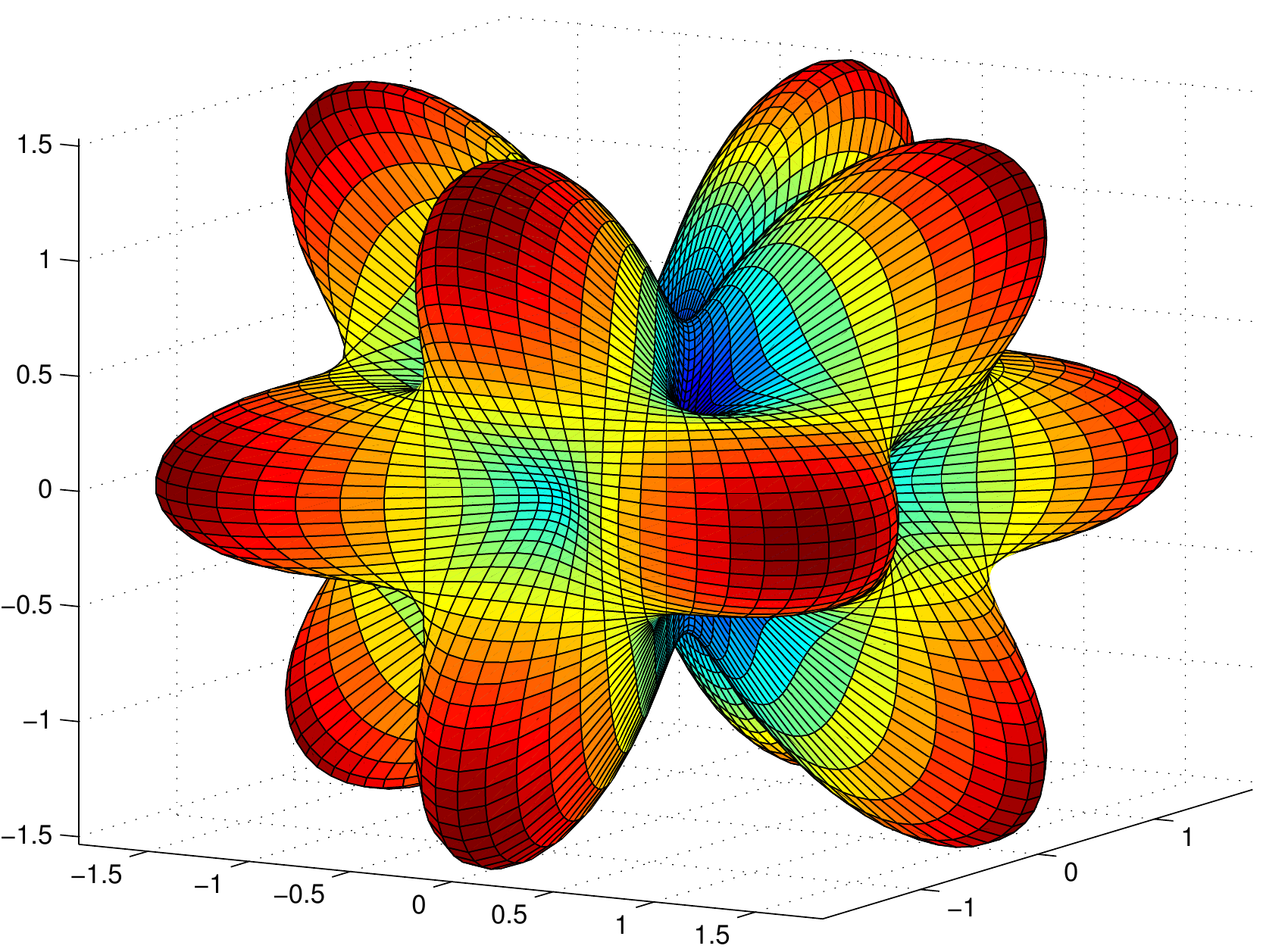}} 
\caption{3-D polar plots in a tetragonal anisotropic system obtained using $[P^6]^{\prime}$. For (a) the polynomial coefficients are chosen such that both $\langle 100 \rangle$ and $\langle 111 \rangle$ are preferred. For (b) the polynomial coefficients are chosen such that $\langle 100 \rangle$, $\langle 111 \rangle$ and $\langle 001\rangle$ are preferred. }\label{F:tetra_P6}
\end{figure}

The polynomial $[P^4]^{\prime}$ lacks $c_1^2 c_2^2 c_3^2$ and hence we can not control the energetics 
of $( 100 )$ and $( 111 )$ planes simultaneously.  Using (normalised version of) Eq.~\ref{E:sixN_tensor_pol} 
and with the choice of $n_1 = 0.5$, $n_2 = 2.0$, $n_3 = 5.0$, $n_4 = 14.0$, $n_5 = 1.0$ and $n_6 = -60.0$ 
we obtain the polar plot as shown in Fig.~\ref{F:tetra_P6} (a). In this case we can see that, 
$\langle 100 \rangle$ and $\langle 111 \rangle$ planes can simultaneously form in the equilibrium precipitate 
morphology. On the other hand, if we choose $n_1 = 1.0$, $n_2 = 0.27$, $n_3 = 5.0$, $n_4 = 10$, $n_5 = 6$ 
and $n_6 = -59.0$, as shown in Fig.~\ref{F:tetra_P6} (b), $( 100 )$, $( 001 )$ and $( 111 )$ planes can 
simultaneously develop in the equilibrium precipitate morphology. 

\section{Governing equations and numerical implementation}\label{Ch:Governing_eqn}

Once the free energy is given, the 
variational derivative of the free energy functional with composition gives the relevant
chemical potential ($\mu$)~\cite{AbiHaider,ShamesDym2013}:
\begin{equation}\label{ChemicalPotential}
\mu = \frac{\delta F}{\delta c} = \frac{\partial f_0}{\partial c} - \frac{\partial}{\partial x_i}  \left[ \frac{\partial f}{\partial c_i} \right]  + \frac{\partial^2}{\partial x_i \partial x_j} \left[ \frac{\partial f}{\partial c_{ij}} \right] - \frac{\partial^3}{\partial x_i \partial x_j \partial x_k} \left[ \frac{\partial f}{\partial c_{ijk}} \right],
\end{equation}
where we have used Einstein summation convention, namely, that repeated indices are summed.

We assume the mobility $M$ to be a constant; incorporating mass conservation (that is, composition is a 
conserved order parameter), we obtain the Cahn-Hilliard equation which governs concentration changes 
in such a system: 
\begin{equation}
\label{E:CH_eq}
\frac{\partial c}{\partial t} = M \nabla^2 \left[ {\mu}_0 - \mu_{c_i} + \mu_{c_{ij}} - \mu_{c_{ijk}} \right]
\end{equation}
where $\mu_0 = \frac{\partial f_0}{\partial c}$, $\mu_{c_i} = \frac{\partial}{\partial x_i}  
\left[ \frac{\partial f}{\partial c_i} \right] $, 
$\mu_{c_{ij}} = \frac{\partial^2}{\partial x_i \partial x_j} \left[ \frac{\partial f}{\partial c_{ij}} \right]$, 
and, $\mu_{c_{ijk}} = \frac{\partial^3}{\partial x_i \partial x_j \partial x_k} 
\left[ \frac{\partial f}{\partial c_{ijk}} \right]$. In Table~\ref{Table3}, we list these chemical potential 
terms for each of the polynomials listed in Table~\ref{Table1}. 

From Table~\ref{Table3}, it is clear that the second order polynomials, be in gradient, or curvature or aberration, 
lead to linear terms in the evolution equation; such evolution equations are solved using semi-implicit Fourier spectral
technique~\cite{AbiHaider,ChenShen1998,ArijitThesis}. The fourth and sixth order polynomials in gradients lead to 
highly nonlinear terms in the ECH equation; they are solved using explicit Fourier spectral technique
the details of which can be found in~\cite{ArijitEtAl,ArijitThesis}.

%\begin{landscape}
\begin{table}[ptbh]
\caption{\label{Table3} The chemical potentials $\mu_{c_i}$, $\mu_{c_{ij}}$ and $\mu_{c_{ijk}}$.}
\begin{tabular}{|c|c|}
\hline \hline
 ${\mathbf{\mu}}$  & {\bf Expression} \\
\hline \hline
$\mu_{c_i}(2)$ &$2p_1 (c_{11} + c_{22}) + 2p_2 c_{33}$ \\
\hline
$\mu_{c_i}(4)$ & $12 m_1 c_1^2 c_{11} + 12 m_1 c_2^2 c_{22} + 16 m_2 c_1 c_2 c_{12} + 4 m_2 c_{11} c_2^2 + 4 m_2 c_{22} c_1^2 + 12 m_3 c_3^2 c_{33}$ \\
 &  $  + 16 m_4 c_3 (c_1 c_{31} +  c_2  c_{23}) + 4 m_4 c_{11} c_3^2 + 4 m_4 c_{22} c_3^2 + 4 m_4 c_{33} (c_1^2 + c_2^2)$ \\
\hline
$\mu_{c_i}(6)$ & $ 6 n_1 (c_{11}+c_{22}) (c_1^2+c_2^2)^2 + 24 n_1 (c_1^2+c_2^2)(c_1 c_1 c_{11} + 2 c_2 c_1 c_{21} + c_2 c_2 c_{22}) + 30 n_2 c_3^4 c_{33}$\\
& $+12n_3 c_1^2 c_2^2(c_{11} + c_{22}) + 16 n_3 c_{12} (c_1^3 c_2 + c_1 c_2^3) + 2 n_3 (c_2^4 c_{11} + c_1^4 c_{22})$ \\
& $+4 n_4 c_3^2 (c_1^2+c_2^2) (c_{11} + c_{22}) + 16 n_4 c_3 (c_1^2+c_2^2) (c_1 c_{31} + c_2 c_{32}) + 2 n_4 c_{33} (c_1^2+c_2^2)^2$\\
& $+ 8 n_4 c_3^2 (c_1 c_1 c_{11} + 2 c_2 c_1 c_{21} + c_2 c_2 c_{22})  + 2 n_5 (c_{11}+c_{22}) c_3^4$\\
& $+ 16 n_5 c_3^3 (c_1 c_{31} + c_2 c_{23}) +12 n_5 c_3^2 c_{33} (c_1^2+c_2^2)$\\
& $+2 n_6 (c_{11} c_2^2 c_3^2 + c_{22}  c_1^2 c_3^2  + c_{33} c_1^2 c_2^2) + 8 n_6 (c_1 c_2 c_{12} c_3^2 + c_1 c_3 c_{31} c_2^2 +c_2 c_3 c_{23} c_1^2) $ \\
\hline \hline
$\mu_{c_{ij}}$ &  $2 q_1 (c_{1111} +c_{2222}) + 2 q_2 c_{3333}$ $+ 4 q_{3} c_{1122} + 4 q_4 (c_{3311} + c_{2233}) +2 q_5 (c_{1313} + c_{2323}) +  2 q_6 c_{1212}$ \\
\hline \hline
$\mu_{c_{ijk}}$ &  $2 r_1 (c_{111111} + c_{222222}) + 2 r_2 c_{333333} + (2 r_{3} + 2 r_{10}) (c_{111122}+ c_{222211})$\\
&  $+ (4 r_4 + 8 r_6 + 2 r_{11}) c_{112233} + (4 r_5 + 2 r_7) (c_{111133}+ c_{222233} ) + (2 r_8 + 4 r_9) (c_{333311} + c_{333322})  $ \\
\hline \hline
\end{tabular}
\end{table}
%\end{landscape}

The numerical implementation is carried out on the non-dimensionalised evolution equations;
the non-dimensionalisation is the same as that described in~\citep{AbiHaider} and leads
to a non-dimensional values of unity for the constants $A$ (in the bulk free energy
density) and $M$ (mobility); the composition $c$ is scaled to lie between 0 and 1. 
The far-field composition in the matrix is denoted by $c_\infty$ and is chosen to be 0.2 
(in 2-D simulations) and 0.1 (in 3-D simulations). The grid spacing for spatial variables 
$\Delta x = \Delta y = \Delta z = 0.5 $ (in 1- and 2-D simulations) and 
$\Delta x = \Delta y = \Delta z = 1.0 $ (in 3-D simulations).
The time step used in these simulations are $\Delta t = 10^{-5}$ (for $P^4$ and $P^6$) and 
$\Delta t = 10^{-1}$ (for $R$). The 2-D simulations are carried out on a $256 \times 256$ grid 
while the 3-D simulations are carried out on $100 \times 100 \times 100$ grid. In the next section, 
at the appropriate places, we list only the 
independent tensor coefficients (described in Tables.~\ref{Table1} and~\ref{Table3}) used in the 
simulations; the 
dependent parameters (such as $r_{14}$ for example) are obtained using the relationships 
listed in Table.~\ref{ConstraintsTable}.

\section{Results} \label{Result}

In all simulations presented in this section, the $a$, $b$ and $c$ tetragonal crystallographic axes are aligned with 
the $x$, $y$ and $z$ of the simulation cell. We first present the results from 1-D 
simulations which help us generate the Wulff plots for given anisotropic interfacial free energies;
these Wulff plots are consistent with the free energy polar plots shown in the formulation section.
Then, we present the 2- and 3-D precipitate morphologies and analyse them using the Wulff construction
on the free energy polynomial plot; this analysis is qualitative and shows that our precipitate morphologies
obtained during the simulations are indeed equilibrium ones.

\subsection{Wulff plots obtained from 1D simulations}

\begin{figure}[htpb]
\centering
\subfigure[]{\includegraphics[height=2.6in,width=2.6in]{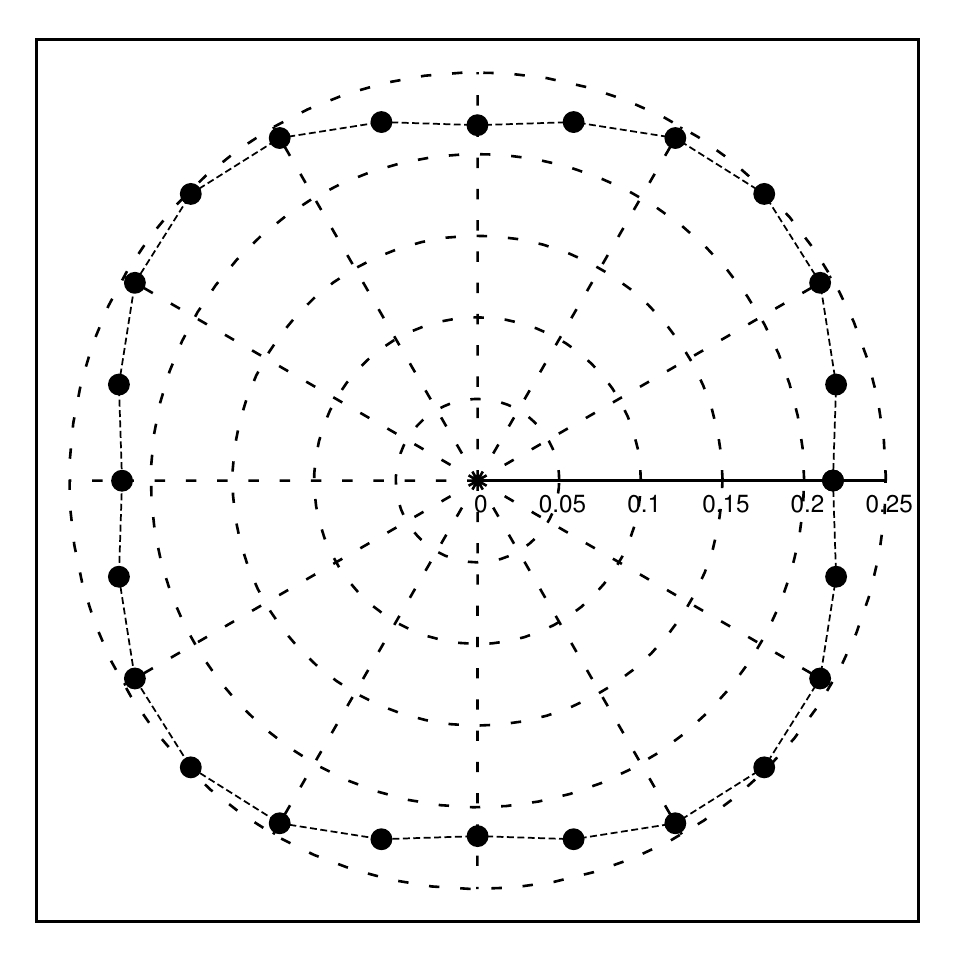}}
\subfigure[]{\includegraphics[height=2.6in,width=2.6in]{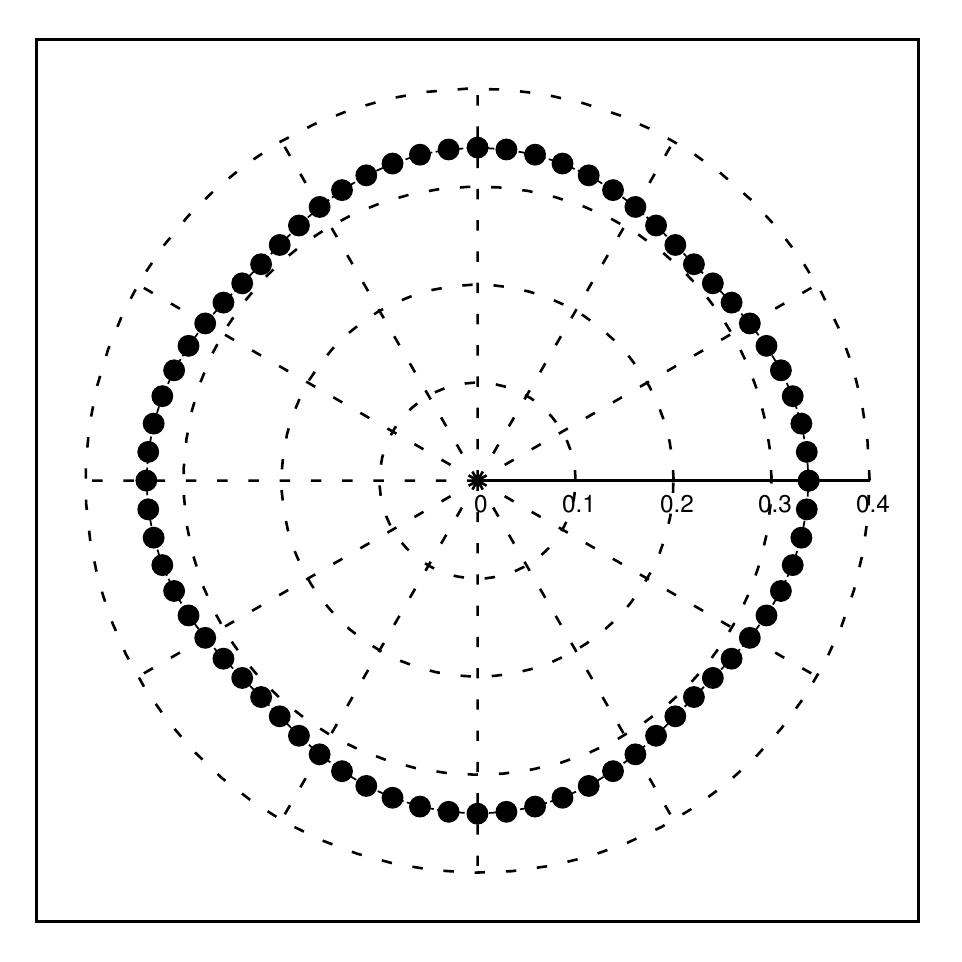}}
\caption{ The (a) $xy$ and (b) $xyz$-sections of the Wulff plot obtained using $P^6$.}\label{F:wulff}
\end{figure}

Using 1-D simulations of planar interfaces with different interface orientations, the
variation of interfacial energy with interface orientation can be plotted -- the so-called 
Wulff plots~\cite{PorterEasterling}. As an example, we show results from a set of calculations in 
which only $P^6$ tensor coefficient was assumed to be non-zero. We have generated the $xy$ plane section of the Wulff plots for 
systems that show tetragonal symmetry (specifically, one in which the $ \langle 100 \rangle$, $
\langle 001 \rangle$ and $\langle 111 \rangle$  directions are preferred). Four-fold $xy$ section perpendicular to $\langle 001 \rangle$ is shown in Fig.~\ref{F:wulff} (a) and $xyz$ section perpendicular to $\langle 110 \rangle$ is shown in Fig.~\ref{F:wulff} (b). For (a), we have used $n_1 = 100.0$ and $n_3 = 500.0$ (as only xy-section is considered here), and 
for (b) we have used  $n_1 = 100.0$, $n_2 = 27.0$, $n_3 = 500.0$, $n_4 = 1000.0$, $n_5 = 600.0$ and $n_6 = -5900.0$. In these calculations we fix the components of second rank gradient free energy coefficients ($P^2$ tensor) at $p_1 = 1.0$ 
and $p_2 = 0.27$. Similar Wulff plot sections for other planes and for other systems are 
possible. However, for the sake of brevity, we only show these two in this paper.  

\subsection{Equilibrium precipitate morphology}

In this subsection, we present the equilibrium precipitate morphologies; we start with 2-D results. In tetragonal systems, 
the $xy$ 2D section is different from the $yz$ 2D section. Of course, the $xz$ 
2D section is the same as the $yz$ section. Hence, in 2D, we have carried out simulations for both the cases, namely,
$xy$ and $yz$ section. After 2D results, we describe more complex precipitate morphologies in 3-D for different choices 
of higher order free energy polynomials. 

\subsubsection{Morphologies in 2D}

\begin{figure}[htbp]
\centering
\subfigure[]{\includegraphics[trim=1.0cm 1.0cm 1.0cm 1.0cm,width=0.3\textwidth]{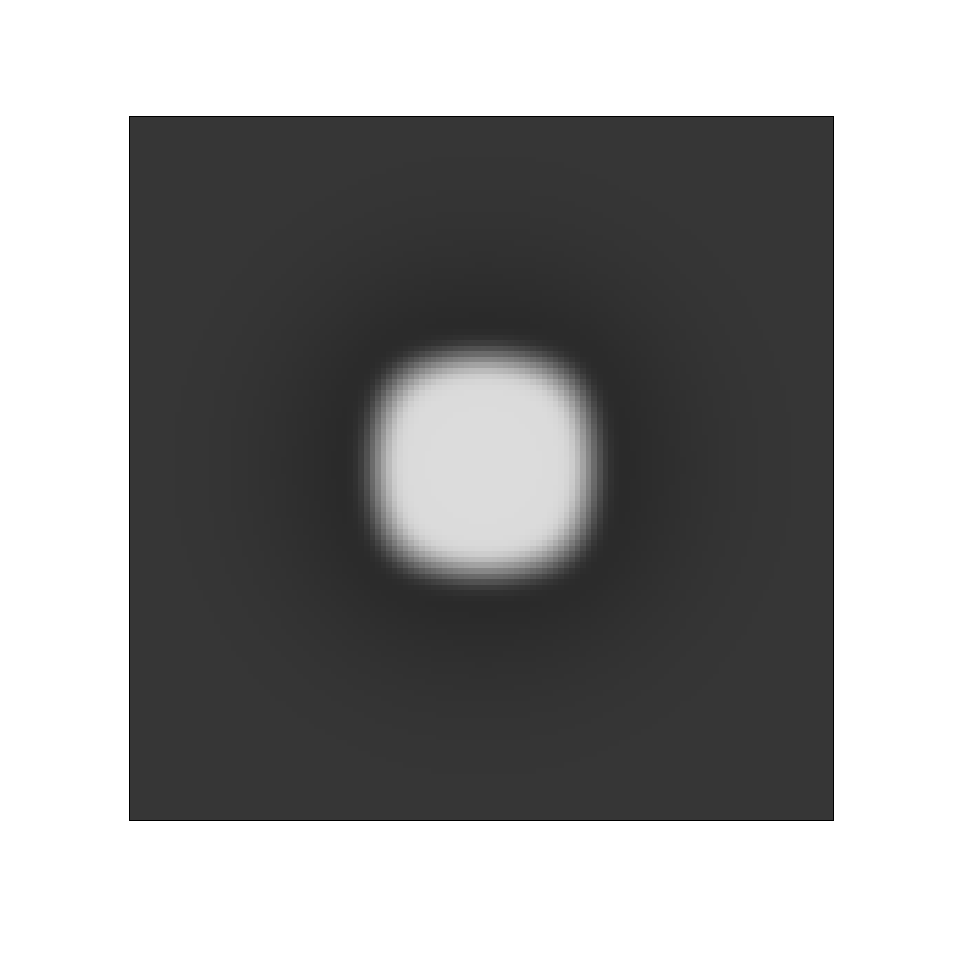}}
\subfigure[]{\includegraphics[trim=1.0cm 1.0cm 1.0cm 1.0cm,width=0.3\textwidth]{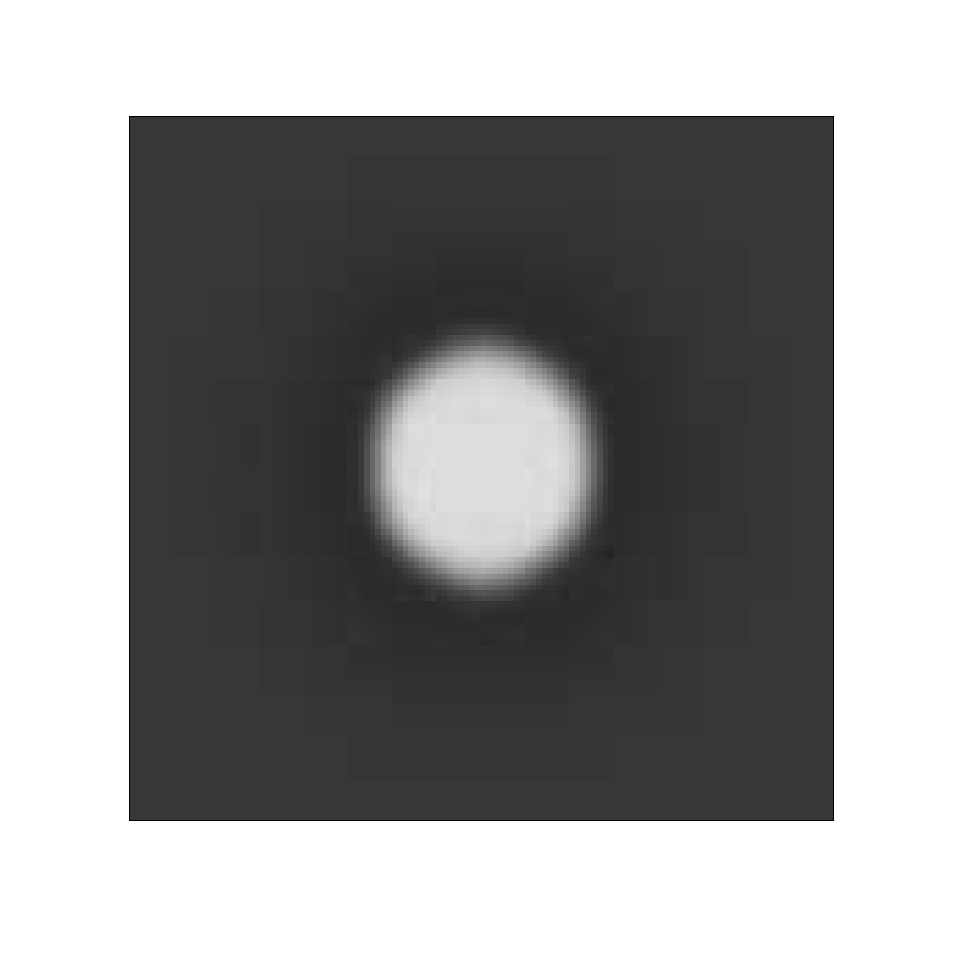}}
\subfigure[]{\includegraphics[width=.3\textwidth]{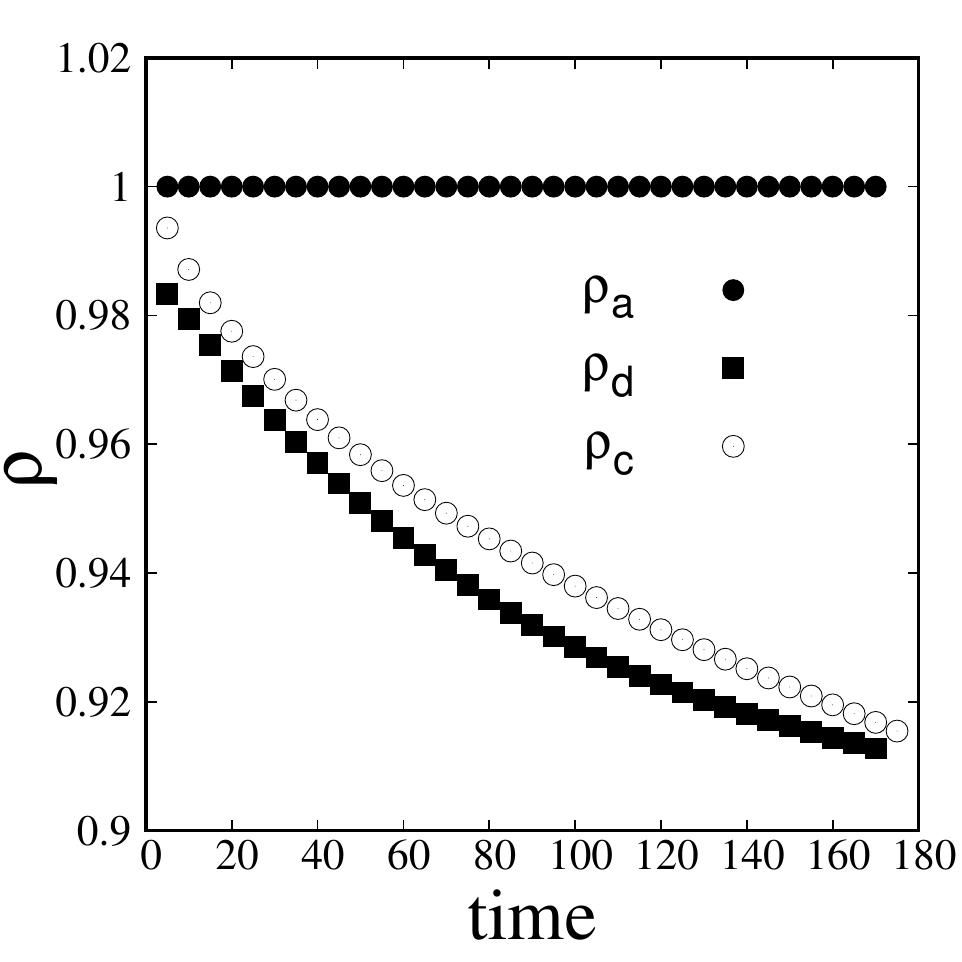}}
\caption{ Precipitate morphology obtained using $P^6$. (a) represents 
morphology with facets along x-axis and (b) represents morphology with corners along c-axis. (c) is
the aspect ratios ($\rho$) of these precipitates during the growth.}\label{2DMorphologies_tetP6}
\end{figure}

In Fig.~\ref{2DMorphologies_tetP6} (a) and (b), we show the precipitate morphology after 170 time units using non-zero $P^6$ --
for the $xy$ and $yz$ planes of the tetragonal system respectively. The simulation was started with a circular precipitate
of size twelve at the centre of the simulation cell. In the $xy$ plane, tetragonal symmetry leads to four fold symmetry --
which is the same as the 2D cubic systems and hence as seen in Fig.~\ref{2DMorphologies_tetP6} (a), the precipitate 
morphology is squarish.
On the other hand, the  Fig.~\ref{2DMorphologies_tetP6} (b), which corresponds to the $yz$ plane leads to a 
lens shaped morphology -- by virtue of $a \neq c$ in tetragonal symmetry. The microstructure in 
Fig.~\ref{2DMorphologies_tetP6} (a) is obtained using $n_1 = 500.0$ and $n_3 = 5000.0$, and in Fig.~\ref{2DMorphologies_tetP6} (b) is obtained using 
$n_1 = 500.0$, $n_2 = 1200$, $n_4 = 5000$ and $n_5 = 10$.  As we show in the next subsection 
(where 3-D precipitate morphologies are described), for a different choice of parameters, in the $yz$ plane,
it is possible to obtain a rectangular morphology (instead of the lens-shape).

In Fig.~\ref{2DMorphologies_tetP6} (c), we also show the aspect ratios ($\rho$) as a function of time.  
For the precipitate in the $xy$ plane, we have plotted two aspect 
ratios: $\rho_{a}$ is the aspect ratio of the size of the precipitate along $x$ axis to that along the $y$-axis; 
$\rho_{d}$ is the aspect ratio of the size of the precipitate along the $x$ axis to that along the diagonal direction 
($xy$ direction). As we can see, the $\rho_a$ starts at unity (since our initial precipitate shape is circular)
and remains at unity (since the shape evolves into a square); $\rho_d$ which starts at unity (circular shape)
becomes smaller than one -- though it does not reach the value of $1/\sqrt{2} = 0.707$ meant for the perfect
square. For the precipitate in the $yz$ plane, we define the aspect ratio $\rho_c$ which is the ratio of the 
size of the precipitate along the $y$ axis to that along the $z$ direction. The $\rho_c$, again, starts from 
unity (circular precipitate) and decreases below unity -- representing the development of the lens shaped morphology. 

\subsubsection{Morphologies in 3D}

In this section, we present results from our 3D studies. By appropriate choice of the constants in the higher
order polynomials, we show that we can obtain precipitates with prism, plate and di-pyramid morphologies,
and, their truncated (more complex) variants. 

\begin{figure}[htpb]
\centering
\subfigure[]{\includegraphics[trim=0.5cm 0.5cm 0.5cm 0.5cm,height=1.6in,width=1.6in]{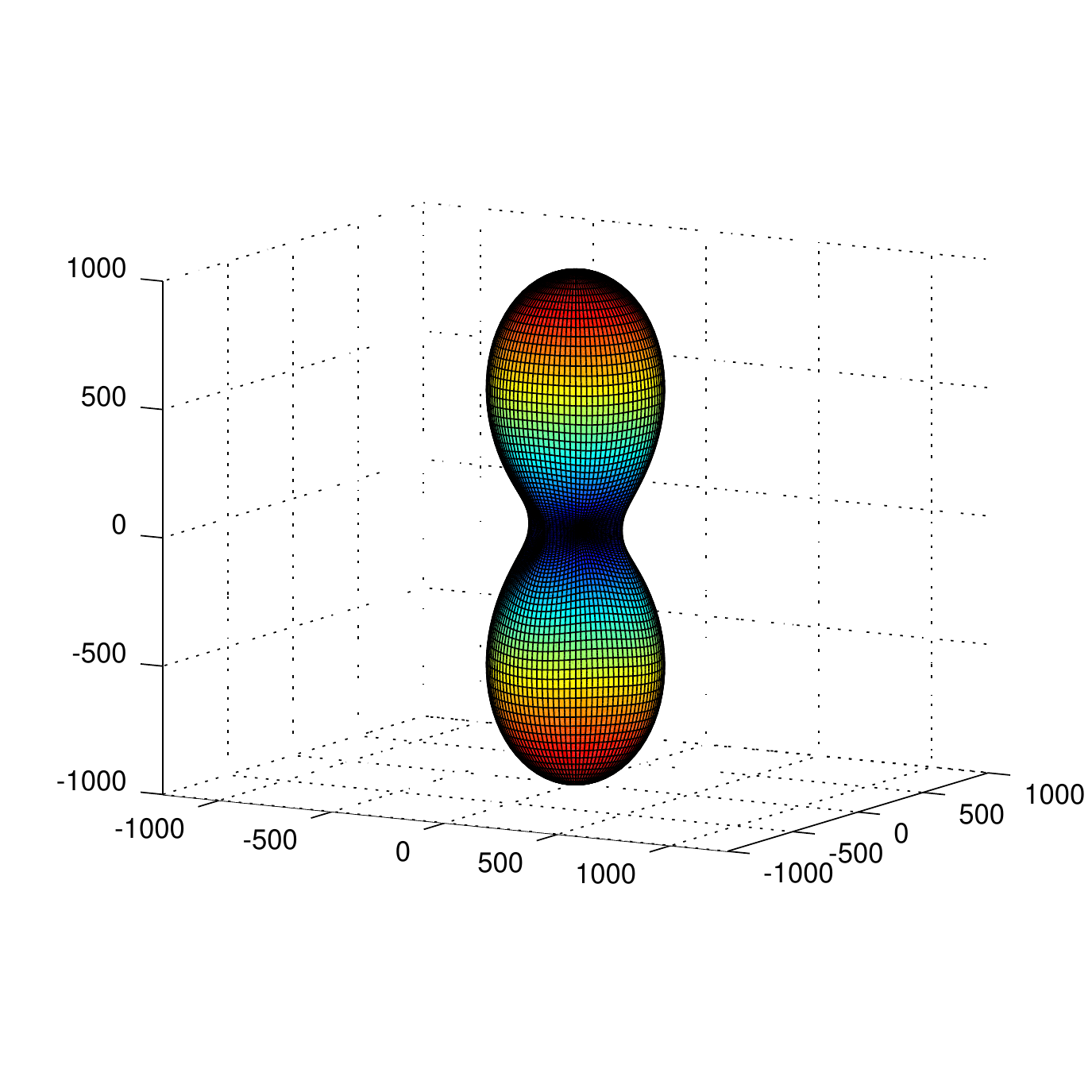}} 
\subfigure[]{\includegraphics[trim=0.5cm 0.5cm 0.5cm 0.5cm,height=1.6in,width=1.6in]{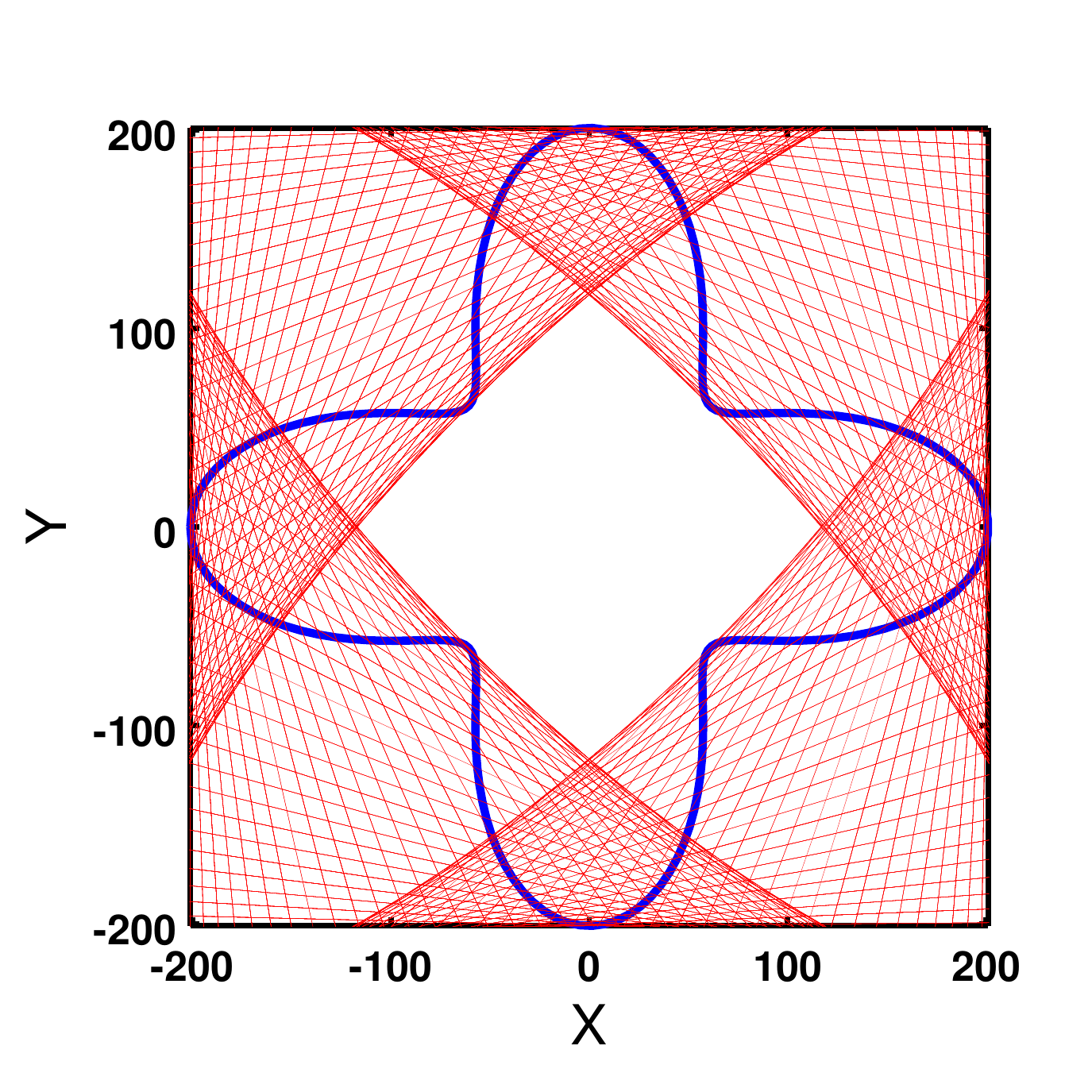}}
\subfigure[]{\includegraphics[trim=0.5cm 0.5cm 0.5cm 0.5cm,height=1.6in,width=1.6in]{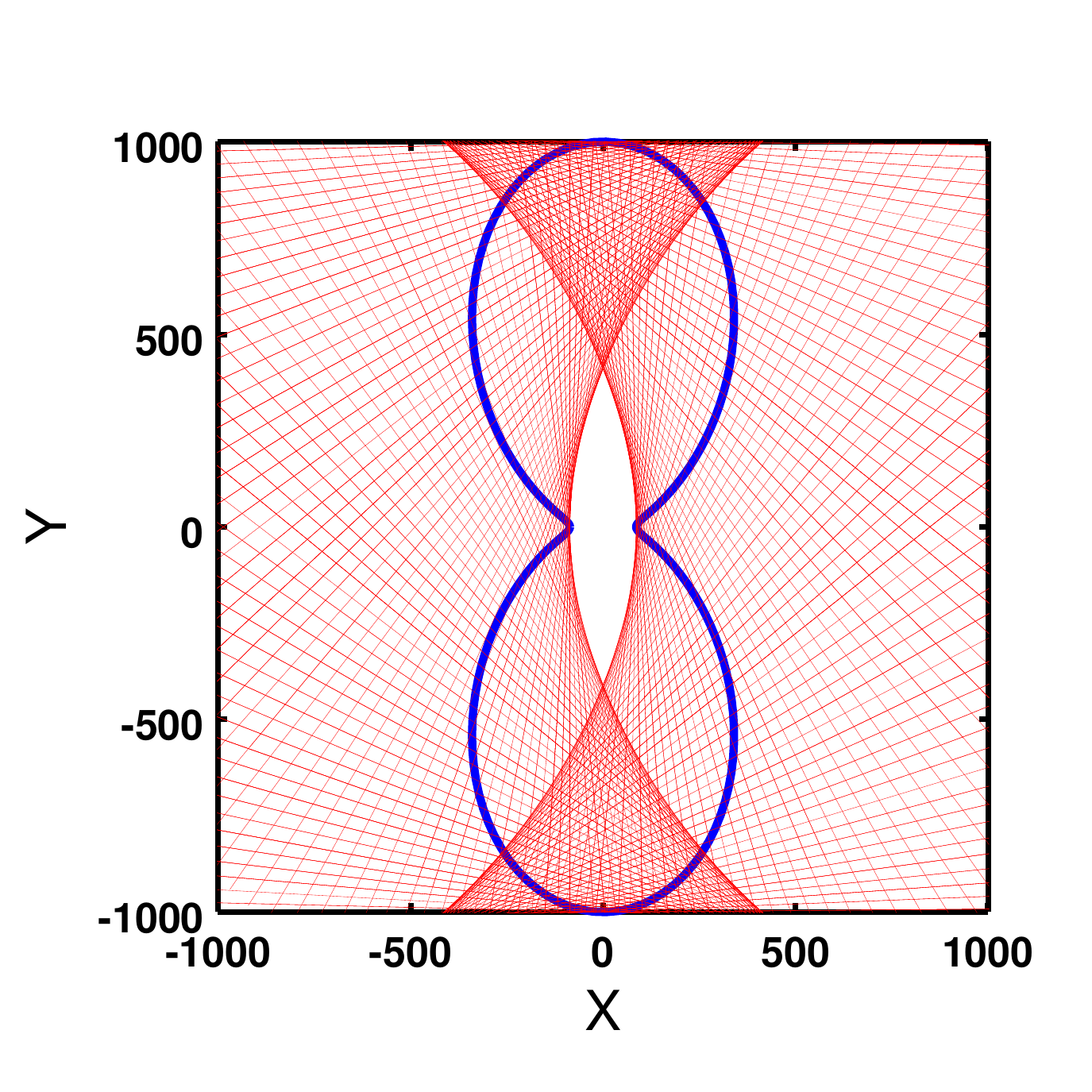}}\\
\subfigure[]{\includegraphics[trim=0.5cm 0.5cm 0.5cm 0.5cm,height=1.6in,width=1.6in]{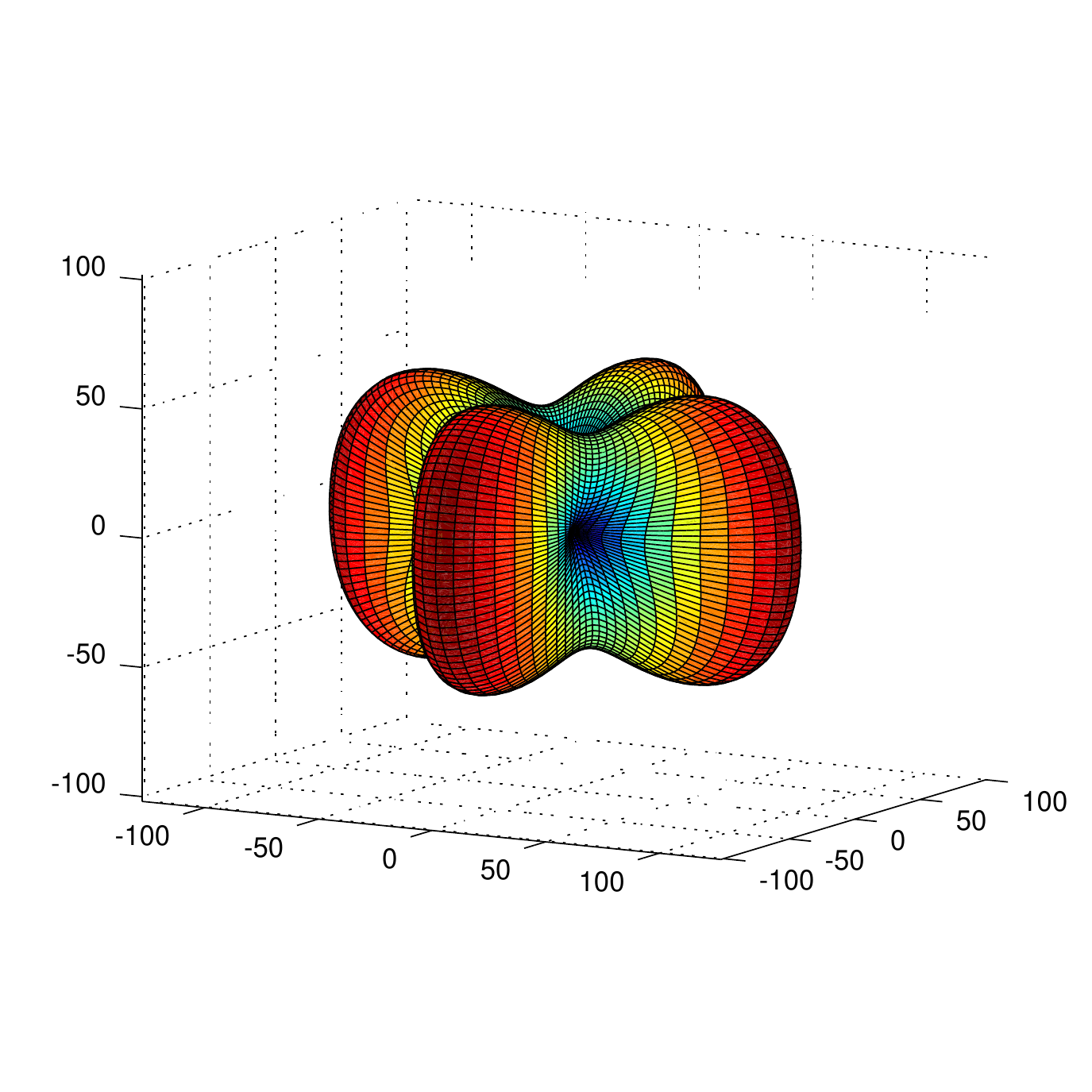}} 
\subfigure[]{\includegraphics[trim=0.5cm 0.5cm 0.5cm 0.5cm,height=1.6in,width=1.6in]{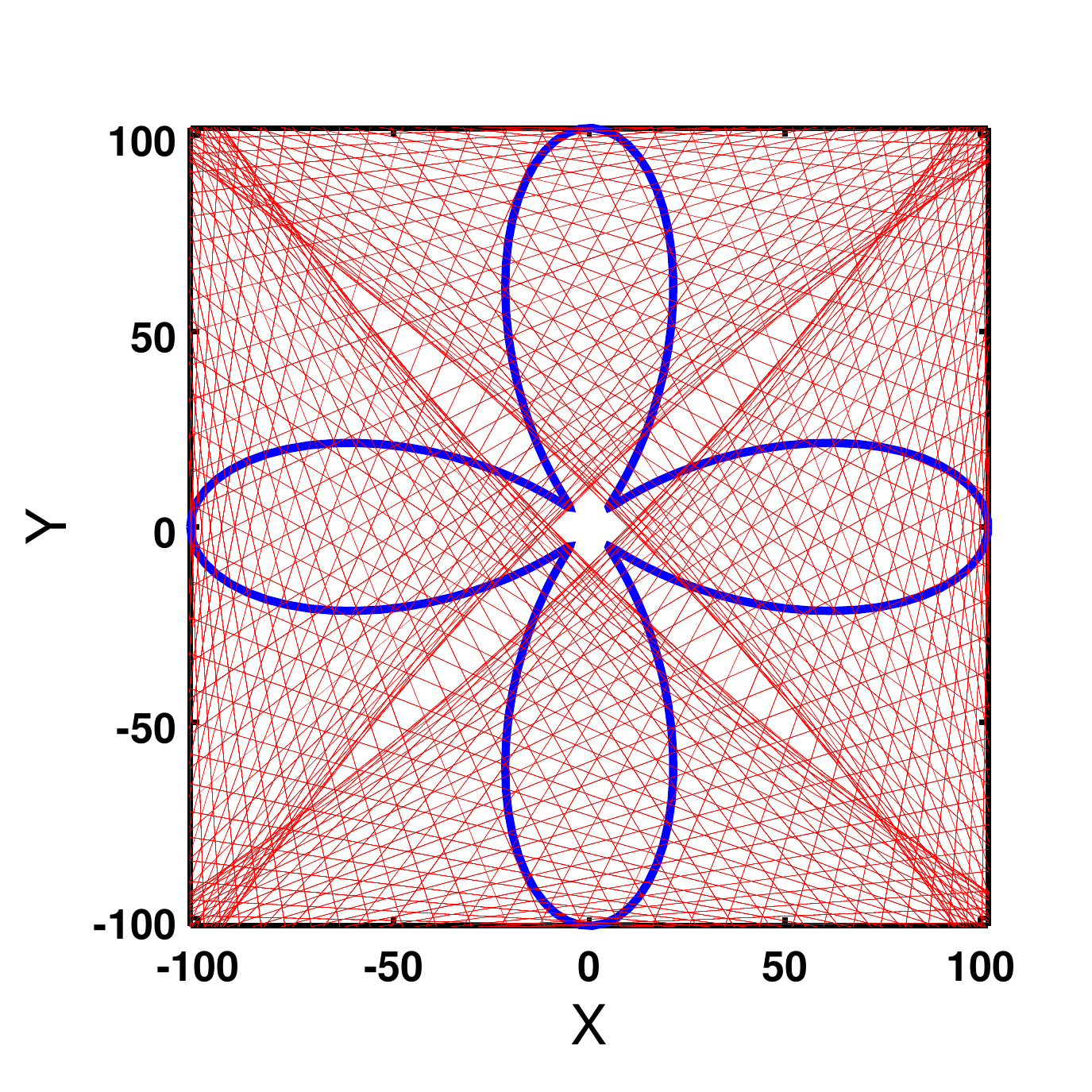}}
\subfigure[]{\includegraphics[trim=0.5cm 0.5cm 0.5cm 0.5cm,height=1.5in,width=2.121in]{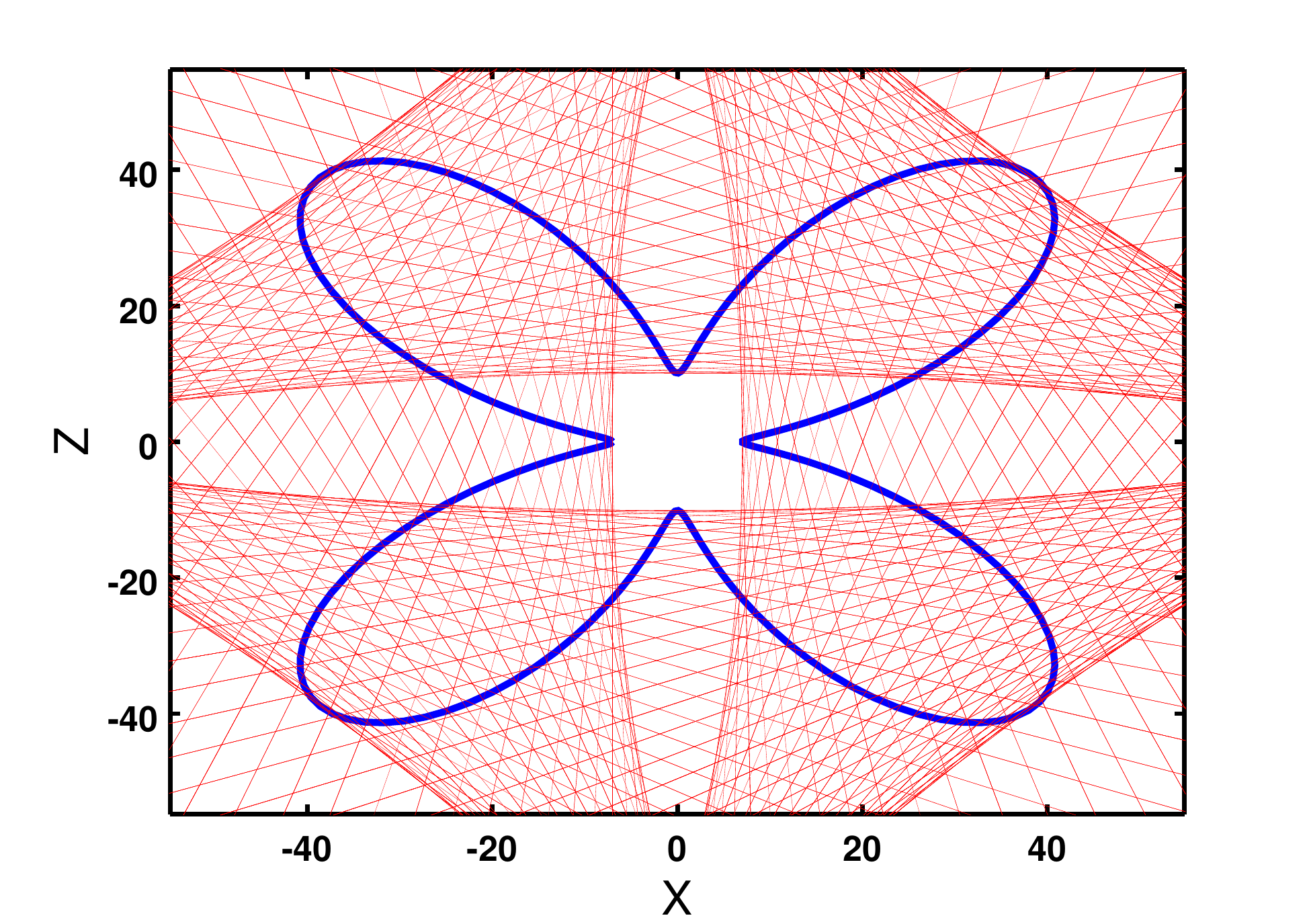}}\\
\subfigure[]{\includegraphics[trim=0.5cm 0.5cm 0.5cm 0.5cm,height=1.6in,width=1.6in]{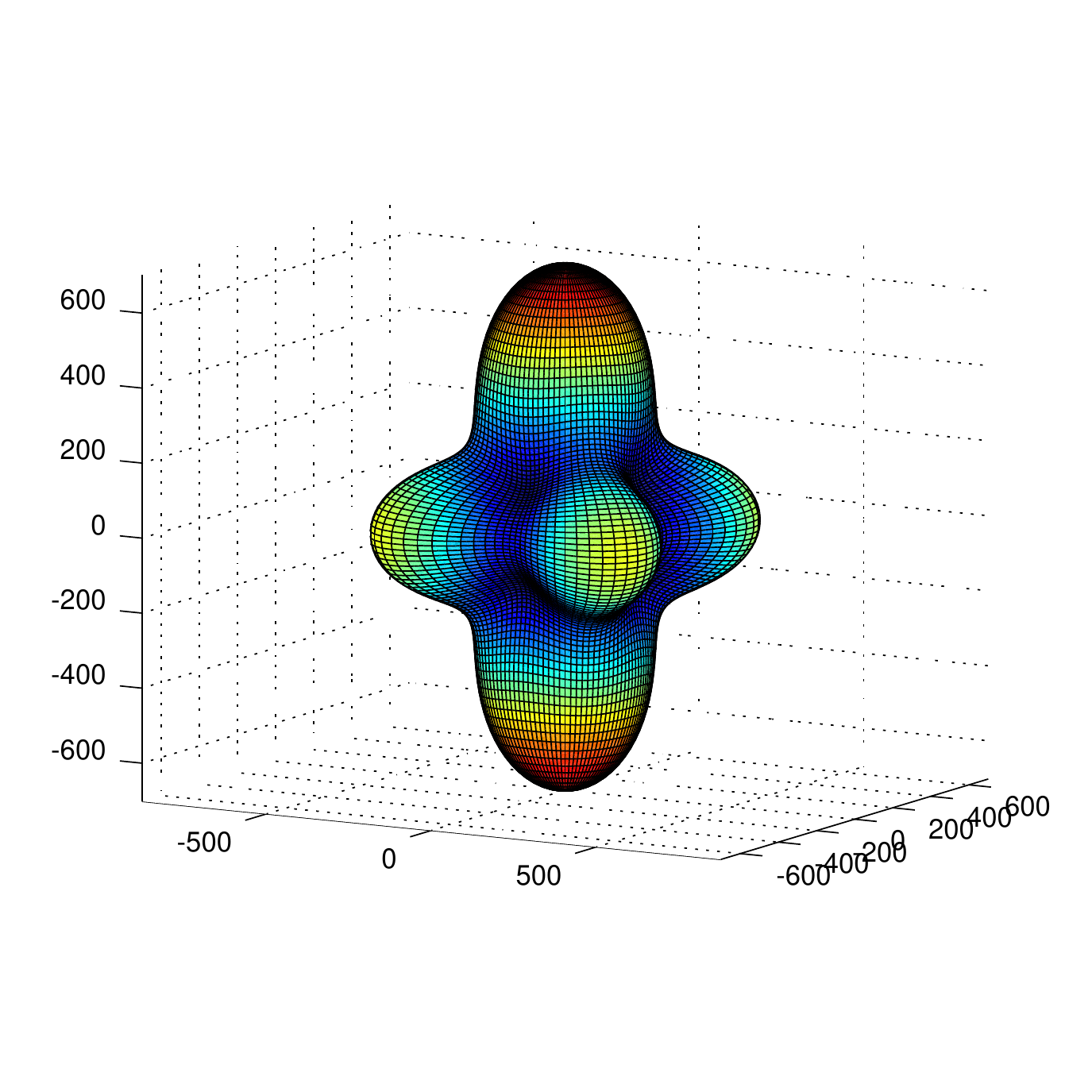}} 
\subfigure[]{\includegraphics[trim=0.5cm 0.5cm 0.5cm 0.5cm,height=1.6in,width=1.6in]{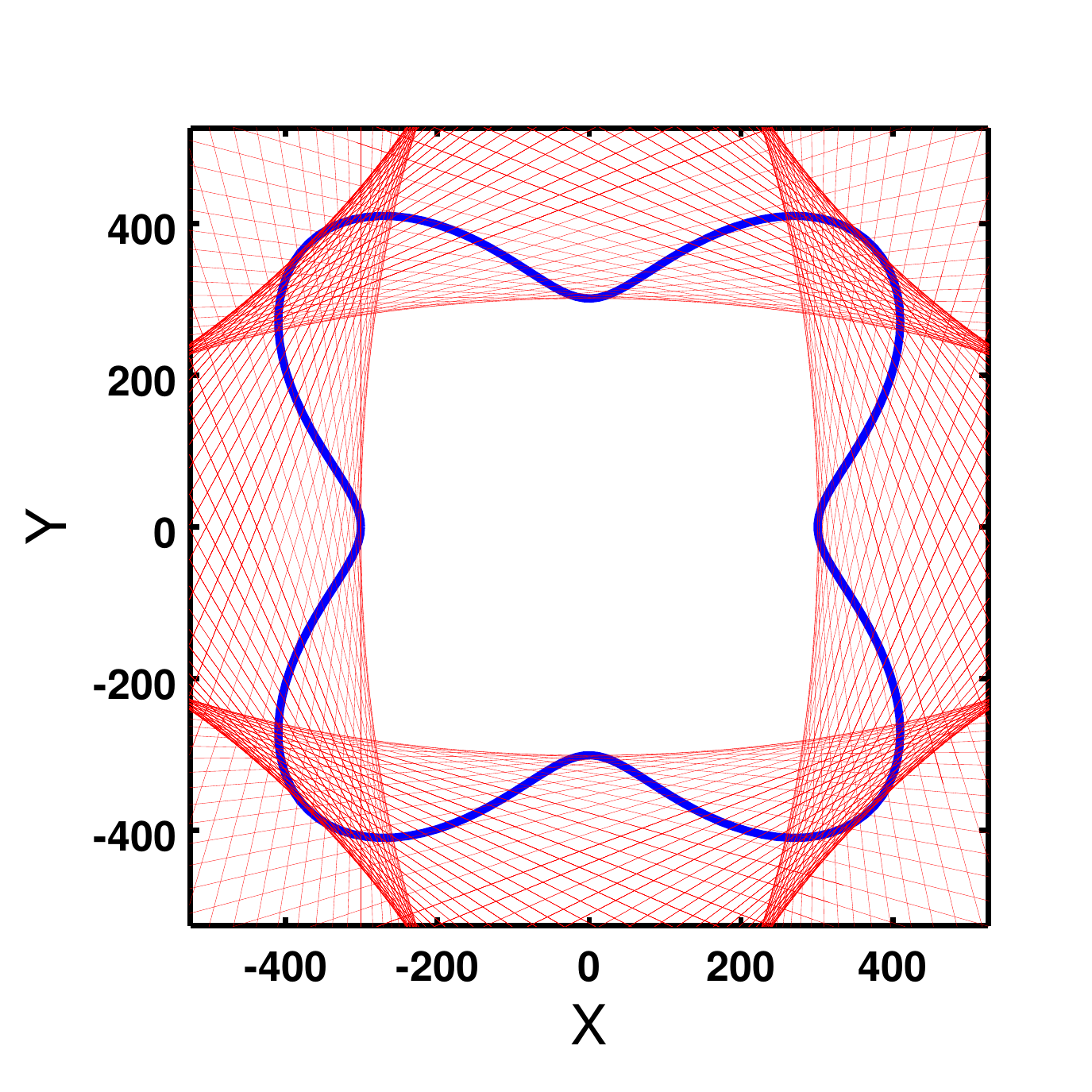}}
\subfigure[]{\includegraphics[trim=0.5cm 0.5cm 0.5cm 0.5cm,height=1.6in,width=1.6in]{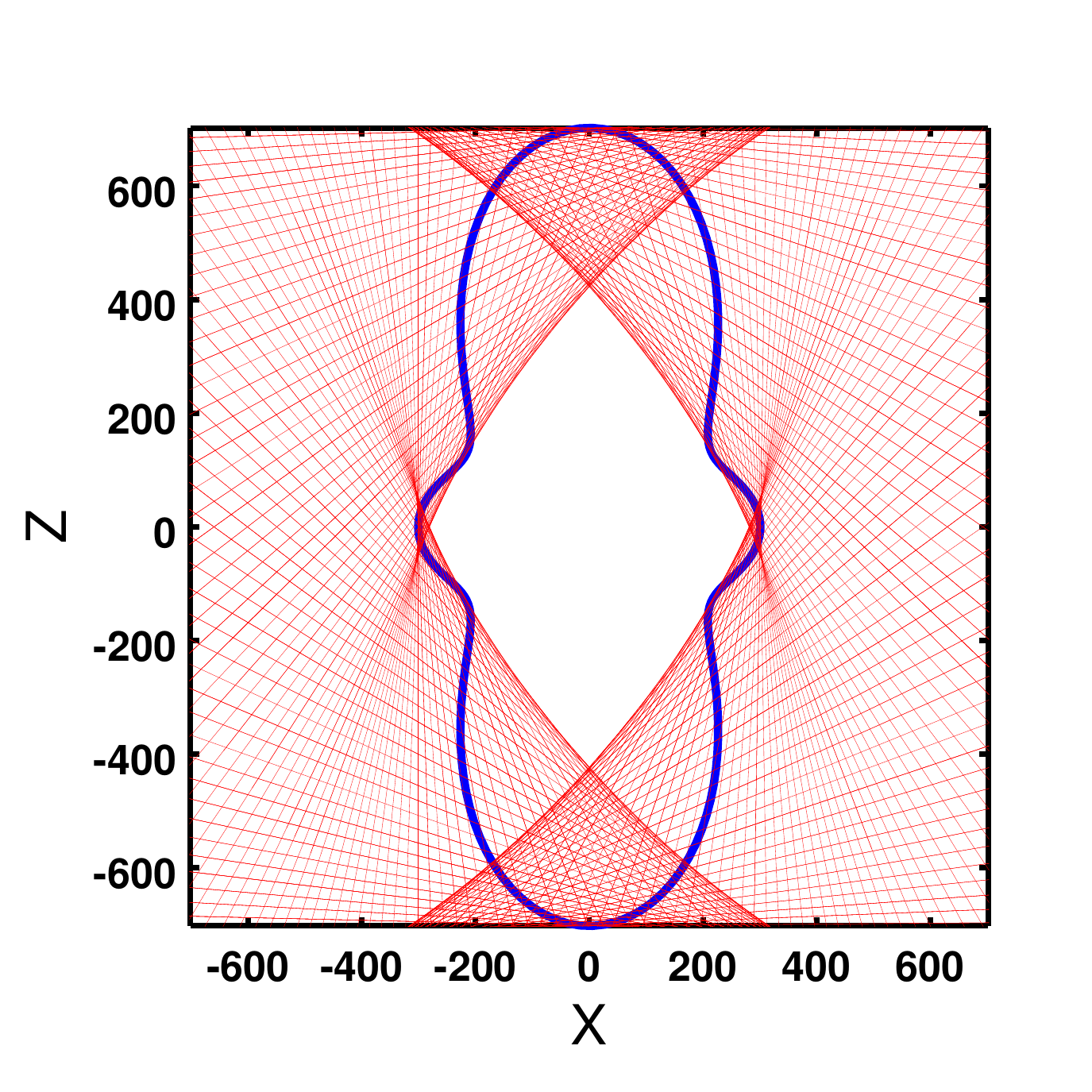}}
\caption{(a) 3-D polar plot and corresponding (b) $xy$ and (c) $xz$ section along with equilibrium Wulff shape for tetragonal prism. (d) 3-D polar plot and corresponding (e) $xy$ and (f) $xz$ section along with equilibrium Wulff shape for tetragonal plate. (g) 3-D polar plot and corresponding (h) $xy$ and (i) $xz$ section along with equilibrium Wulff shape for tetragonal di-pyramid.}\label{F:3DpolP4}
\end{figure}

In Fig.~\ref{F:3DpolP4} (a), (d) and (g), we show the 3-D polar plots for tetragonal prism, plate and di-pyramid
morphologies -- obtained using $P^2$ and $P^4$ (and the other higher order tensor terms
are assumed to be zero). Fig.~\ref{F:3DpolP4} (a) is obtained using $m_1 = 200.0$, $m_2 = 
1000.0$, $m_3= -30.0$, $m_4 = 291.0$, and $p_1 = p_2 = 1$; Fig.~\ref{F:3DpolP4} (d) is obtained using
$m_1 = 200.0$, $m_2 = 10.0$, $m_3= -90.0$, $m_4 = 100.0$,$p_1 = 2.0$ and $p_2 = 0.1$; 
Fig.~\ref{F:3DpolP4} (g) is obtained using $m_1 = 300.0$, $m_2 = 700.0$, 
$m_3= 750.0$, $m_4 = 100.0$, $p_1 = 1$ and $p_2 = 2$. 

The $xy$ and $xz$ sections of the prism, plate and di-pyramid are shown using blue lines in Fig.~\ref{F:3DpolP4} 
(b) and (c); Fig.~\ref{F:3DpolP4} (e) and (f); 
and, Fig.~\ref{F:3DpolP4} (h) and (i), respectively. The Wulff construction on these
sections of the free energy polynomial are shown by the red lines; the inner envelope indicating the
equilibrium shape in those sections is clearly seen in these plots. However, note that the equilibrium shapes
are qualitative in the sense that they are constructed using the free energy polynomial; for quantitative
shapes, the same figures have to be constructed using interfacial free energy plots. 

In the case of tetragonal prism, a diamond shape is seen on the $xy$ section of the polar plot 
due to the free energy minima that appear along $\langle 110 \rangle$ directions; on the other hand, the
$xz$ section drawn perpendicular to $(110)$ cut plane (see Fig.~\ref{F:3DpolP4} (c)), shows elongated c-axis 
with rounded corners. 
In the case of tetragonal plate also, a diamond shape is seen on the $xy$ section of the polar plot; 
on the other hand, on $xz$ section (Fig.~\ref{F:3DpolP4} (f)),
the precipitate is rectangular with elongation along the $z$-axis. Here, we draw the attention of the reader
to the fact that in the Wulff shape, $x$-axis of the figure is along the diagonal and hence is $\sqrt{2}$ times
the crystallographic $a$-axis. This is important in calculating the aspect ratio of the plates.
In the case of tetragonal di-pyramid, a squarish shape is seen in the $xy$ section of the polar plot;
we can see equilibrium $(011)$ facets in the $xz$ cross section -- shown in Fig.~\ref{F:3DpolP4} (i).

\begin{figure}[htpb]
\centering
\subfigure[]{\includegraphics[height=2.0in,width=2.0in]{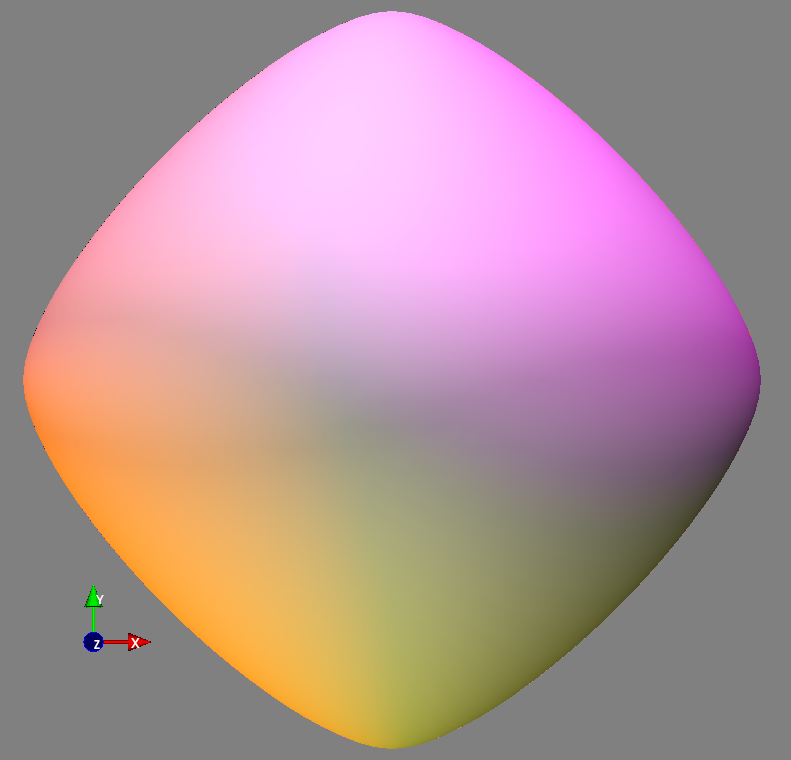}}
\subfigure[]{\includegraphics[height=2.0in,width=2.0in]{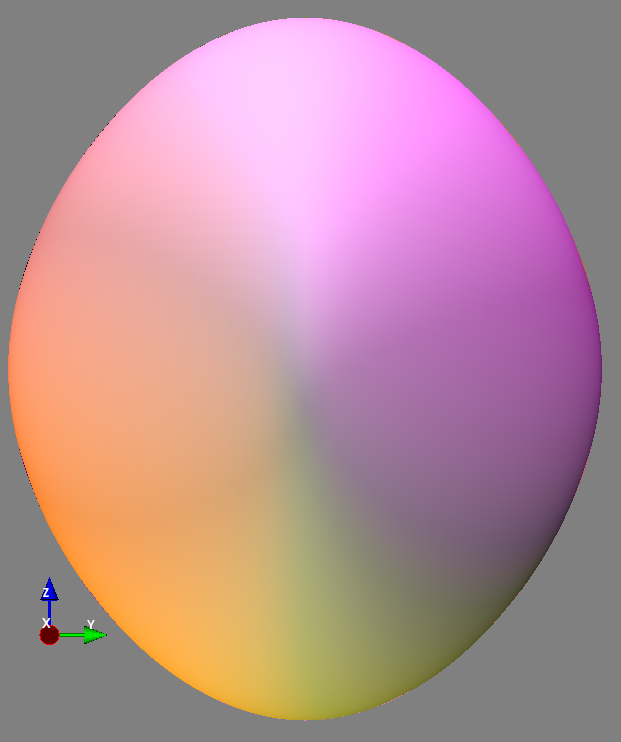}}\\
\subfigure[]{\includegraphics[height=2in,width=2in]{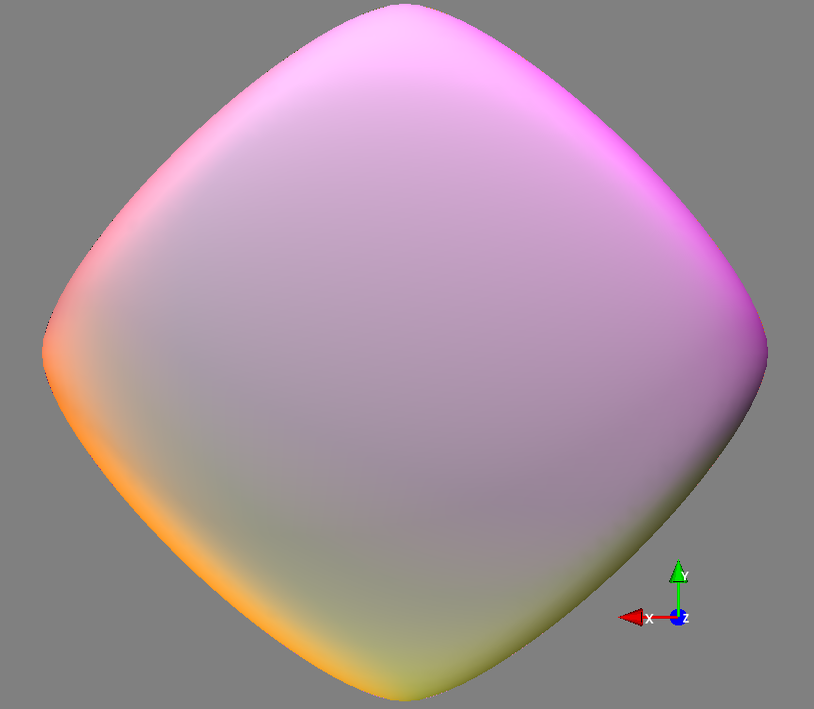}} 
\subfigure[]{\includegraphics[height=2in,width=2in]{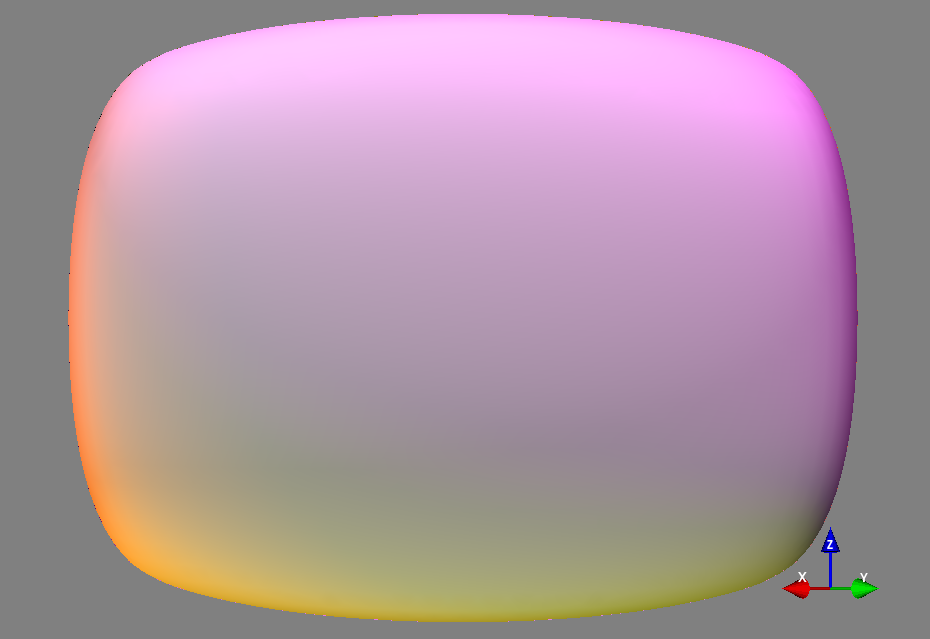}}\\
\subfigure[]{\includegraphics[height=2in,width=2in]{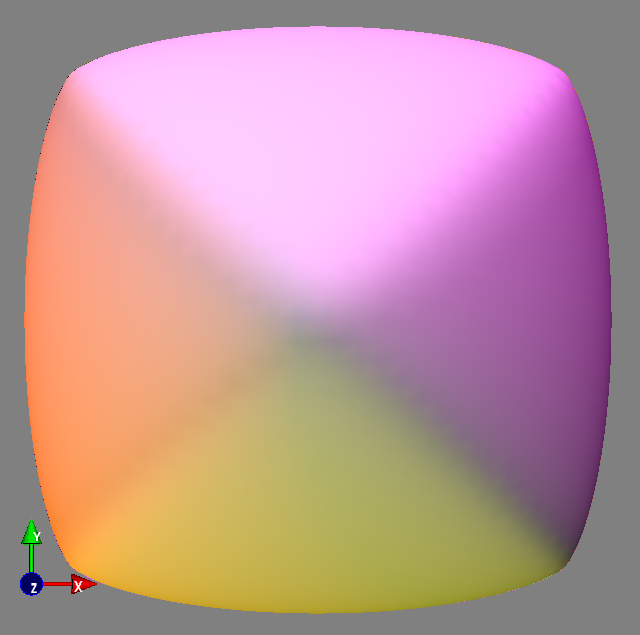}} 
\subfigure[]{\includegraphics[height=2in,width=2in]{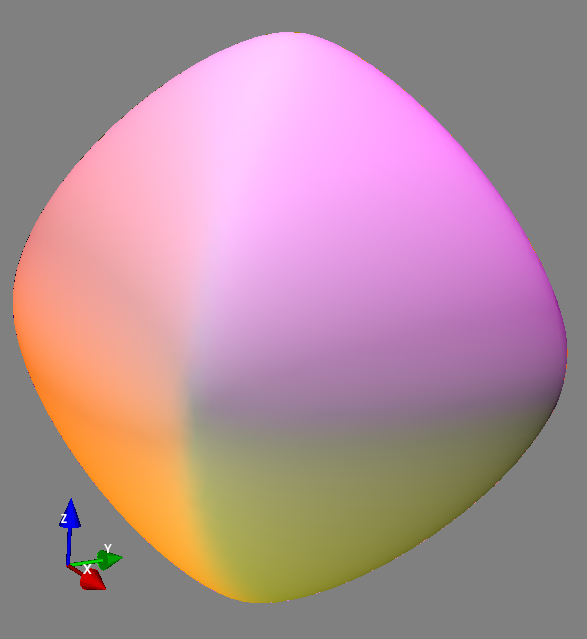}}
\caption{View from (a) $\langle001\rangle$ and (b) $\langle100\rangle$ directions of a tetragonal prism morphology, from (c) $\langle001\rangle$ and (d) $\langle110\rangle$ directions of a tetragonal plate morphology, and from (e) $\langle001\rangle$ and (f) $\langle1\bar{1}0\rangle$ directions of a tetragonal di-pyramid morphology.}
\label{F:3DmorphP4}
\end{figure}

In Fig.~\ref{F:3DmorphP4}, the 3-D precipitate morphologies obtained using our numerical simulations are shown.
In all these simulations we start with a spherical precipitate of size twelve at the centre of the simulation
cell.

The parameters that give rise to the polar plot Fig.~\ref{F:3DpolP4} (a) lead to tetragonal prism
(four fold symmetry in the $ab$-plane with an elongated c-axis with rounded corners) as seen in Fig.~\ref{F:3DmorphP4} 
(a) and (b). The morphologies correspond to 300 time units; Fig.~\ref{F:3DmorphP4} (a)
is the view from $\langle001\rangle$ direction and Fig.~\ref{F:3DmorphP4} (b) is the view from $\langle100\rangle$ direction.
The parameters that give rise to the polar plot Fig.~\ref{F:3DpolP4} (d) lead to tetragonal plate
(four fold symmetry in the $ab$-plane with a shortened c-axis) as seen in Fig.~\ref{F:3DmorphP4} 
(c) and (d). The morphologies correspond to 360 time units; Fig.~\ref{F:3DmorphP4} (c)
is the view from $\langle001\rangle$ direction and Fig.~\ref{F:3DmorphP4} (d) is the view from $\langle110\rangle$ direction.
The parameters that give rise to the polar plot Fig.~\ref{F:3DpolP4} (g) lead to tetragonal di-pyramid
(a shape bounded by eight (111) type of planes) as seen in Fig.~\ref{F:3DmorphP4} 
(e) and (f). The morphologies correspond to 121 time units; Fig.~\ref{F:3DmorphP4} (e)
is the view from $\langle001\rangle$ direction and Fig.~\ref{F:3DmorphP4} (f) is the view from 
$\langle1\bar{1}0\rangle$ direction. Thus, we see that the qualitative Wulff plots obtained by
us are consistent with the equilibrium morphologies seen in the simulations.

\begin{figure}[tbph]
\centering
\subfigure[]{\includegraphics[height=1.2in,width=1.6in]{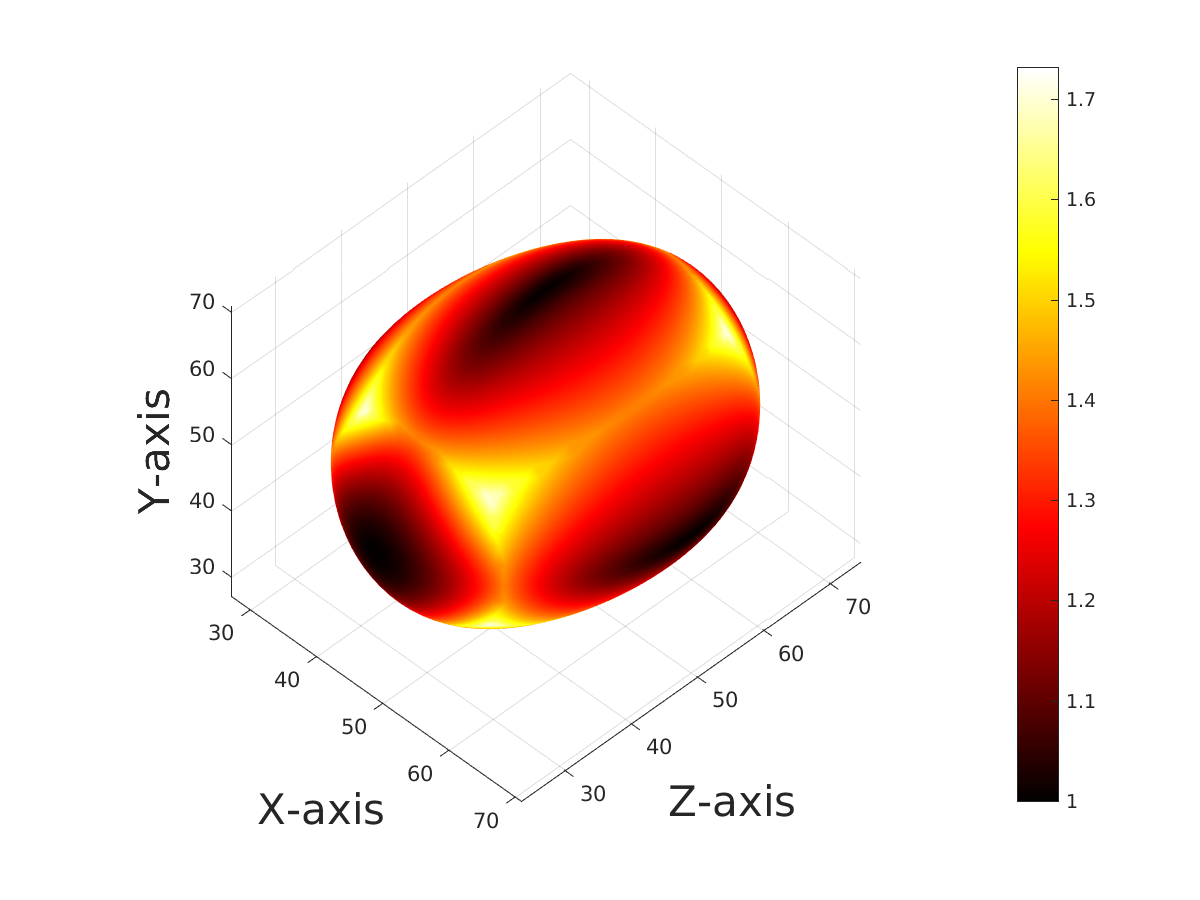}}
\subfigure[]{\includegraphics[height=1.2in,width=1.6in]{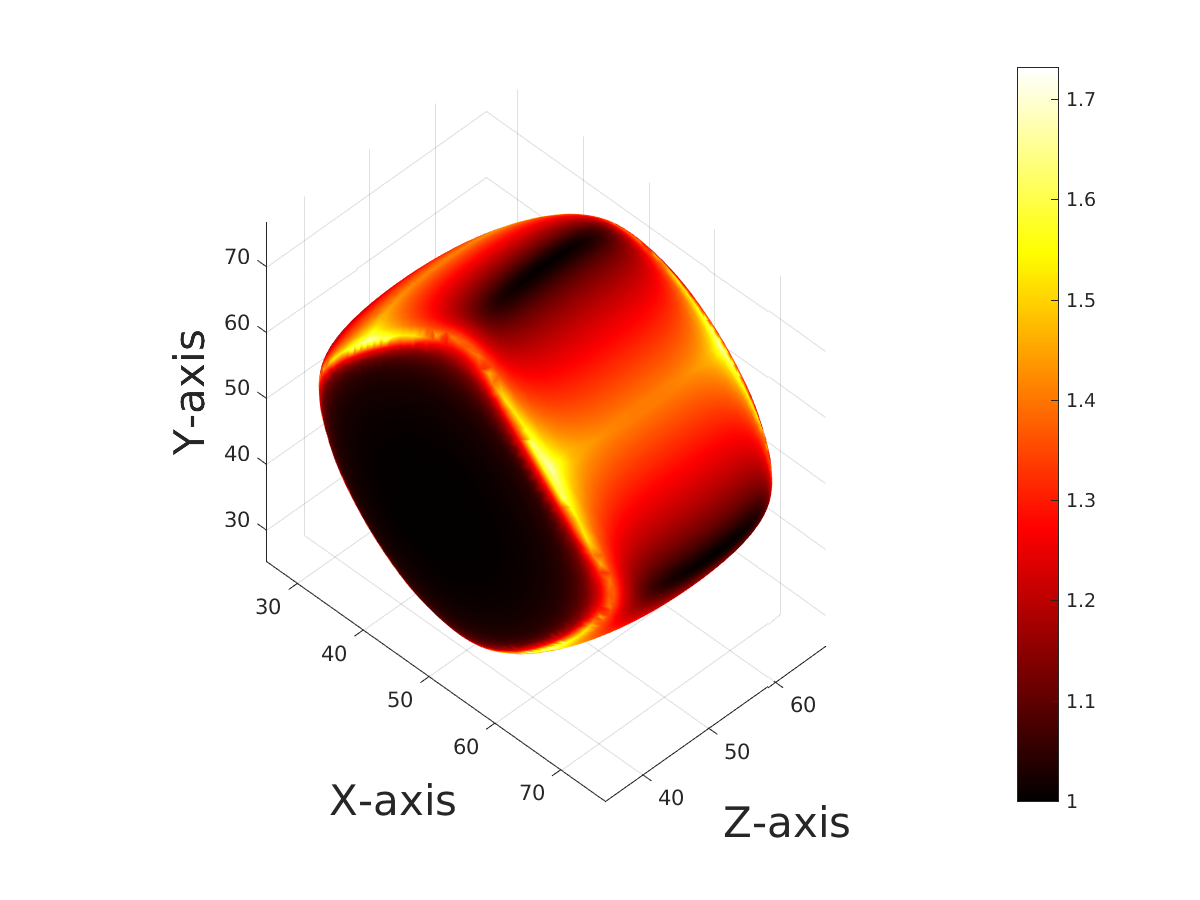}}
\subfigure[]{\includegraphics[height=1.2in,width=1.6in]{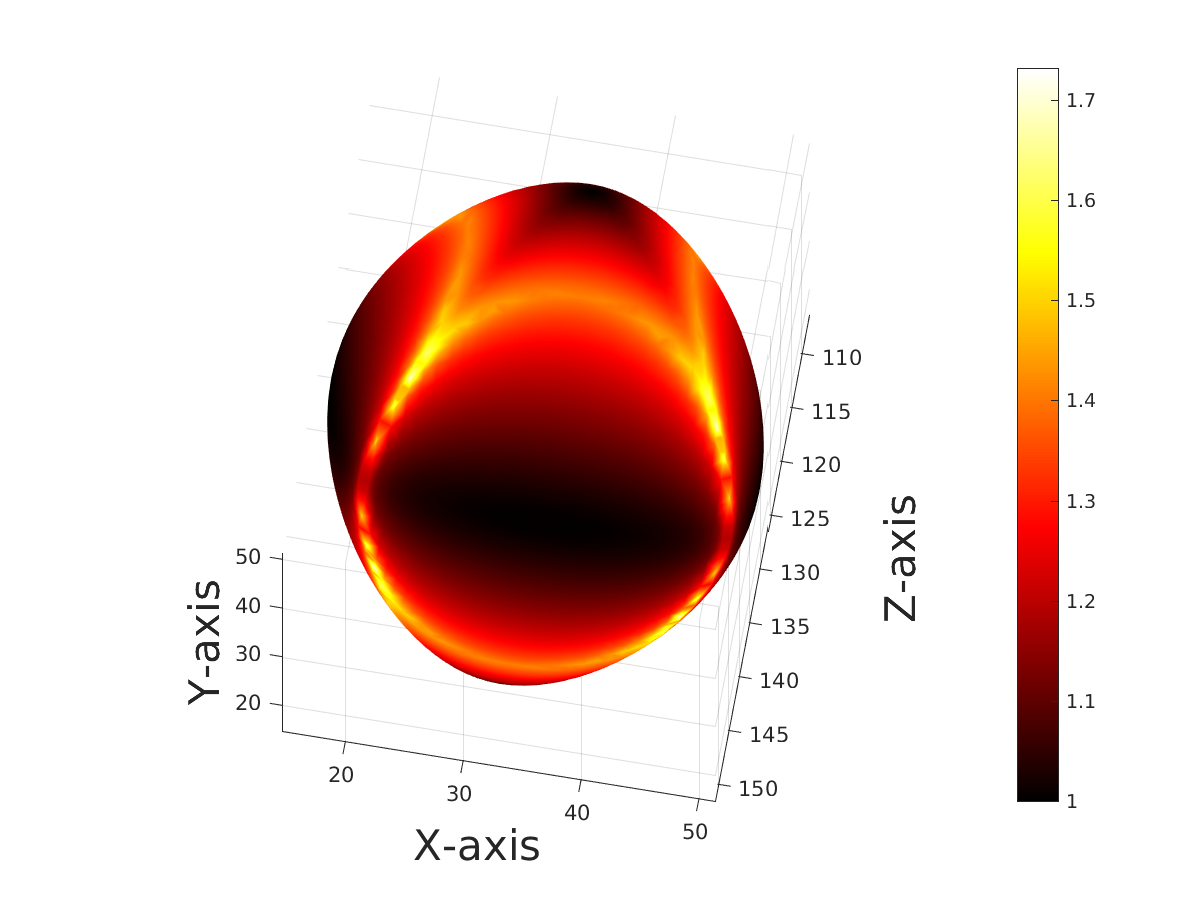}}
\caption{Surface normal distribution -- (a) for tetragonal prism morphology;
(b) for tetragonal plate morphology; and, for tetragonal di-pyramid morphology.}
\label{F:3DsurfP4}
\end{figure}

In order to better understand the 3D morphologies, we calculate and plot the surface normals at 
the precipitate-matrix interface (by identifying
the $c=0.5$ surface as the interface). In Fig.~\ref{F:3DsurfP4} we present the 3-D surface normal distribution
for the prism, plate and di-pyramid ((a), (b) and (c) respectively) for the precipitates shown in Fig.~\ref{F:3DmorphP4}.
Formation of $(110)$ facets are visible in Fig.~\ref{F:3DsurfP4} (a); 
we can also see that the planes perpendicular to the c-axes are rounded.
Formation of plates (that is, the thickness is small as compared to the dimensions in the $ab$-plane) with
$(001)$ and $(110)$ facets are clearly visible in Fig.~\ref{F:3DsurfP4} (b).
The formation of sharp corners and $(011)$ facets can be clearly seen in Fig.~\ref{F:3DsurfP4} (c).

Further characterisation of the morphologies and their evolution during the growth can be achieved using
 the aspect ratio calculations. Unlike in 2-D and squarish shapes, 
in these cases (namely, prism, plate and di-pyramids) a large number of aspect ratios can be defined. 
We have decided to define the following aspect
ratios. For all the cases, in the $ab$-plane (which possesses a four-fold symmetry), we define the aspect
ratio $\rho$ as the ratio of size along $\langle 100 \rangle$ 
to that of $\langle 110 \rangle$ -- $\rho^{pl}_{110}$, $\rho^{pr}_{110}$ and $\rho^{dip}_{110}$ for plate, prism 
and di-pyramid morphologies respectively. In order to understand the tetragonality of the morphology, we 
define $\rho^{pl}_{001}$, $\rho^{pr}_{001}$ and $\rho^{dip}_{001}$ 
as the ratios of the sizes of the plates, prisms and di-pyramids, respectively, along $\langle 100 \rangle$ 
to that along $\langle 001 \rangle$. 

\begin{figure}[tpbh]
\centering
\includegraphics[trim=0.5cm 0.5cm 0.5cm 0.5cm,height=2.3in,width=2.3in]{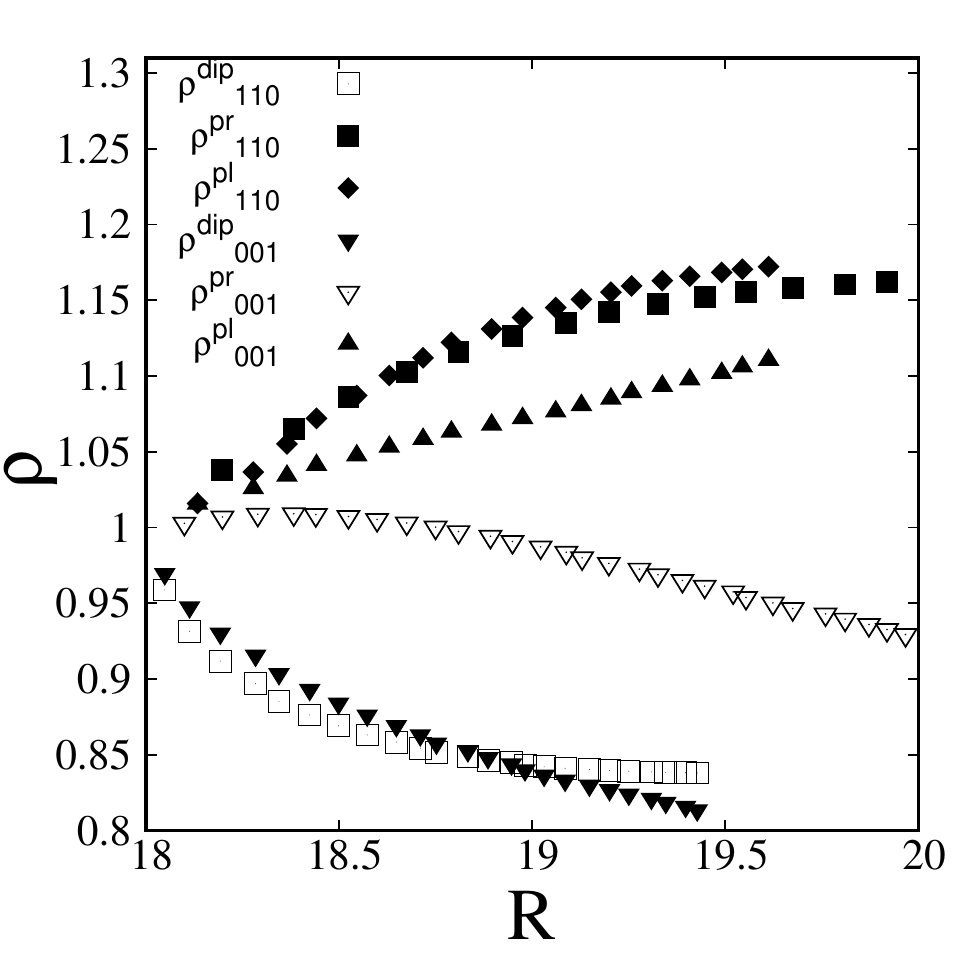}
\caption{Variation of aspect ratios ($\rho$) with effective radii for tetragonal precipitate morphologies obtained using $P^4$ polynomial.}\label{F:aspect_tetraP4}
\end{figure}

The aspect ratios ($\rho$) for morphologies shown in Fig.~\ref{F:3DmorphP4} are shown in Fig.~\ref{F:aspect_tetraP4}.
We show the variation of $\rho$ with effective radius $R$ (that is, the radius of a spherical precipitate with the same
volume) of the precipitate. In in $ab$-plane, in the case of plate and prism morphologies
$\langle 110 \rangle$ facets form, and, in the case of di-pyramid morphology, 
$\langle 100 \rangle$ facets form. Hence, both $\rho^{pl}_{110}$ and $\rho^{pr}_{110}$ attain values above unity (namely,
1.182 and 1.169, respectively), while for the di-pyramid morphology, $\rho^{dip}_{001}$ attains a value less than unity 
(namely, 0.83). 

In the case of prism and di-pyramid precipitate morphologies, $\rho^{pr}_{001}$ and $\rho^{dip}_{001}$ attain values 
less than unity (0.856 for prism and 0.814 for di-pyramid morphology). This is because the precipitate is elongated along
c-axis is as compared to a- and b-axes. For the plate morphology, $\rho^{pr}_{001}$ attains values greater than unity (1.110).
This is because the dimensions in the $ab$-plane are larger as compared to the thickness of the precipitate.

\subsubsection{3-D equilibrium precipitate morphologies with more than one family of planes}

In these systems, there could be precipitate morphologies that consist of two sets of planes, namely,
$( 1 0 0 )$ and $( 1 1 1 )$ or morphologies that consists of three sets of planes, namely.  
$\{ 1 0 0 \}$, $\{ 0 0 1 \}$ and $ \{ 1 1 1 \}$. These lead to truncated paralleopiped morphologies. 
 
\begin{figure}[tpbh]
\centering
\subfigure[]{\includegraphics[height=2in,width=2in]{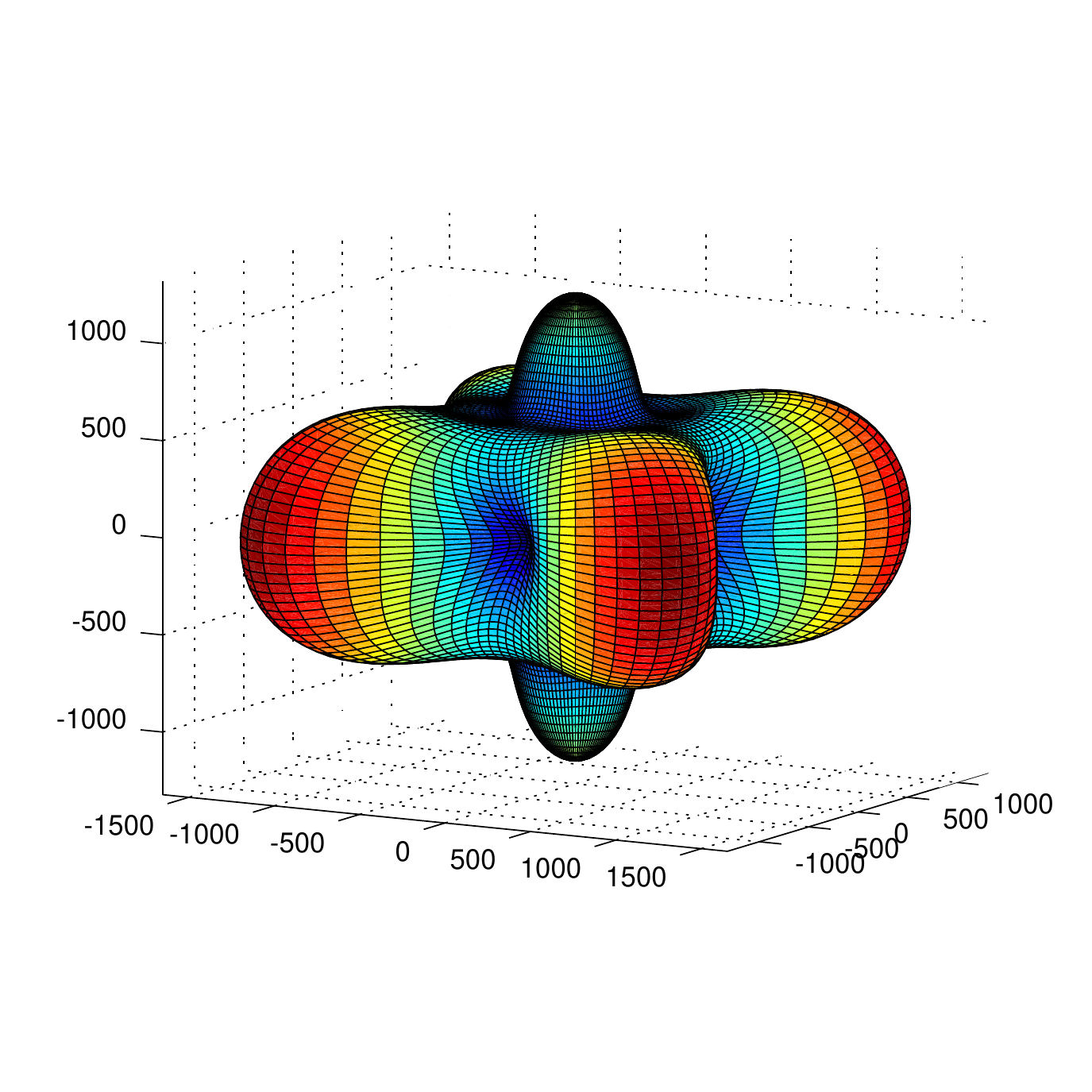}} 
\subfigure[]{\includegraphics[height=2in,width=2in]{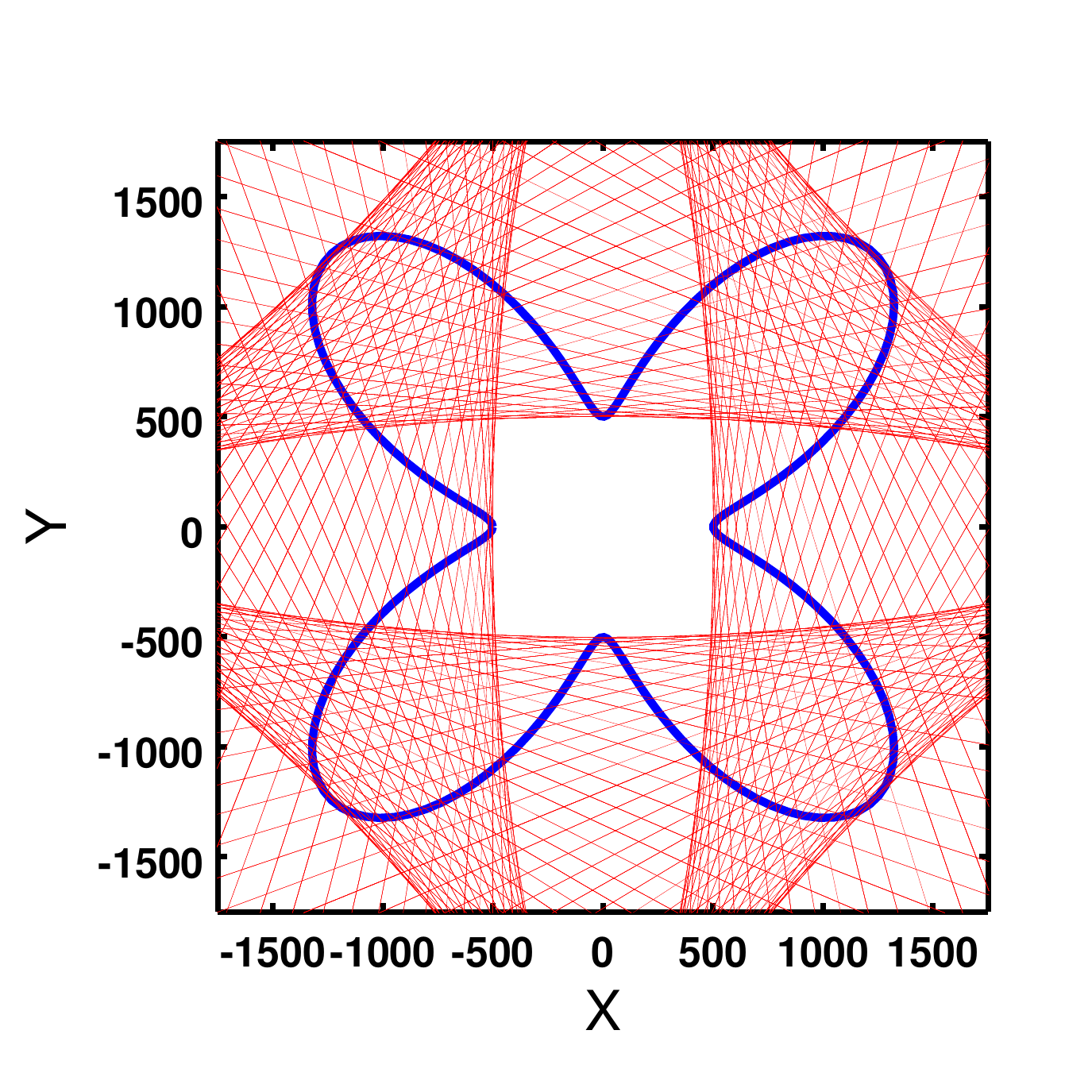}}\\
\subfigure[]{\includegraphics[height=2in,width=2in]{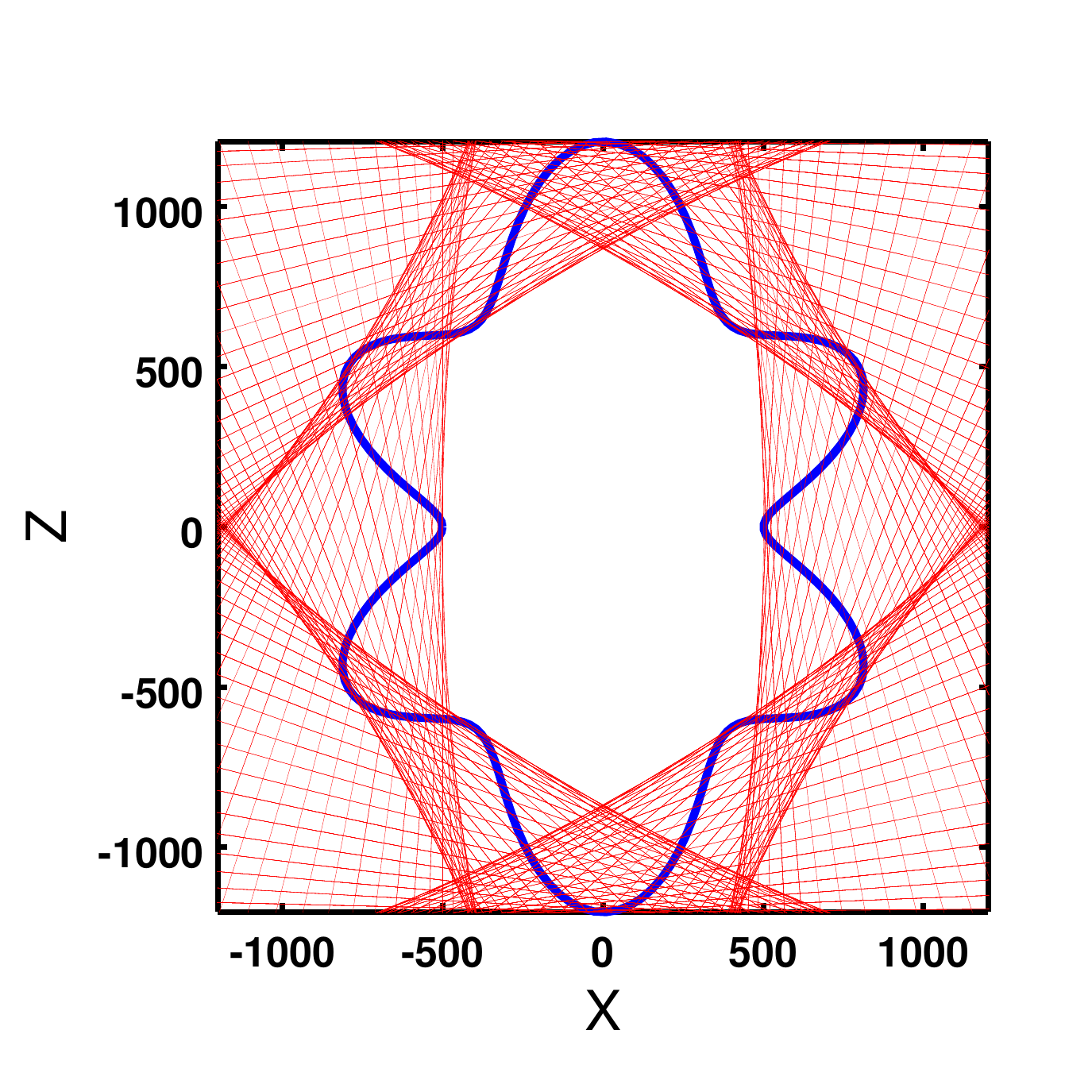}}
\subfigure[]{\includegraphics[height=2in,width=2in]{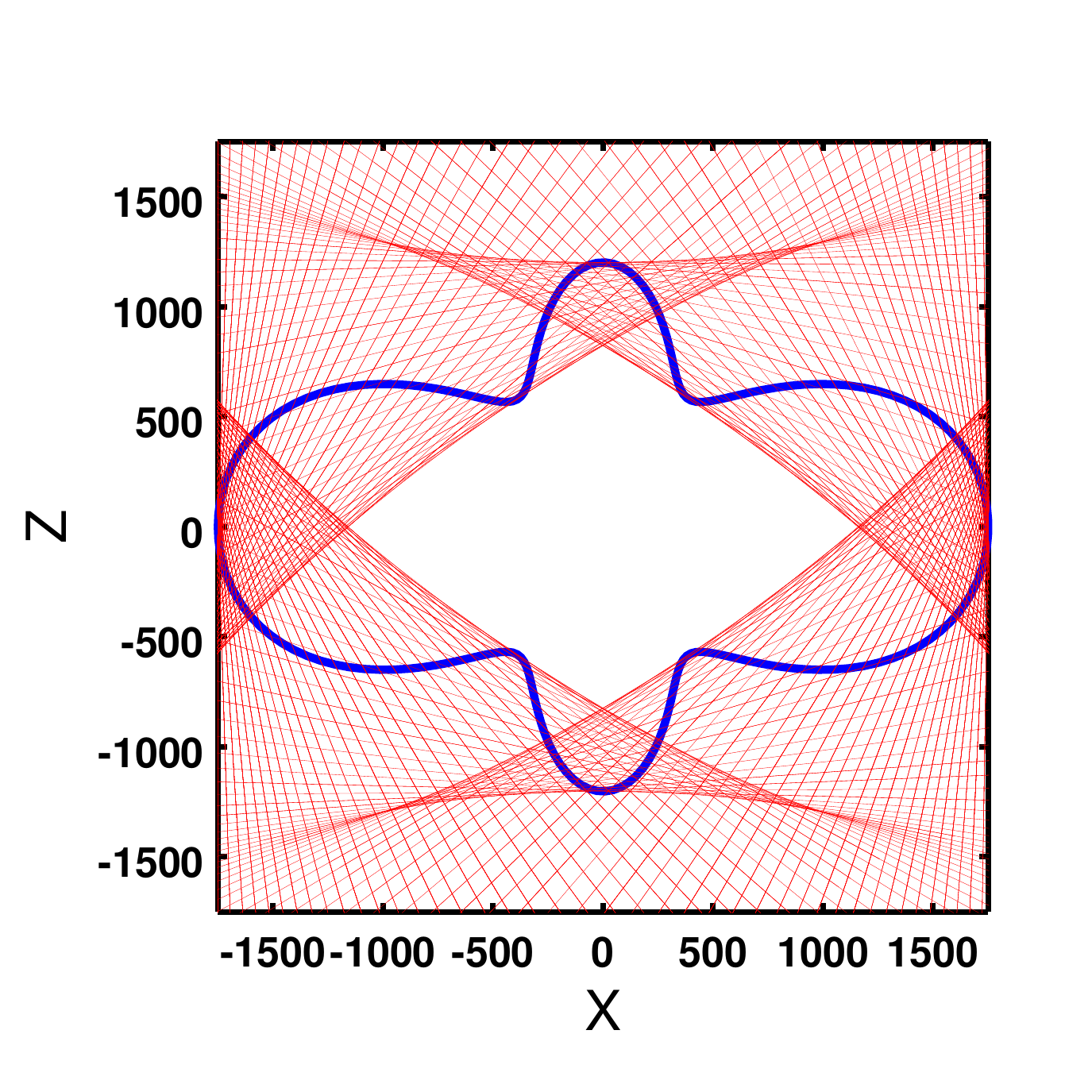}}
\caption{Morphology with two types of facets obtained using $P^6$: (a) 3-D polar plot; (b) $xy$ section (at the centre), 
and $xz$ section (at the centre) perpendicular to (c) $\langle 100\rangle$ and (d) $\langle 110\rangle$ 
directions; in all the 2D sections, the Wulff constructions are also shown.}\label{F:3DpolP6_2facet}
\end{figure}

\begin{figure}[tpbh]
\centering
\subfigure[]{\includegraphics[height=1.5in,width=1.5in]{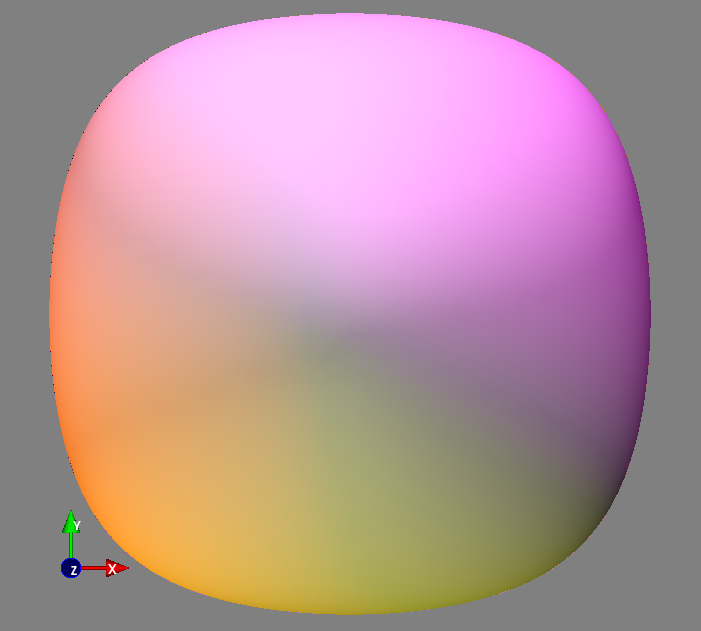}}
\subfigure[]{\includegraphics[height=1.5in,width=1.0in]{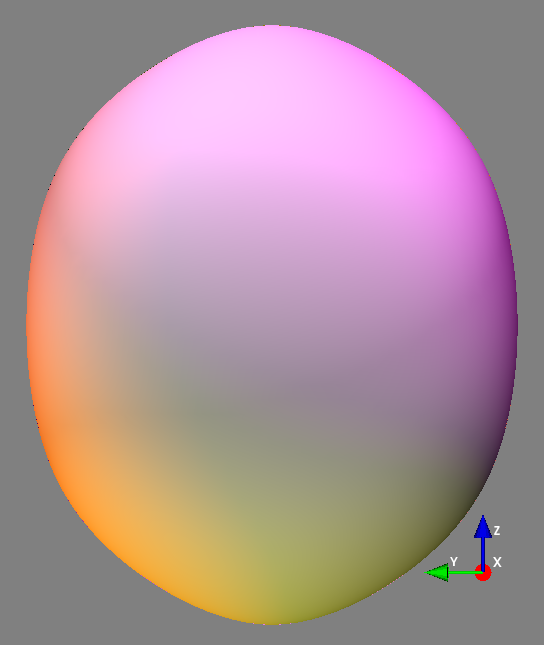}}
\subfigure[]{\includegraphics[height=1.5in,width=2.0in]{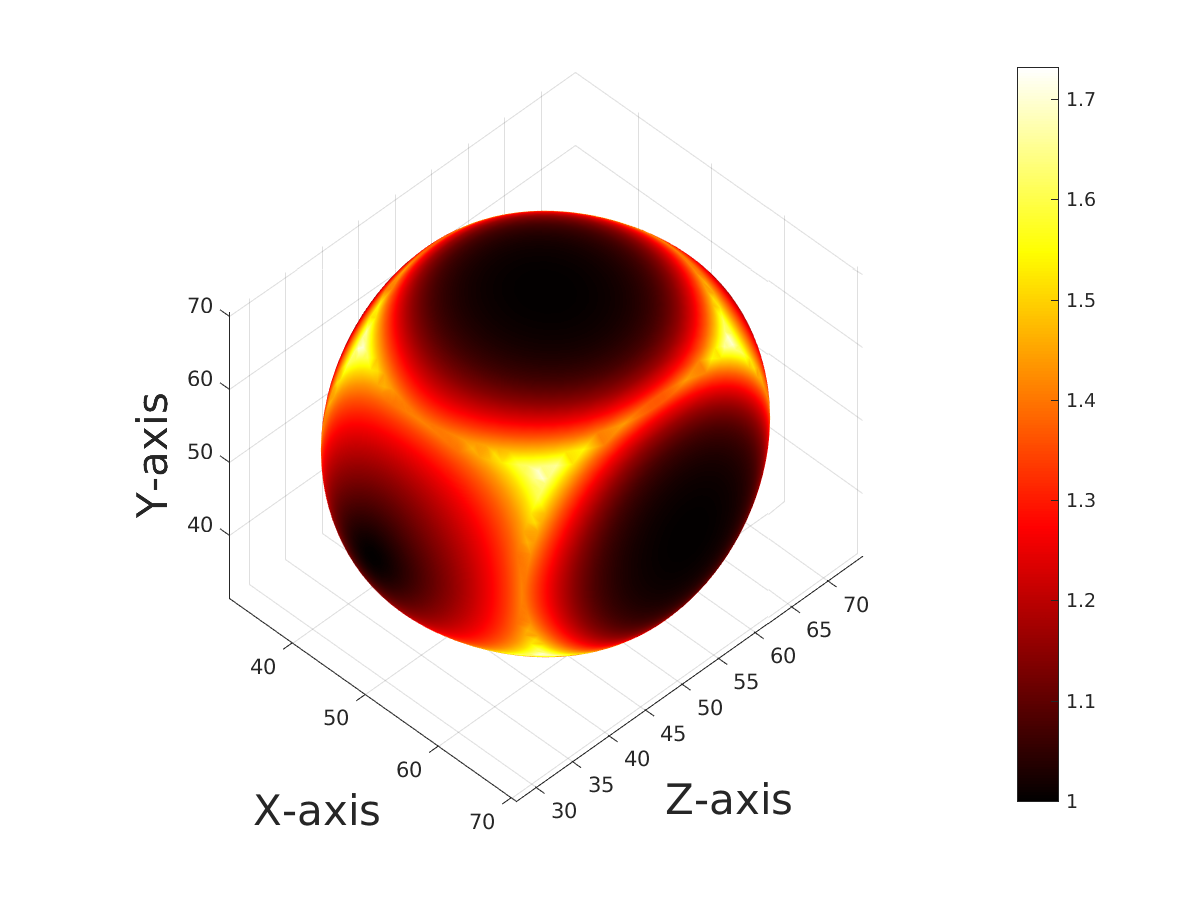}}
\caption{ View from (a) $\langle001\rangle$ and (b) $\langle100\rangle$ of a precipitate with
two types of facets obtained by incorporating sixth-rank gradient tensor coefficients. The  
surface normal distribution for the morphology is shown in (c). }\label{F:3D_grad_6_2facet}
\end{figure}

The free energy polynomial assuming that only $P^6$ and $P^2$ are non-zero is shown in Fig.~\ref{F:3DpolP6_2facet} (a);
specifically, we have used $n_1 = 500.0$, $n_2 = 1200.0$, $n_3= 5000.1$, $n_4 = 5000.0$, $n_5 = 10.0$,
$n_6 = -5000$, $p_1=1$ and $p_2=2.4$. The $xy$ section, $xz$ section perpendicular to $\langle 100\rangle$ 
and the $xz$ section perpendicular to $\langle 110\rangle$ are shown (in blue) in 
Fig.~\ref{F:3DpolP6_2facet} (b), (c) and (d), respectively. The red lines are the Wulff construction lines.
Even though the Wulff construction on these 2-D sections show that there is minima along $\langle 100 \rangle$,
$\langle 011 \rangle$ and $\langle 111 \rangle$ directions, the equilibrium morphology in this system 
consists only of $(100)$ and $(111)$ facets. This is because the $(100)$ planes have
relatively lower energy as compared to $(011)$ planes. 
Our numerical simulations in which a spherical precipitate of size twelve placed at the centre of the simulation cell,
indeed leads to the expected morphology after $250$ time units as shown in Fig.~\ref{F:3D_grad_6_2facet}: (a) is
the view of the precipitate from $\langle 001 \rangle$ and (b) is the view of the precipitate from $\langle 100 \rangle$.
The Fig.~\ref{F:3D_grad_6_2facet} (c) shows the surface normal plot of this precipitate (generated by identifying the
interface at c = 0.5); the colour bar indicates $\sqrt{h^2 + k^2 + l^2}$ for $\langle h k l \rangle$ 
orientation. So, a numerical value of 1.0, 1.414 and 1.732 represent the planes containing 
$\langle 100 \rangle$, $\langle 110 \rangle$ and $\langle 111 \rangle$ directions respectively. Note
that the surface normal plot also shows that $(001)$ planes are missing in this morphology.

\begin{figure}[tpbh]
\centering
\subfigure[]{\includegraphics[height=2in,width=2in]{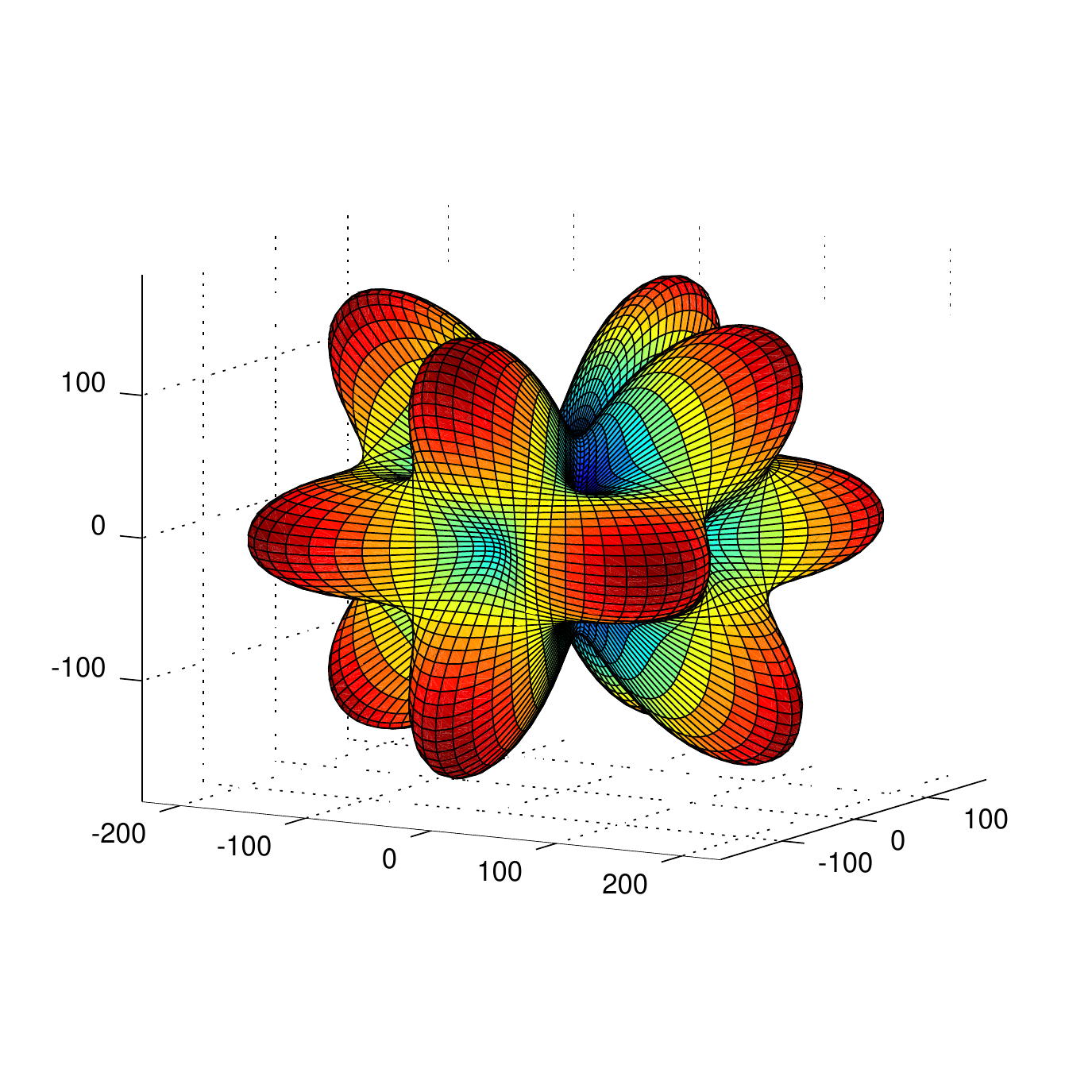}} 
\subfigure[]{\includegraphics[height=2in,width=2in]{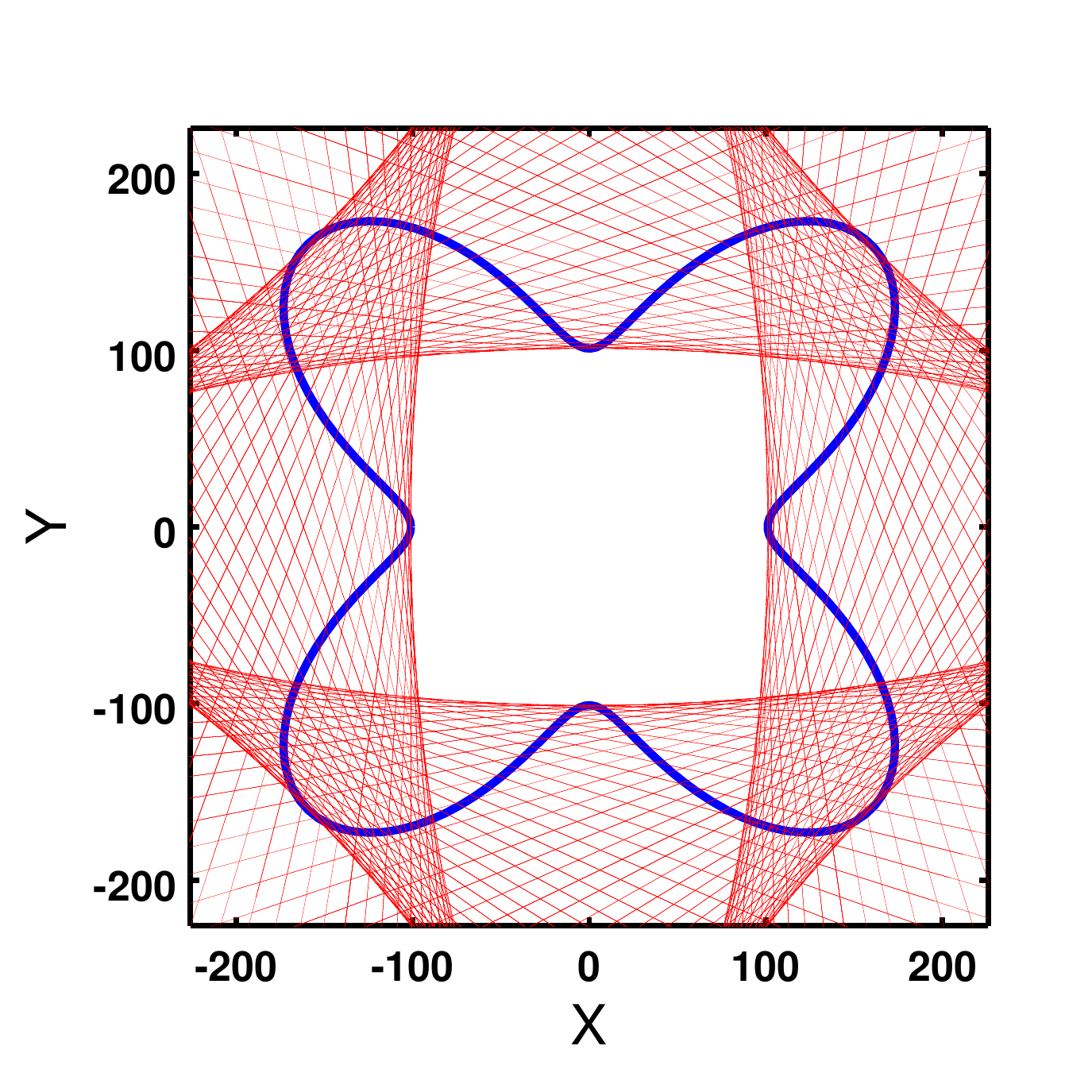}} \\
\subfigure[]{\includegraphics[height=2in,width=2in]{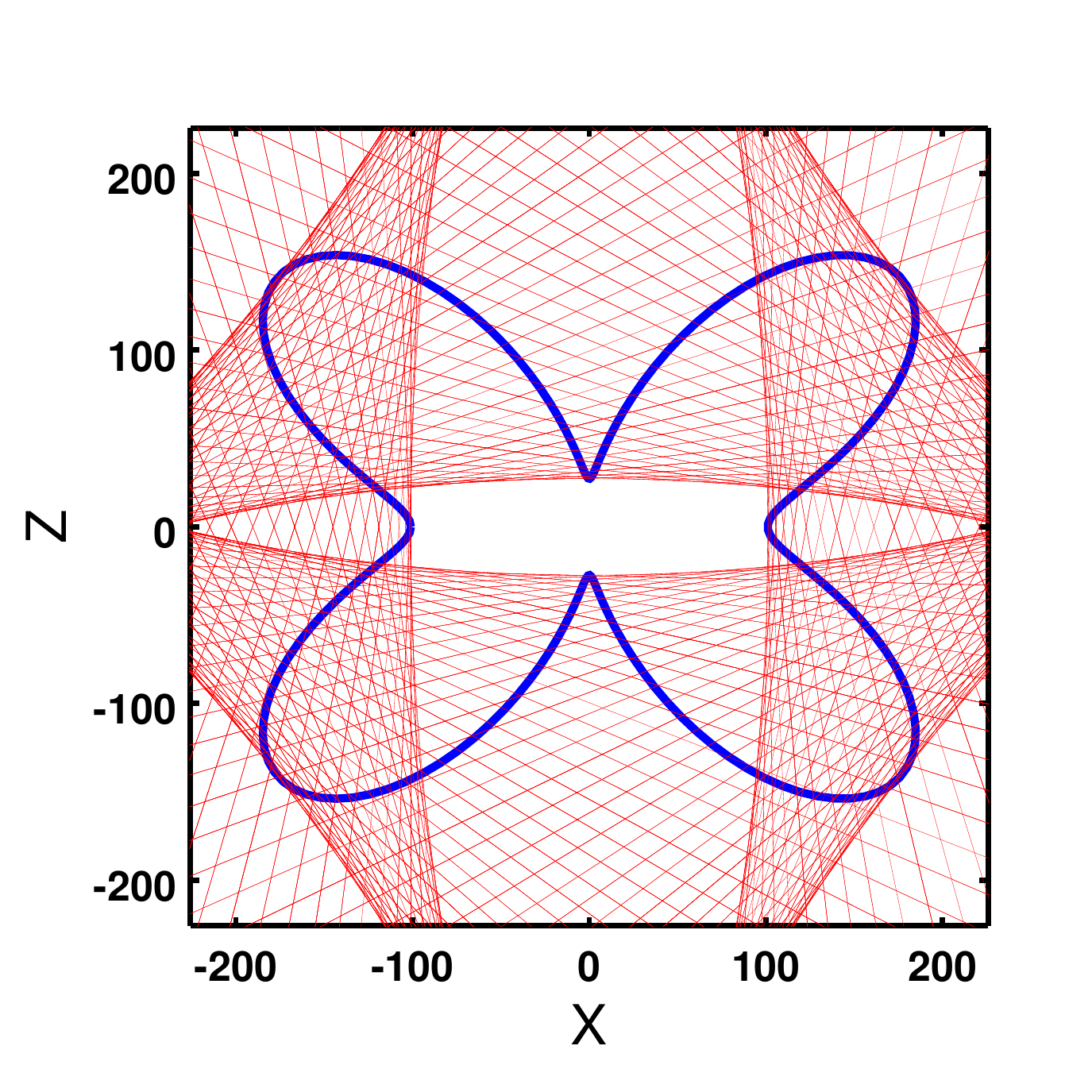}}
\subfigure[]{\includegraphics[height=2in,width=2in]{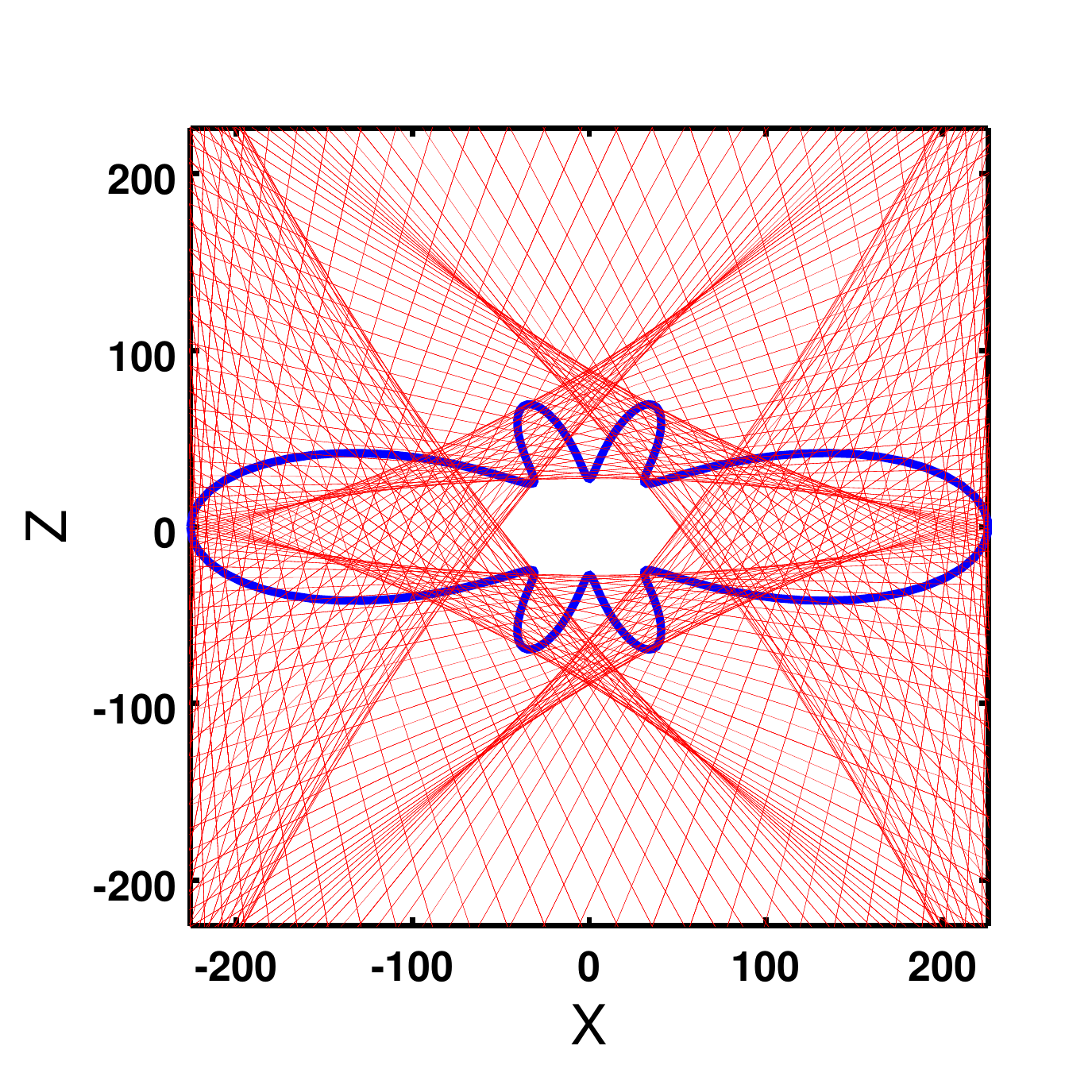}}
\caption{Morphology with three types of facets obtained using $P^6$: (a) 3-D polar plot; (b) $xy$ section (at the center);
$xz$ sections perpendicular to (c) $\langle 100\rangle$ and (d) $\langle 110\rangle$ directions. The Wulff
constructions are also shown in all the 2-D sections.}\label{F:3DpolP6_3facet}
\end{figure}

\begin{figure}[tpbh]
\centering
\subfigure[]{\includegraphics[height=1.5in,width=1.5in]{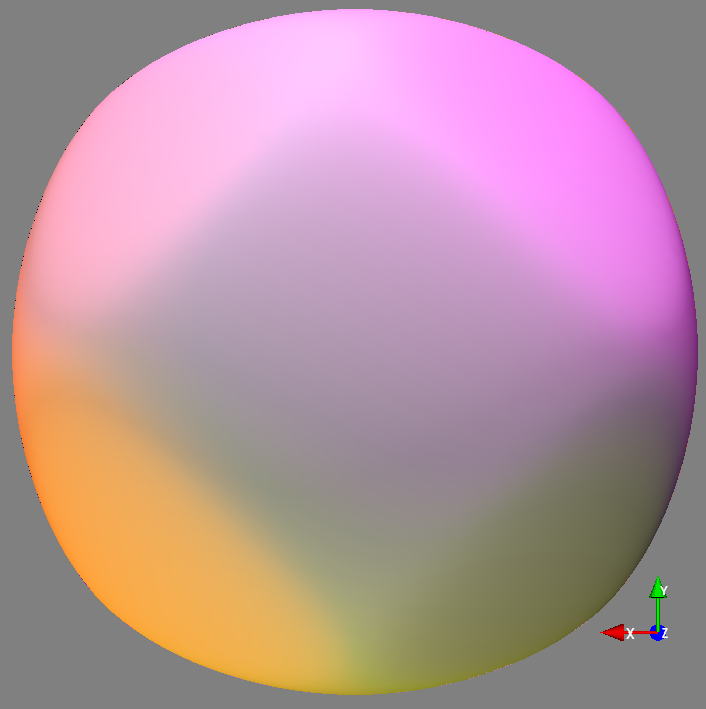}}
\subfigure[]{\includegraphics[height=1.5in,width=1.5in]{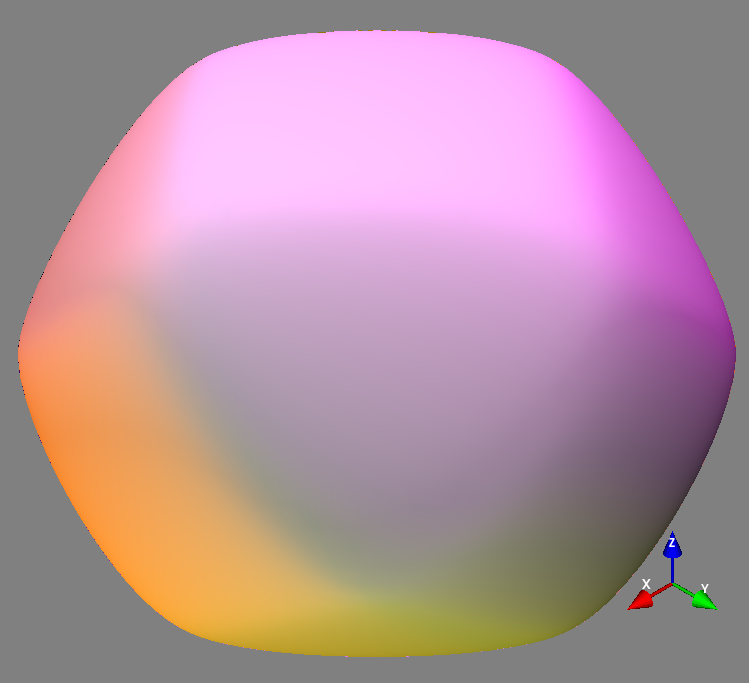}} 
\subfigure[]{\includegraphics[height=1.5in,width=2.0in]{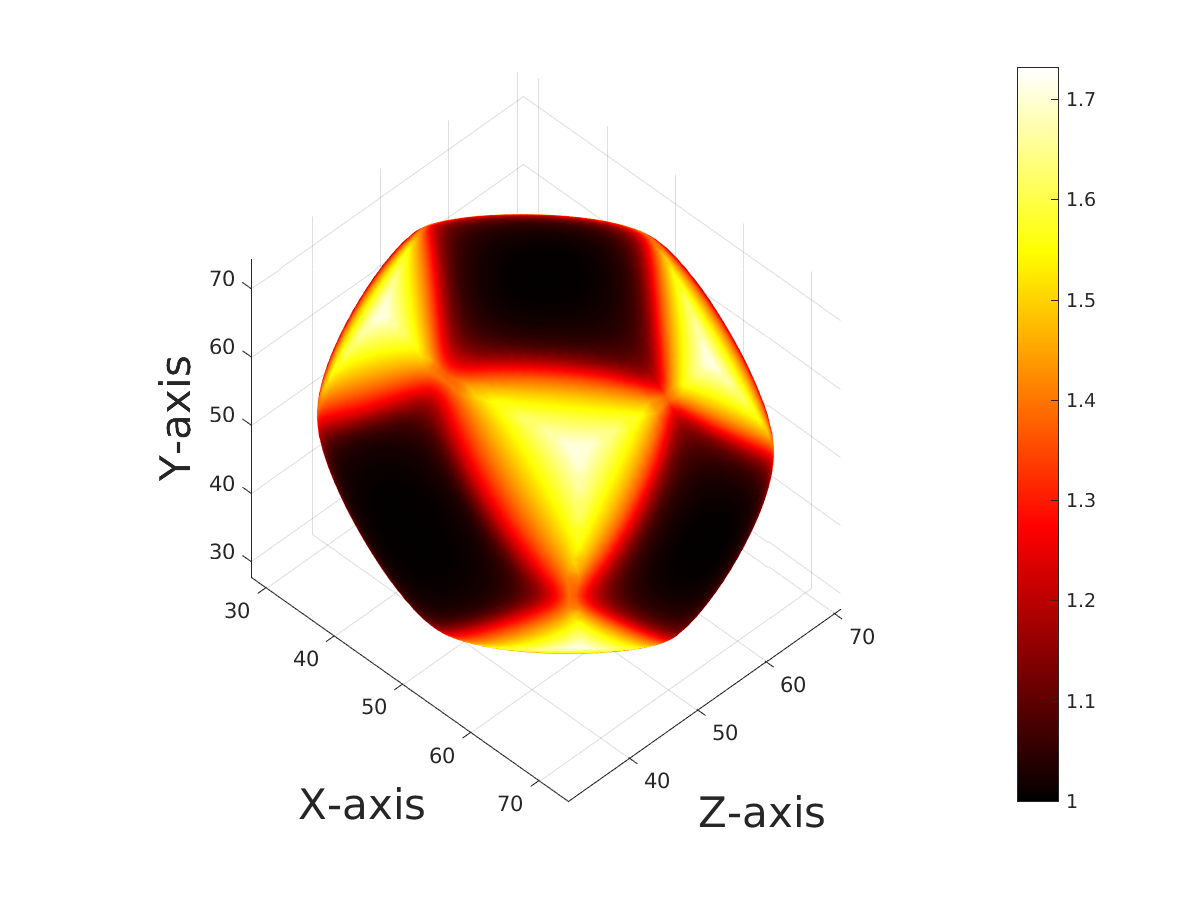}}
\caption{Tetragonal morphology with three facets obtained using sixth-rank gradient tensor terms. (a) represents a view 
from $\langle001\rangle$ direction and (b) represents a view from $\langle111\rangle$ direction.  The  
surface normal distribution for the prism morphology is shown in (c). }\label{F:3D_grad_6_3facet}
\end{figure}

If the parameters are chosen as $n_1 = 100.0$, $n_2 = 27.0$, $n_3= 500.0$, 
$n_4 = 1000.0$, $n_5 = 600$, $n_6 = -5900.0$, $p_1 = 1$ and $p_2 = 0.27$, 
the resultant free energy polynomial is as shown in Fig.~\ref{F:3DpolP6_3facet}. From
this polar plot as well as the Wulff construction on the 2-D sections ($xy$ and $xz$-sections 
perpendicular to $\langle 100 \rangle$ and $\langle 110 \rangle$), it is clear that
in this case, in addition to $(100)$ and $(111)$ planes, $(001)$ planes are also expected to form.
The morphology obtained from numerical simulations of a precipitate of size twelve
placed at the centre of the simulation cell and evolved to 360 time units is shown in Fig.~\ref{F:3D_grad_6_3facet}:
(a) is the view of the precipitate from $\langle 001 \rangle$ and (b) is the view of the precipitate from 
$\langle 111 \rangle$. From these figures as well as the surface normal plot shown in Fig.~\ref{F:3D_grad_6_3facet} (c),
it is clear that this morphology does consist of three families of planes -- $(100)$, $(001)$ and $(111)$.

\section{Conclusions} \label{Section6}

A wide variety of morphologies are observed in technologically important tetragonal systems 
(such as TiO$_2$ and Sn, for example).
We have developed a family of Extended Cahn-Hilliard (ECH) models for systems of tetragonal interfacial
free energy anisotropy; we have identified the non-zero and independent constants that enter the model 
(for fourth and sixth rank tensors) of gradient, curvature and aberration terms  as well as the constraints 
on these. We have numerically implemented the ECH model; and, we show that by appropriate choice of constants,
using our numerical implementation, it is possible to obtain prisms, plates, di-pyramids and their truncated 
variants (with two or three different crystallographic planes bounding the precipitate). Our formulation leads to 
precipitate morphologies that are consistent with the Wulff construction. We have characterised the precipitate 
morphologies using aspect ratios and surface normal plots. It is possible to extend our model by incorporating
elastic stress effects and we believe that such an extension will lead to interesting insights in morphological
evolution during solid-solid phase transformations in tetragonal systems. 

\section*{Acknowledgements}

We thank Industrial Research and Consultancy Centre, IIT Bombay 
for financial support (09IRCC16); and, PARAM-YUVA at CDAC, Pune, Nebula, Dendrite (DST-FIST HPC facility), Spinode,
and Leopard, GPU Centre of Excellence (GCoE), IIT Bombay for computational facilities. One of us (AR) 
thank DST for partial financial support (14DST017).

\bibliographystyle{tfq}
\bibliography{tPHMguide}

\end{document}